\documentclass[12pt]{article}
\usepackage{a4wide,axodraw,cite,graphicx} 

\usepackage{latexsym,amssymb,epsf} 

\usepackage{a4wide}
\newcommand{\newc}{\newcommand}
\newc{\gev}{\,GeV}
\newc{\sgn}{\mr{sgn}\,}
\newc{\ra}{\rightarrow}
\newc{\rpv}{$\mathrm{\not\!R_p}$}
\newc{\met}{$\not\!\!E_T$}
\newc{\rp}{$\mathrm{R_p}$}
\newc{\real}{\mathcal{R}e}
\newc{\alsm}{{\displaystyle \sum_{\alpha=1,2}}}
\newc{\besm}{{\displaystyle \sum_{\beta=1,2}}}
\newc{\al}{\alpha}
\newc{\be}{\beta}
\newc{\ga}{\gamma}
\newc{\de}{\delta}
\newc{\cw}{\cos\theta_w}
\newc{\ssw}{\sin^2\theta_w}
\newc{\ccw}{\cos^2\theta_w}
\newc{\cbe}{\cos\beta}
\newc{\sbe}{\sin\beta}
\newc{\sh}{\hat{s}}
\newc{\sa}{\sin\al}
\newc{\ca}{\cos\al}
\newc{\bv}{$\mathrm{\not\!B}$}
\newc{\lv}{$\mathrm{\not\!L}$}
\newc{\ie}{{\it i.e.\/}\ }
\newc{\lam}{\lambda}
\newc{\cht}{\tilde{\chi}}
\newc{\upt}{\tilde{u}}
\newc{\elt}{\tilde{\ell}}
\newc{\hgt}{\tilde{H}}
\newc{\nut}{\tilde{\nu}}
\newc{\dnt}{\tilde{d}}
\newc{\psb}{\bar{\psi}}
\newc{\rtt}{\sqrt{2}}
\newc{\mut}{\tilde{\mu}}
\newc{\mr}{\mathrm}
\newc{\bath}{\bar{\theta}}
\newc{\tht}{\theta}
\newc{\JC}{{\bf J}}
\newc{\lra}{\longrightarrow}
\newc{\eg}{{\it e.g.\,}}
\newc{\barr}{\begin{eqnarray}}
\newc{\earr}{\end{eqnarray}}
\newc{\beq}{\begin{equation}}
\newc{\eeq}{\end{equation}}
\newc{\me}{\mathcal{M}}
\newc{\dbm}{\partial_\mu}
\newc{\sgm}{\sigma_\mu}
\begin{document}
\begin{titlepage}
\vspace{-2cm}
\begin{flushright}
OUTP2000-32P\\
RAL-TR-2000-035\\
hep-ph/0007226\\
\end{flushright}
\vspace{+2cm}

\begin{center}
 {\Large{\bf Resonant Slepton Production in Hadron--Hadron Collisions}}\\
        \vskip 0.6 cm {\large{\bf 
               H. Dreiner$^*$\footnote{E-mail address: H.K.Dreiner@rl.ac.uk},
               P. Richardson$^{\dagger}$\footnote{E-mail address: 
                                       p.richardson1@physics.ox.ac.uk},
               and M. H. Seymour$^*$\footnote{E-mail address: 
                                M.Seymour@rl.ac.uk}}}
        \vskip 1cm
{{\it $^*$ Rutherford Appleton Laboratory, Chilton, Didcot OX11 0QX, U.K.}}\\
{{\it $^{\dagger}$ Department of Physics, Theoretical Physics,
                    University of Oxford,}}\\
{\it 1 Keble Road, Oxford OX1 3NP, United Kingdom}\\
\vskip 0.2 cm
\end{center}
\setcounter{footnote}{0}
\begin{abstract}\noindent
  We consider the resonant production of sleptons via \rpv\  in 
  hadron--hadron collisions followed by supersymmetric gauge decays of
  the sleptons. 
  We look at decay modes which lead to the production of a like-sign
  dilepton pair.
  The dominant production mechanism giving this signature is the resonant
  production of a charged slepton followed by a decay to a charged lepton 
  and a neutralino which then decays via \rpv. 
   The discovery potential of this process at 
  Run II of the Tevatron
  and the LHC is investigated using the HERWIG Monte Carlo event generator.
  We include the backgrounds from the MSSM.
  We conclude with a discussion of the
  possibility of extracting the lightest neutralino and slepton masses.
\end{abstract}
\end{titlepage}
%
%
\section{Introduction}

  In the R-parity violating (\rpv) extension of the Minimal
  Supersymmetric Standard Model (MSSM)
  \cite{Dreiner:1997uz:Bhattacharyya:1997vv:Barbier:1998fe}
  supersymmetric particles can be produced on resonance. While the
  cross sections for these resonant production processes are
  suppressed by the \rpv\  Yukawa couplings, the kinematic reach is
  greater than that of supersymmetric particle pair production.

  The \rpv\  extension to the MSSM contains the following additional terms
  in the superpotential 
\beq
  {\bf W_{\not R_p}} = 
  \frac{1}{2}\lam_{ijk}\varepsilon^{ab}L_{a}^{i}L_{b}^{j}\overline{E}^{k}
  + \lam_{ijk}'\varepsilon^{ab}L_{a}^{i}Q_{b}^{j}\overline{D}^{k} +
  \frac{1}{2}\lam_{ijk}''\varepsilon^{\al\be\gamma}\overline{U}_{\al}^{i}
  \overline{D}_{\be}^{j}\overline{D}_{\gamma}^{k} + \kappa_iL_{i}H_2,
\label{eqn:super} 
\eeq
  where $i,j=1,2,3$ are the generation indices, $a,b=1,2$ are the $SU(2)_L$
  indices
  and $\al,\be,\gamma=1,2,3$ are the $SU(3)_C$ indices. $L^i$ ($Q^i$) are the
  lepton (quark) $SU(2)$ doublet superfields, $\overline{E}^{i}$
  ($\overline{D}^{i}, \overline{U}^{i}$) are the electron (down and up
  quark) $SU(2)$ singlet superfields, and $H_n,$ $n=1,2$, are the Higgs
  superfields. For a recent summary of
  the bounds on the couplings in Eqn.\,\ref{eqn:super} see
  \cite{Allanach:1999ic}. 

  The terms in Eqn.\,\ref{eqn:super} lead to different resonant
  production mechanisms in various collider experiments. The first term
   leads to resonant sneutrino production in $\mr{e^+e^-}$ collisions 
  \cite{Dimopoulos:1988jw,Barger:1989rk,Dreiner:1992ur,Giudice:1996dm,
	Accomando:1997wt,Erler:1997ww,Kalinowski:1997bc},
  while the third term gives resonant
  squark production in hadron--hadron collisions 
  \cite{Dimopoulos:1990fr,Dreiner:1991pe,Dreiner:1999qz,Allanach:1999bf,
	Datta:1997us,
	Yang:1997uw,Oakes:1998zg,Berger:1999zt}.
  The second term gives both resonant 
  squark production in $\mr{ep}$ collisions \cite{Butterworth:1993tc}
  and resonant slepton production in hadron--hadron collisions,
  which we will consider here.

%
%
\begin{table}[hbp]
\renewcommand{\arraystretch}{1.2}
\begin{center}
\begin{tabular}{|c|c|c|}
\hline
 & Charged Sleptons   & Sneutrinos  \\
\hline
 Supersymmetric  & $\mr{\elt_{i\al} 	\ra	\ell^{-}_{i} 	\cht^0	}$ 
 	 	 & $\mr{\nut_i 		\ra 	\nu_i 		\cht^0	}$ \\
  Gauge Decays	 & $\mr{\elt_{i\al} 	\ra	\nu_i	 	\cht^-	}$
 		 & $\mr{\nut_i 		\ra	\ell^-_i 	\cht^+	}$\\
\hline
 \rpv\  Decays	 & $\mr{\elt_{i\al} 	\ra	\bar{u}_j 	d_k	}$ 
		 & $\mr{\nut_i 		\ra	\bar{d}_j 	d_k	}$ \\
		 & $\mr{\elt_{i\al} 	\ra	\bar{\nu}_j 	\ell^-_k}$ 
 		 & $\mr{\nut_i 		\ra	\ell^+_j 	\ell^-_k}$ \\
\hline
 Weak Decays	 & $\mr{\elt_{i\al} 	\ra	\nut_i		W^-	}$
 		 & $\mr{\nut_i		\ra	\elt_{i\al}	W^+	}$ \\
		 & $\mr{\elt_{i2}	\ra	\elt_{i1}	Z_0	}$
		 & \\
\hline
 Higgs Decays	 & $\mr{\elt_{i\al}	\ra	\nut_i		H^-	}$
 		 & $\mr{\nut_i		\ra	\elt_{i\al}	H^+	}$ \\
		 & $\mr{\elt_{i2}	\ra	\elt_{i1}	h_0,H_0,A_0}$
 		 & \\
\hline	
\end{tabular}
\caption{Decay modes of charged sleptons and sneutrinos. The index 
	 $\al=1,2$ gives the mass eigenstate of the slepton.}
\label{tab:decaymodes}
\end{center}
\end{table}
  
  A systematic study of \rpv\  signatures at hadron colliders was first
  performed in \cite{Dreiner:1991pe}.  Resonant slepton production in
  hadron--hadron collisions has previously been considered in
  \cite{Dimopoulos:1990fr,Dreiner:1991pe,
  Kalinowski:1997zt,Hewett:1998fu,
  Allanach:1999bf,Dreiner:1999qz,Dreiner:1998gz:Dreiner:2000qf,
  Moreau:1999bt:Moreau:2000ps:Moreau:2000bs,Abdullin:1999zp}.
  The signature of this
  process depends on the decay mode of the resonant slepton. The
  various possible decay channels are given in
  Table\,\ref{tab:decaymodes}. Most of the previous studies have
  considered only the \rpv\  decays of the resonant slepton to either
  leptons via the first term in Eqn.\,\ref{eqn:super}
  \cite{Dimopoulos:1990fr,Kalinowski:1997zt,Hewett:1998fu,Allanach:1999bf},
  or to quarks via the second term in Eqn.\,\ref{eqn:super}
  \cite{Dimopoulos:1990fr,Hewett:1998fu,Allanach:1999bf,Dreiner:1999qz}.
  There has been no study of the supersymmetric gauge decays of the
  resonant charged sleptons. The cross sections for these processes
  were first presented in \cite{Dimopoulos:1990fr} where there was a
  discussion of the possible experimental signatures, however the
  signal we are considering was not discussed and there was no
  calculation of the Standard Model background.  In several workshop
  contributions we have presented first studies of the supersymmetric
  gauge decays of charged sleptons
  \cite{Allanach:1999bf,Dreiner:1998gz:Dreiner:2000qf,Abdullin:1999zp}.
  Here we present a complete analysis of both the Tevatron and LHC case.
  The supersymmetric gauge decays of sneutrinos have been studied in
  \cite{Moreau:1999bt:Moreau:2000ps:Moreau:2000bs,Abdullin:1999zp}.
  These studies were performed using a hadron-level Monte Carlo
  simulation for both the signal and background processes, in addition
  the analyses of
  \cite{Moreau:1999bt:Moreau:2000ps:Moreau:2000bs,Abdullin:1999zp}
  used a detector simulation and looked at the trilepton signature for
  resonant sneutrino production.

  We will consider a specific signature, \ie like-sign dilepton production,
  for these processes rather than any
  one given resonant production mechanism.
  We would expect like-sign dilepton production
  to have a low background from 
  Standard Model (SM) processes. This is an extension of our analysis of 
  \cite{Allanach:1999bf,Dreiner:1998gz:Dreiner:2000qf,Abdullin:1999zp}
  to include additional
  signal processes and Standard Model backgrounds.
  We also consider the background from sparticle pair production
  and the possibility of
  measuring the sparticle masses which was not considered in 
  \cite{Allanach:1999bf,Dreiner:1998gz:Dreiner:2000qf,Abdullin:1999zp}. 
  We use a parton-shower
  Monte Carlo simulation, including hadronization but no detector simulation,
  for both the signal and background processes.

  In Section~\ref{sec:signal} we will consider the signal processes in more 
  detail, followed by a discussion of the various different processes which
  contribute to the background in 
  Section~\ref{sec:backgrounds}. We will also discuss the
  various cuts which can be used to reduce the background. In
  Section~\ref{sec:results} we will then consider the discovery potential
  at both Run II of the Tevatron and at the LHC. We also
  consider the possibility of reconstructing the neutralino and
  slepton masses using their decay products.

%
%
\section{Signal}
\label{sec:signal}

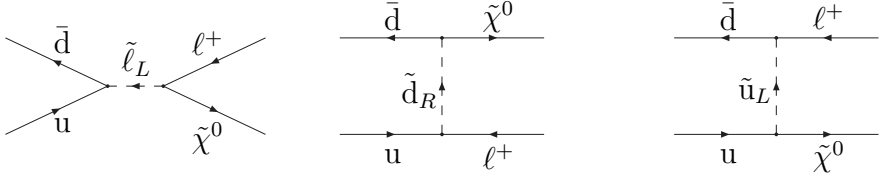
\begin{figure}
\begin{center}
\begin{picture}(360,60)(0,30)
\SetScale{0.7}
\SetOffset(0,-25)
\ArrowLine(5,76)(60,102)
\ArrowLine(60,102)(5,128)
\DashArrowLine(90,102)(60,102){5}
\ArrowLine(90,102)(145,76)
\ArrowLine(145,128)(90,102)
\Text(80,52)[]{$\mathrm{\tilde{\chi}^{0}}$}
\Text(80,88)[]{$\mathrm{\ell^{+}}$}
\Text(25,89)[]{$\mathrm{\bar{d}}$}
\Text(25,57)[]{$\mathrm{u}$}
\Text(53,82)[]{$\mathrm{\tilde{\ell}}_{L}$}
\Vertex(60,102){1}
\Vertex(90,102){1}
\SetScale{0.7}
\ArrowLine(240,128)(185,128)
\ArrowLine(240,128)(295,128)
\ArrowLine(185,76)(240,76)
\ArrowLine(295,76)(240,76)
\DashArrowLine(240,76)(240,128){5}
\Text(190,98)[]{$\mathrm{\tilde{\chi}^{0}}$}
\Text(190,45)[]{$\mathrm{\ell^{+}}$}
\Text(150,45)[]{$\mathrm{u}$}
\Text(150,98)[]{$\mathrm{\bar{d}}$}
\Text(160,70)[]{$\mathrm{\tilde{d}}_{R}$}
\Vertex(240,128){1}
\Vertex(240,76){1}
\ArrowLine(420,128)(365,128)
\ArrowLine(475,128)(420,128)
\ArrowLine(365,76)(420,76)
\ArrowLine(420,76)(475,76)
\DashArrowLine(420,76)(420,128){5}
\Text(315,45)[]{$\mathrm{\tilde{\chi}^{0}}$}
\Text(315,98)[]{$\mathrm{\ell^{+}}$}
\Text(277,98)[]{$\mathrm{\bar{d}}$}
\Text(277,45)[]{$\mathrm{u}$}
\Text(287,70)[]{$\mathrm{\tilde{u}}_{L}$}
\Vertex(420,128){1}
\Vertex(420,76){1}
\end{picture}
\end{center}
\caption{Production of $\tilde{\chi}^{0}\ell^{+}$.}
\label{fig:cross}
\end{figure}

\begin{figure}
\begin{center}
\begin{picture}(360,35)(0,0)
\SetScale{0.7}
\SetOffset(0,-30)
\ArrowLine(5,78)(60,78)
\ArrowLine(105,105)(60,78)
\ArrowLine(84,53)(129,26)
\ArrowLine(129,80)(84,53)
\DashArrowLine(60,78)(84,53){5}
\Text(25,63)[]{$\mathrm{\tilde{\chi^{0}}}$}
\Text(55,70)[]{$\mathrm{\ell^{+}}$}
\Text(75,20)[]{$\mathrm{d}$}
\Text(75,54)[]{$\mathrm{\bar{u}}$}
\Text(45,40)[]{$\mathrm{\tilde{\ell}}_L$}
\Vertex(60,78){1}
\Vertex(84,53){1}
\ArrowLine(185,78)(240,78)
\ArrowLine(285,105)(240,78)
\ArrowLine(264,53)(309,26)
\ArrowLine(309,80)(264,53)
\DashArrowLine(240,78)(264,53){5}
\Text(150,63)[]{$\mathrm{\tilde{\chi}^{0}}$}
\Text(200,54)[]{$\mathrm{\ell^{+}}$}
\Text(200,20)[]{$\mathrm{d}$}
\Text(180,70)[]{$\mathrm{\bar{u}}$}
\Text(170,40)[]{$\mathrm{\tilde{u}}_L$}
\Vertex(240,78){1}
\Vertex(264,53){1}
\ArrowLine(365,78)(420,78)
\ArrowLine(420,78)(465,105)
\ArrowLine(489,26)(444,53)
\ArrowLine(489,80)(444,53)
\DashArrowLine(444,53)(420,78){5}
\Text(277,63)[]{$\mathrm{\tilde{\chi}^{0}}$}
\Text(330,20)[]{$\mathrm{\ell^{+}}$}
\Text(310,70)[]{$\mathrm{d}$}
\Text(330,54)[]{$\mathrm{\bar{u}}$}
\Text(300,40)[]{$\mathrm{\tilde{d}}_R$}
\Vertex(420,78){1}
\Vertex(444,53){1}
\end{picture}
\end{center}
\caption{Feynman diagrams for the decay 
	$\mathrm{{\tilde\chi}^0}\ra\ell^+\mr{d}\mr{\bar{u}}$.
	The neutralino is a Majorana
	fermion and decays to the charge conjugate final state as well.
	There is a further decay mode $\mr{\cht^0\ra\nu d\bar{d}}$.}
\label{fig:decay}
\end{figure}
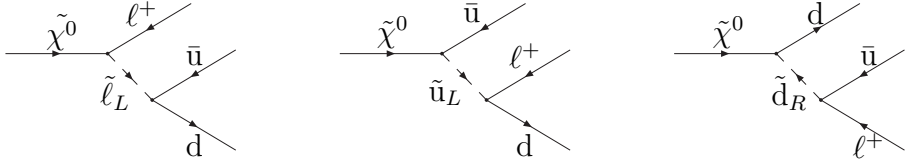

  There are a number of different possible production mechanisms for a
  like-sign
  dilepton pair via resonant slepton production. The dominant production
  mechanism is the production of a charged slepton followed by a 
  supersymmetric gauge decay
  of the charged slepton to a neutralino and
  a charged lepton. This neutralino can then decay via the crossed process
  to give a second charged lepton, which due to the Majorana nature
  of the neutralino
  can have the same charge as the lepton produced in the slepton decay.
  The production of a charged lepton and a neutralino via the LQD term in
  the \rpv\  superpotential, Eqn.\,\ref{eqn:super}, occurs at tree-level 
  via the Feynman
  diagrams given in Fig.\,\ref{fig:cross}. 
  The decay of the neutralino occurs at tree-level via the diagrams
  given in 
  Fig.\,\ref{fig:decay}.
  
  Like-sign dileptons can also be produced in resonant charged slepton
  production with a supersymmetric gauge decay of the slepton to a
  chargino and neutrino, $\mr{\elt}^+\ra\mr{\cht}^+_1\mr{\nu_\ell}$.
  The chargino can then decay $\mr{\cht}^+_1\ra\mr{\ell}^+\mr{\nu_
  \ell}\mr{\cht}^0_1$.  Again, given the majorana nature of the
  neutralino, it can decay to give a like-sign dilepton pair.

  The production of like-sign dileptons is also possible in resonant
  sneutrino production  followed by a supersymmetric 
  gauge decay to a chargino and a charged lepton,
  $\nut\ra\ell^-\cht^+_1$. This can be followed by 
  $\mr{\cht}^+_1\ra \mr{q}\mr{\bar{q}'}\mr{\cht}^0_1$,
  the neutralino can then decay as in Fig.\,\ref{fig:decay} to give 
  a like-sign dilepton pair. 

  All the resonant \rpv\  production mechanisms and the decays of the SUSY
  particles
  have been included in the HERWIG event generator 
  \cite{Corcella:1999qn:Marchesini:1991ch}. The implementation of both 
  R-parity conserving and R-parity violating SUSY is described in
  \cite{SUSYimplement},
  the matrix elements used for the various \rpv\  processes are given in
  \cite{Dreiner:1999qz}. 
 
  We will only consider one of the \rpv\  Yukawa couplings to be
  non-zero at a time. We shall focus on ${\lam'}_{211}$, which leads
  to resonant smuon production. The production cross-section depends
  quadratically on the \rpv\  Yukawa coupling. The low-energy bound is
  given by
\beq
{\lam'}_{211} < 0.059 \times \left(\frac{M_{\dnt_R}}{100 \mr{GeV}}\right),
\label{eqn:bound211}
\eeq
  from the ratio $R_\pi=\Gamma(\mr{\pi\ra e\nu_e})/\Gamma(\mr{\pi\ra
  \mu\nu_\mu})$ \cite{Allanach:1999ic,Barger:1989rk}. The bound on the
  coupling ${\lam'}_{111}$ from neutrino-less double beta decay
  \cite{Allanach:1999ic,Hirsch:1995zi:Hirsch:1996ek:Babu:1995vh} is
  very strict and basically excludes an observable signal. For higher 
  generation couplings $\lam'_{2ij}$ the cross section is suppressed by
  low parton luminosities. We do not consider the production of $\tau$
  leptons.

  As we are considering a dominant ${\lam}'_{211}$ coupling the
  leptons produced in the neutralino decays and the hard processes
  will be muons. We will therefore require throughout that both
  leptons are muons because this reduces the background, where
  electrons and muons are produced with equal probability, with
  respect to the signal.  This typically reduces the Standard Model
  background by a factor of four while leaving the dominant signal
  process almost unaffected.  It will lead to some reduction of the
  signal from channels where some of the leptons are produced in
  cascade decays from the decay of a W or Z boson.
  
%
%

  The signal has a number of features, in addition to the presence of a
  like-sign dilepton pair, which will enable us to extract it above the
  background: 
\begin{itemize}
\item   Provided that the difference between the slepton and
	the neutralino/chargino masses is large enough both 
	the leptons will have a high transverse momentum, $p_T$,
 	and be well isolated.
\item 	As the neutralino decays inside the detector, for this signature,
	there
	will be little missing transverse energy, \met, in the event. Any
	\met\  will come from semi-leptonic hadron decays or from cascade
	decays following the production of a chargino or 
        one of the heavier neutralinos.
\item 	The presence of a third lepton can only come from semi-leptonic
	hadron decays, or in SUSY cascade decays if a chargino or
	one of the heavier
	neutralinos is produced.
\item 	The presence of two hard jets from the decay of the neutralino.
\end{itemize}

  The cross section for the signal processes and the 
  acceptance\footnote{We define the acceptance to be the
	fraction of signal events which
  pass the cuts.}
  will depend upon the
  various SUSY parameters. Due to the large number of parameters we will
  use the standard minimal SUGRA scenario where the soft SUSY breaking
  masses for the
  gauginos ($M_{1/2}$) and scalars ($M_0$), and the tri-linear SUSY
  breaking terms ($A_0$)
  are universal at the GUT scale. In addition we require radiative 
  electroweak symmetry breaking. This leaves five parameters 
  $M_{1/2}$, $M_0$, $A_0$, $\tan\beta$ and $\sgn\mu$. 

%
%
 
\begin{figure}
\begin{center}
\includegraphics[angle=90,width=0.48\textwidth]{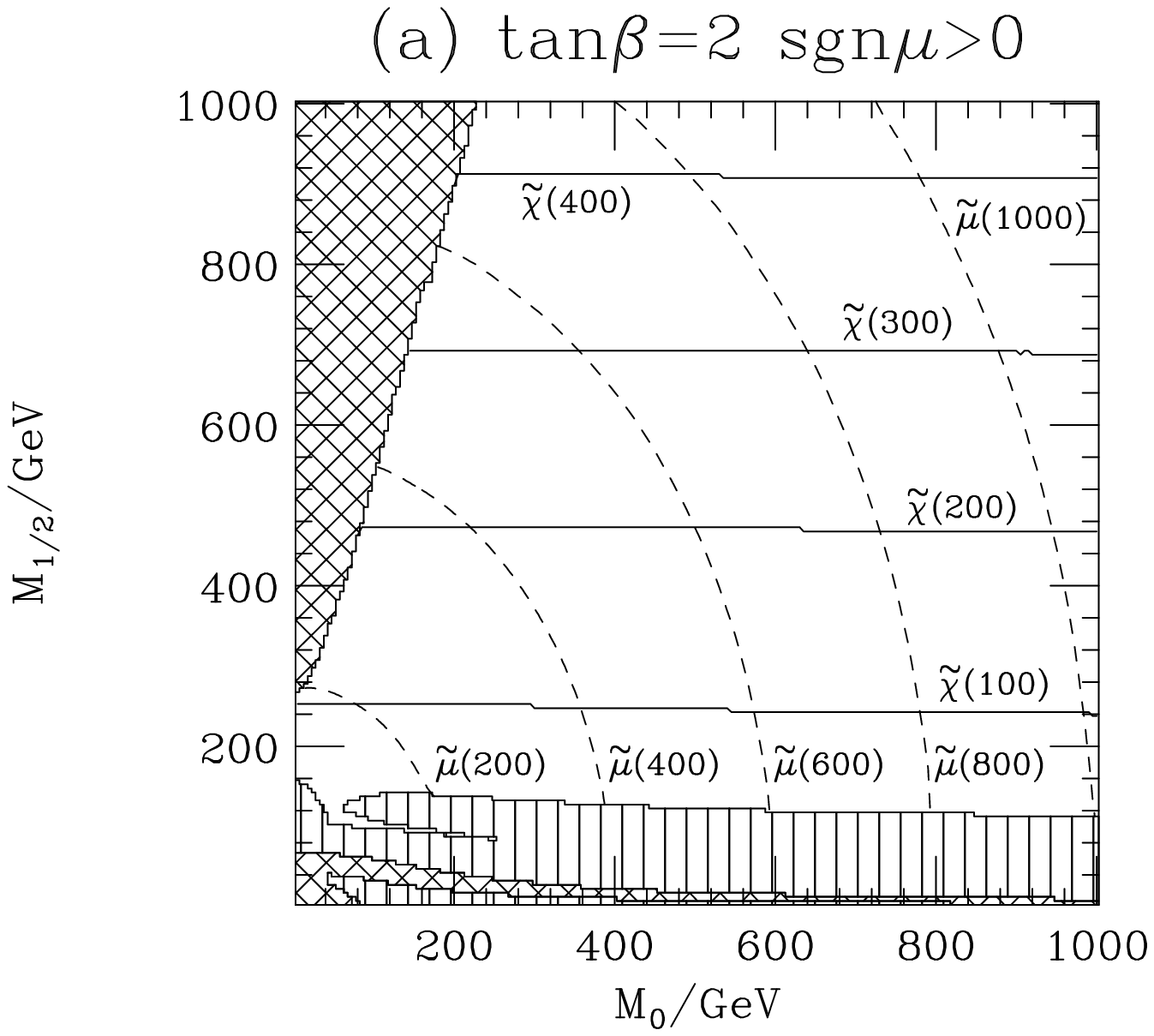}
\hfill
\includegraphics[angle=90,width=0.48\textwidth]{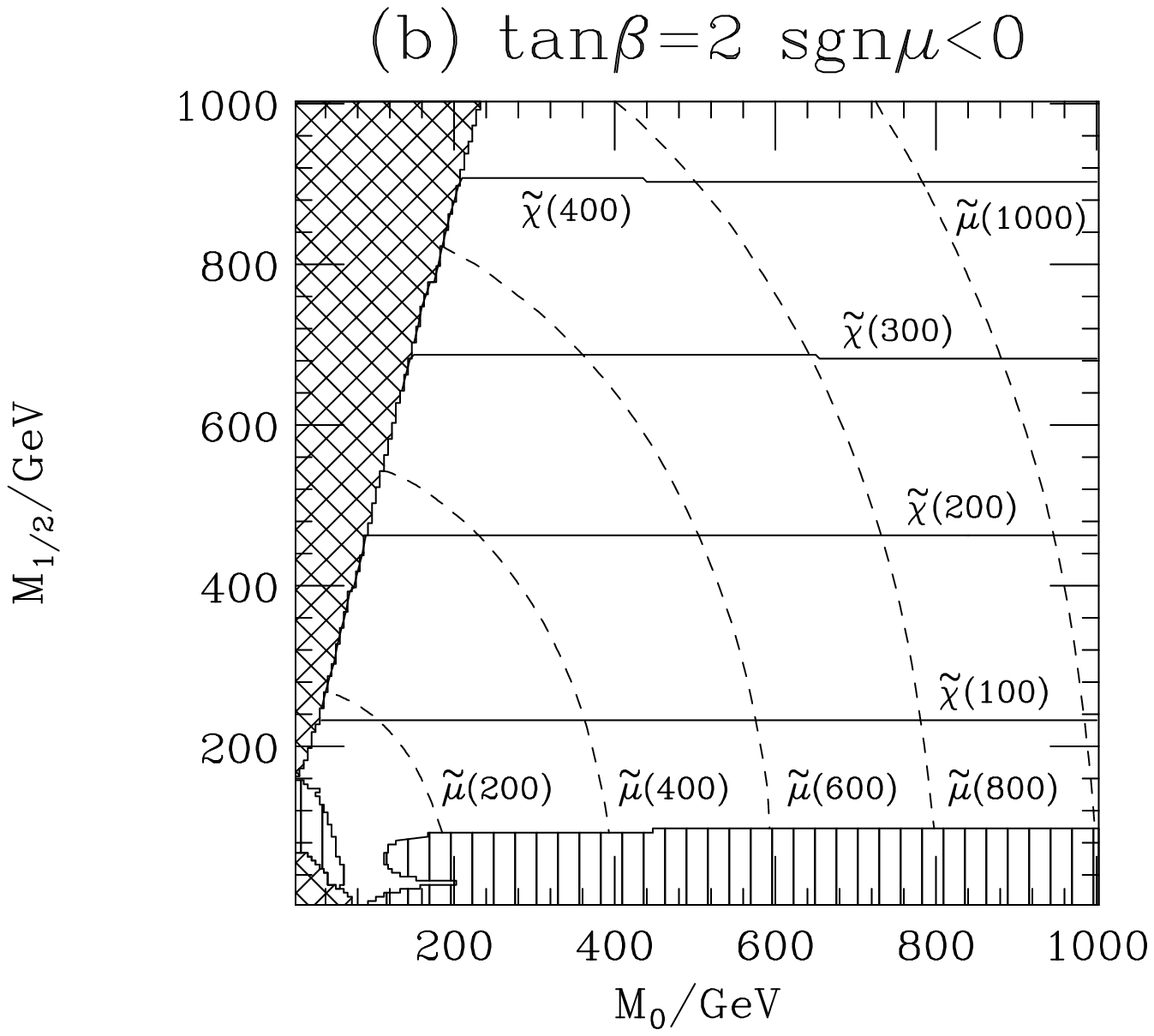}\\
\vskip 7mm
\includegraphics[angle=90,width=0.48\textwidth]{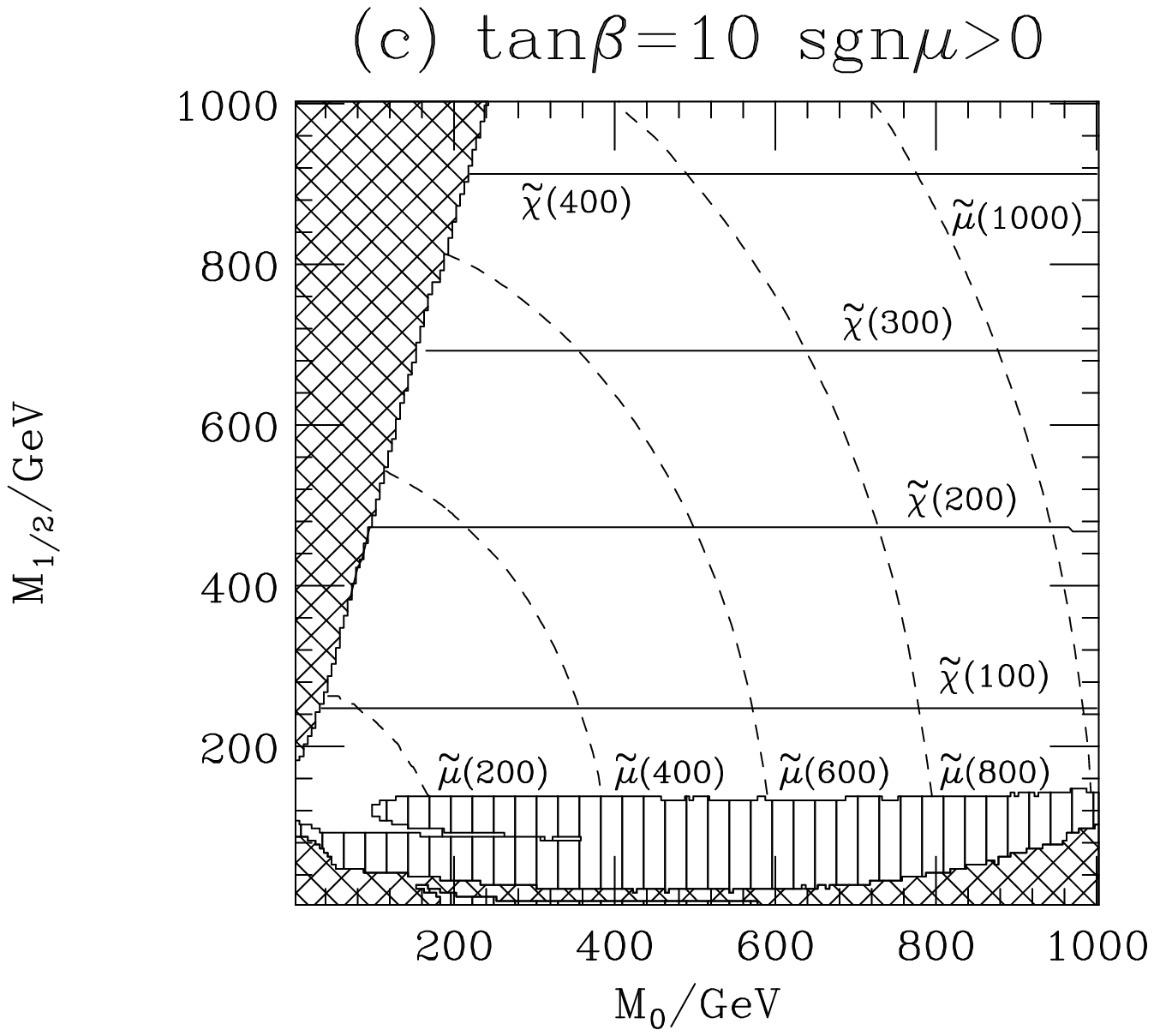}
\hfill
\includegraphics[angle=90,width=0.48\textwidth]{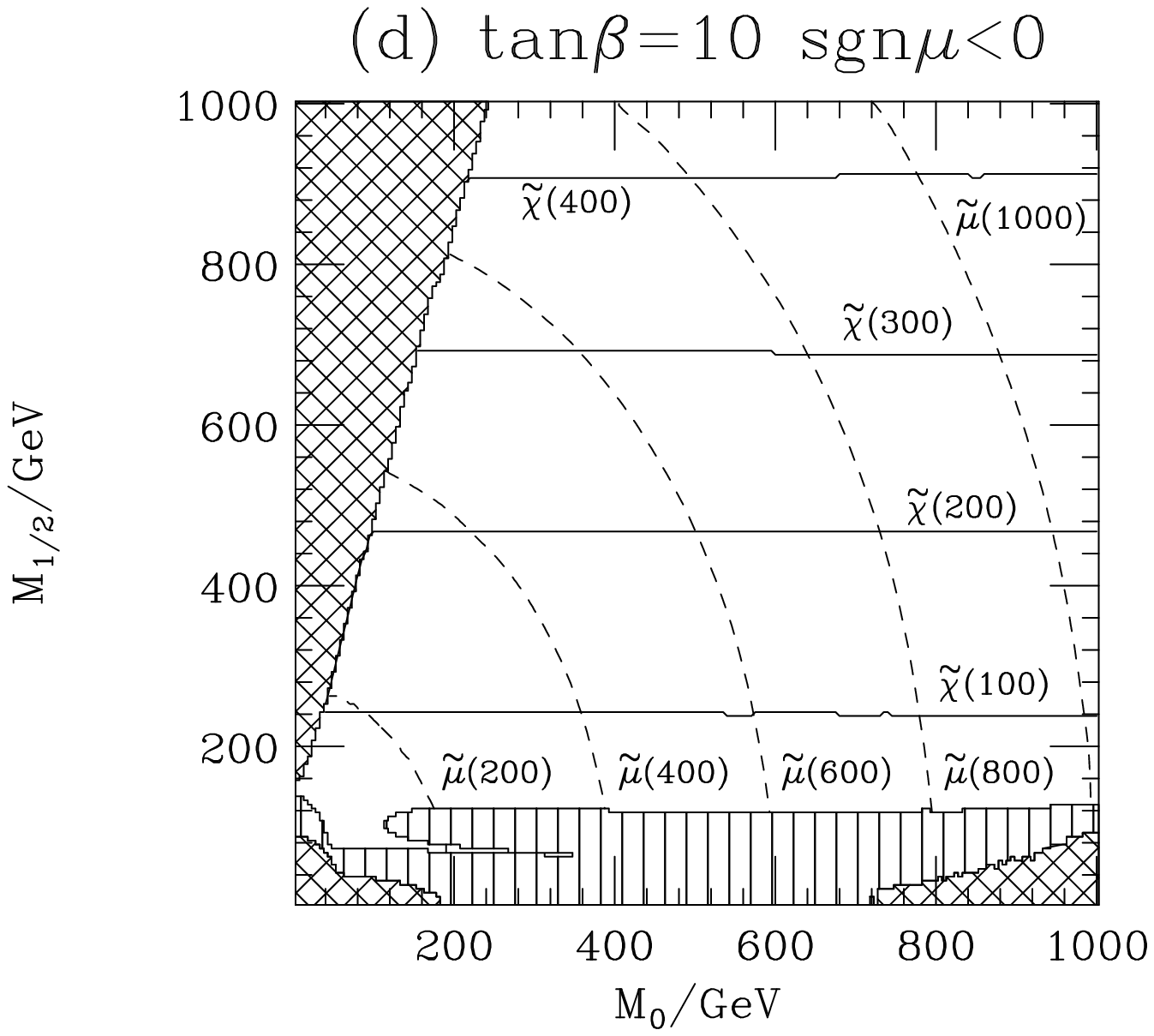}\\
\caption{Contours showing the lightest neutralino mass, solid lines, and the
	$\mr{\tilde{\mu}}_L$ mass, dashed lines, in the $M_0$, $M_{1/2}$
	plane with 
 	$A_0=0\, \mr{\gev}$  for different values
	of $\tan\beta$ and $\sgn\mu$.
	The hatched regions at small $M_0$ are excluded by the requirement
	that the $\mr{\cht^0_1}$ be the LSP.
	The region at large $M_0$ and $\tan\beta$
  	is excluded because there is no radiative 
	electroweak symmetry breaking. 
  	The vertically striped region is excluded by the LEP experiments.
	This was obtained
	using the limits on the chargino \cite{Abbiendi:1998ff} and smuon 
	\cite{Abbiendi:1999is} production cross sections,
	and the chargino mass \cite{Barate:1998gy}. 
        This analysis was performed using ISAJET 7.48 \cite{Baer:1999sp}.} 
\label{fig:SUSYmass}
\end{center}
\end{figure}

  We have performed a scan in $M_0$ and $M_{1/2}$ with $A_0=0\, \mr{\gev}$  
  for two different values of $\tan\beta$ and both values of
  $\sgn\mu$. The masses of the left-handed smuon and the lightest neutralino
  are shown in Fig.\,\ref{fig:SUSYmass}.  There are
  regions in these plots which we have not considered either due to the lack
  of radiative electroweak symmetry breaking, or because the lightest
  neutralino is not the lightest supersymmetric particle (LSP). In the
  MSSM, the LSP must be a neutral colour singlet \cite{Ellis:1984ew}, from
  cosmological bounds on electric- or colour-charged stable
  relics. However if R-parity is violated the LSP can decay and these
  bounds no longer apply. We should therefore consider cases where one
  of the other SUSY
  particles is the LSP. We have only considered the case where the neutralino
  is the LSP for two reasons:
\begin{enumerate}
\item Given the unification of the SUSY breaking parameters at the GUT scale
	it is hard to find points in parameter space where the lightest 
	neutralino 
	is not the LSP without the lightest neutralino becoming heavier
	than the 
	sleptons, which tend to be the lightest sfermions in these models.
	If the neutralino is heavier than the sleptons the resonance will not
 	be accessible for
 	the supersymmetric 
	gauge decay modes we are considering and the slepton will decay
	via \rpv\  modes.

\item The ISAJET code for the running of the couplings and the calculation 
      of the MSSM decay modes only works when the neutralino is the LSP.
\end{enumerate}

%
%
  The plots in Fig.\,\ref{fig:SUSYmass}
  also include the current experimental limits on the SUSY parameters 
  from LEP. This experimentally excluded region comes from two sources: the
  region at large $M_0$ is excluded by the limit on the cross section for
  chargino pair production from \cite{Abbiendi:1998ff} and the limit on the
  chargino mass from \cite{Barate:1998gy}; the region at small $M_0$ is
  excluded by the limit on the production of smuons from 
  \cite{Abbiendi:1999is}.
  There is also a limit on the neutralino production cross section from 
  \cite{Abbiendi:1998ff}, however for most of the SUGRA parameter space this
  is weaker than the limit on chargino pair production. The gap in the
  excluded region between $M_0$ of about $50\, \mr{\gev}$  and
  $100\, \mr{\gev}$ is due to the presence
  of a destructive interference between the $t$-channel
  sneutrino exchange and the $s$-channel photon and Z exchanges in
  the chargino production cross section in $\mr{e^+e^-}$ collisions.

%
%
\begin{figure}
\begin{center}
\includegraphics[angle=90,width=0.48\textwidth]{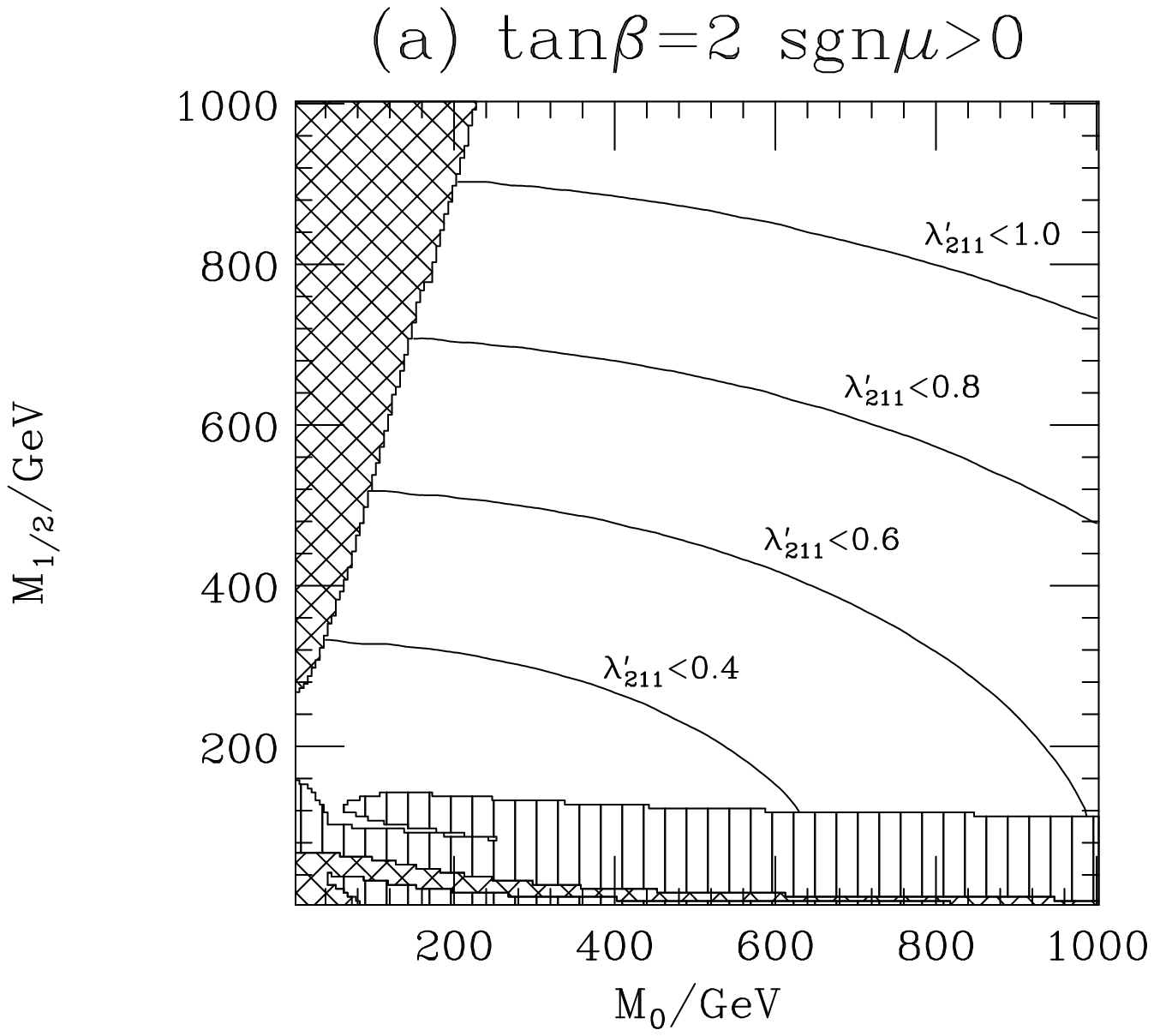}
\hfill
\includegraphics[angle=90,width=0.48\textwidth]{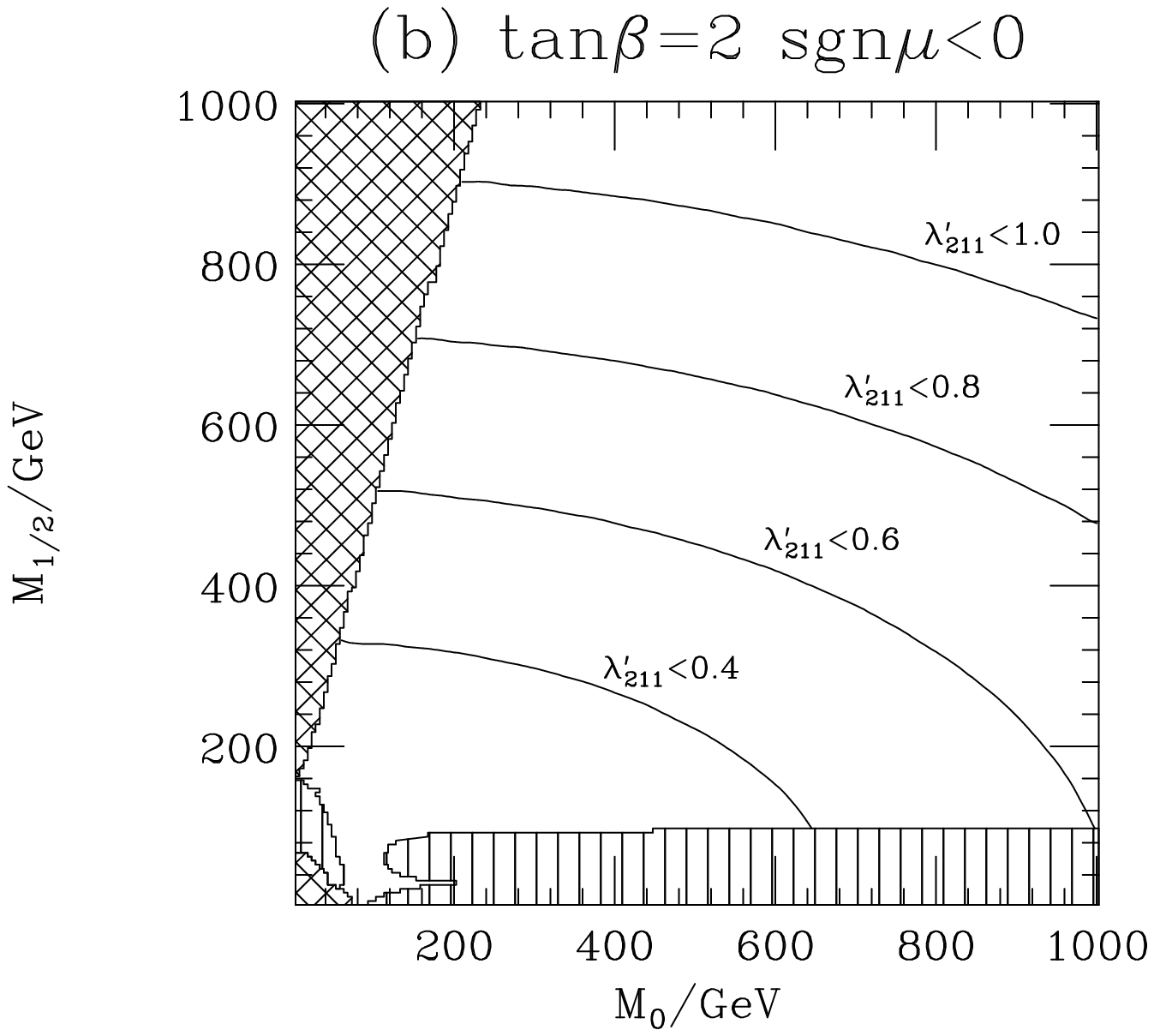}\\
\vskip 7mm
\includegraphics[angle=90,width=0.48\textwidth]{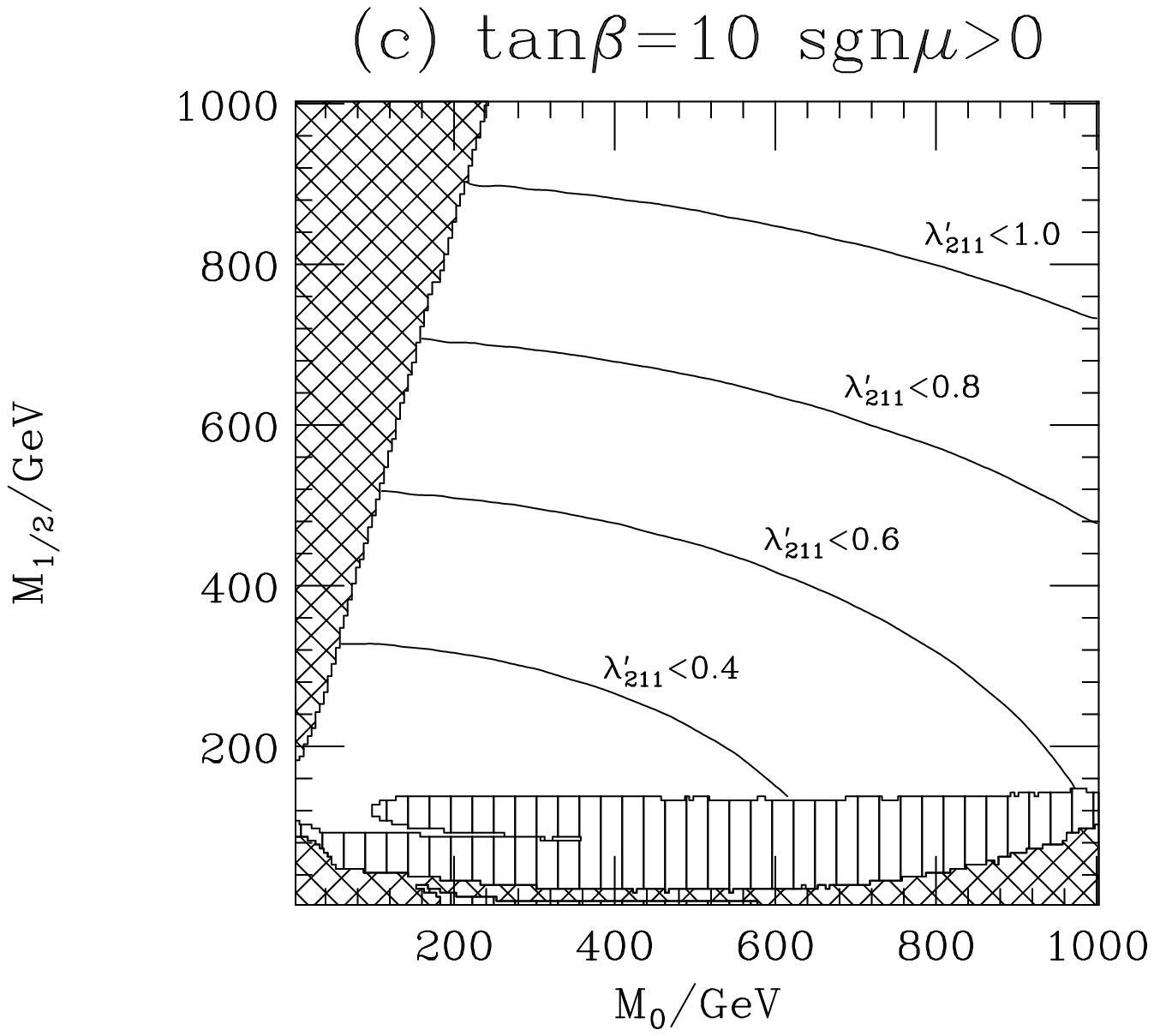}
\hfill
\includegraphics[angle=90,width=0.48\textwidth]{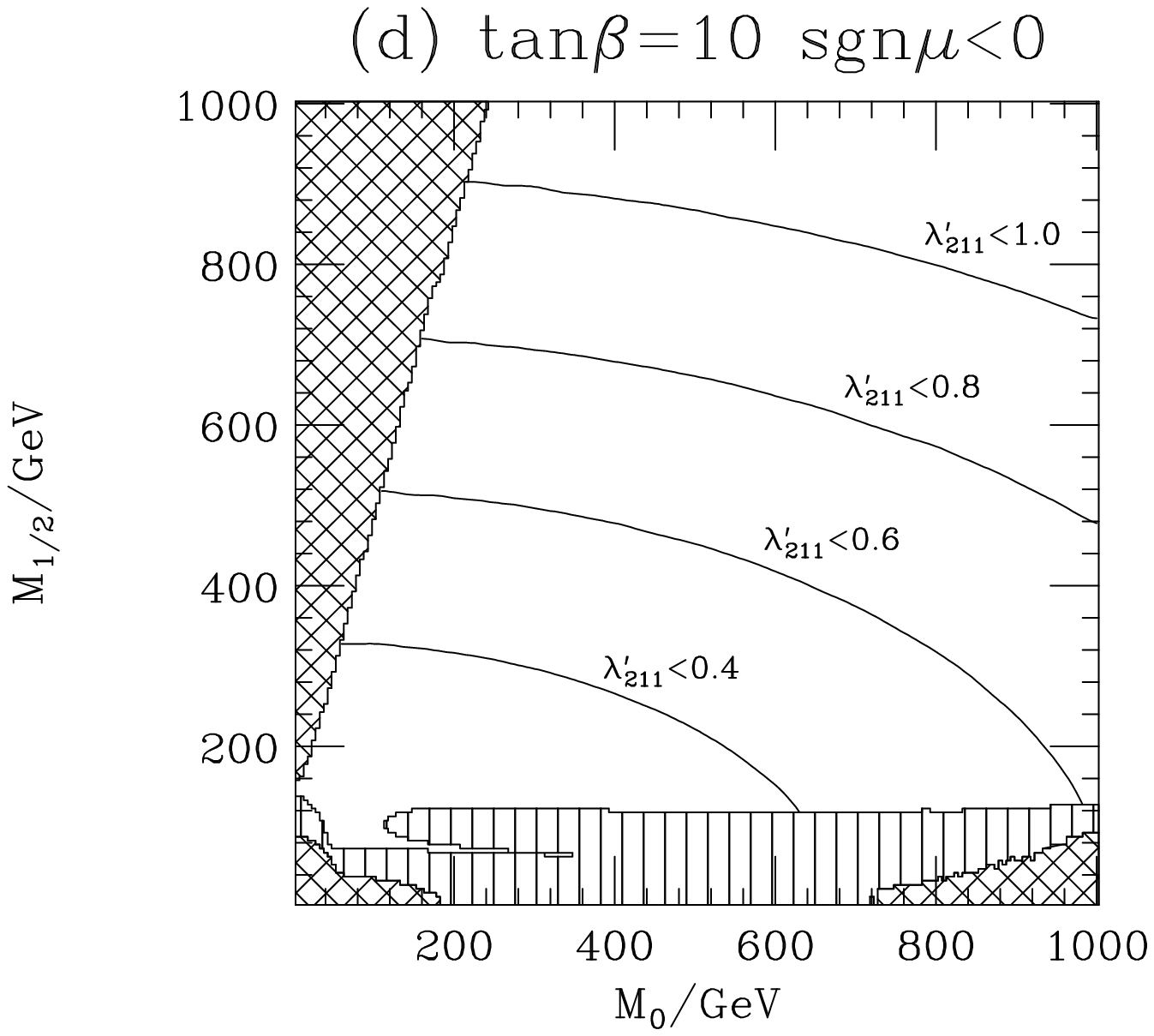}\\
\caption{Contours showing the limit on the \rpv\  Yukawa coupling
 	 ${\lam'}_{211}$ in the 
	 $M_0$, $M_{1/2}$ plane for $A_0=0\, \mr{\gev}$  
	 and different values of $\tan\beta$ and  $\sgn\mu$. The
	 striped and hatched regions are the same as in
	 Fig.\,\ref{fig:SUSYmass}.} 
\label{fig:SUSYlimit}
\end{center}
\end{figure}

  The limit on the coupling ${\lam'}_{211}$ is shown in
  Fig.\,\ref{fig:SUSYlimit}. As can be seen from Figs.\,\ref{fig:SUSYmass}
  and 
  \ref{fig:SUSYlimit} the limit on the coupling is fairly weak for large
  regions of parameter space, even when the smuon is relatively light.
  This is
  due to the squark masses, upon which the limit depends, being larger than
  the slepton masses, in the SUGRA models ({\it c.f.} Eq.\ref{eqn:bound211}).

\begin{figure}
\begin{center}
\includegraphics[angle=90,width=0.48\textwidth]{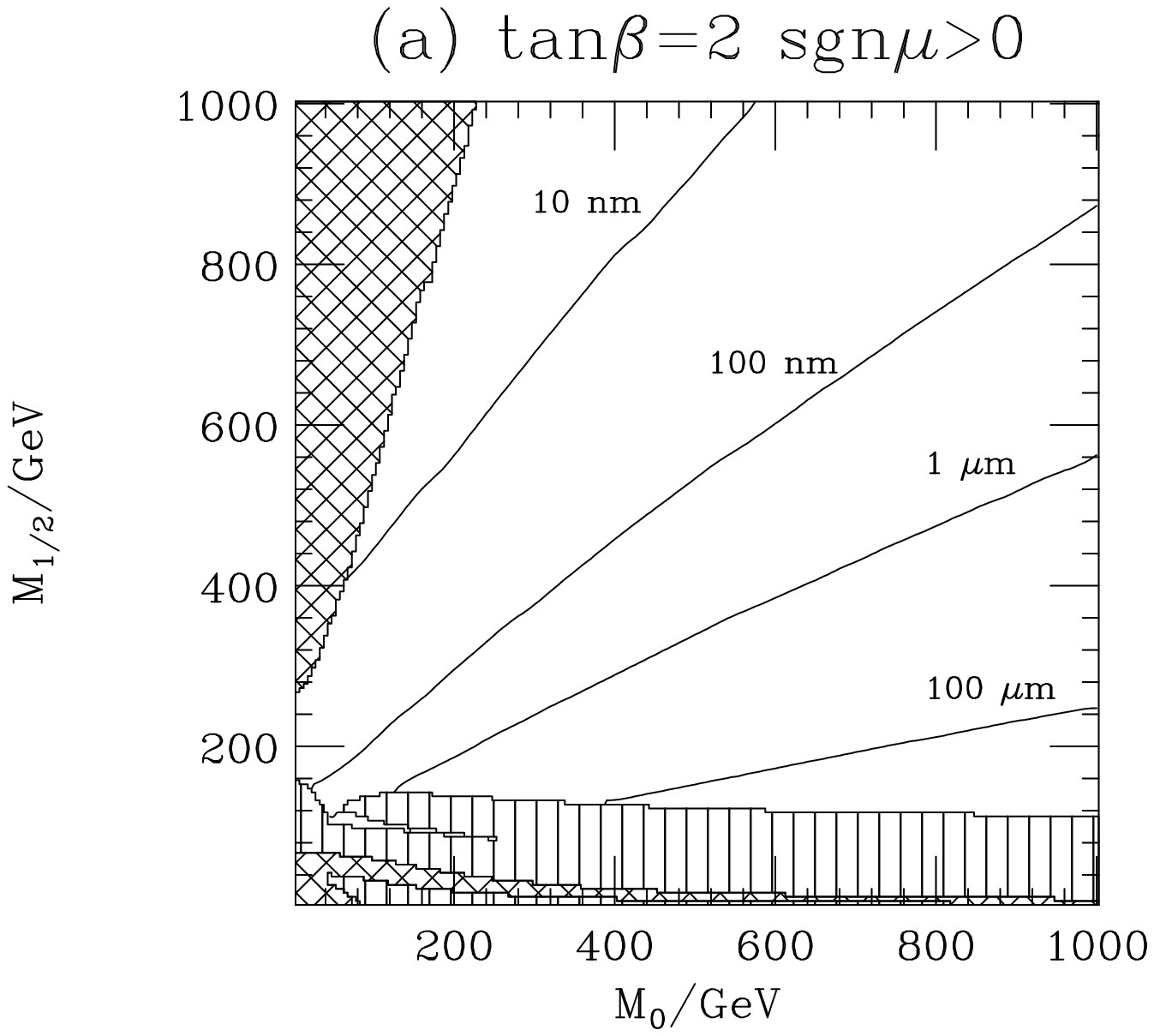}
\hfill
\includegraphics[angle=90,width=0.48\textwidth]{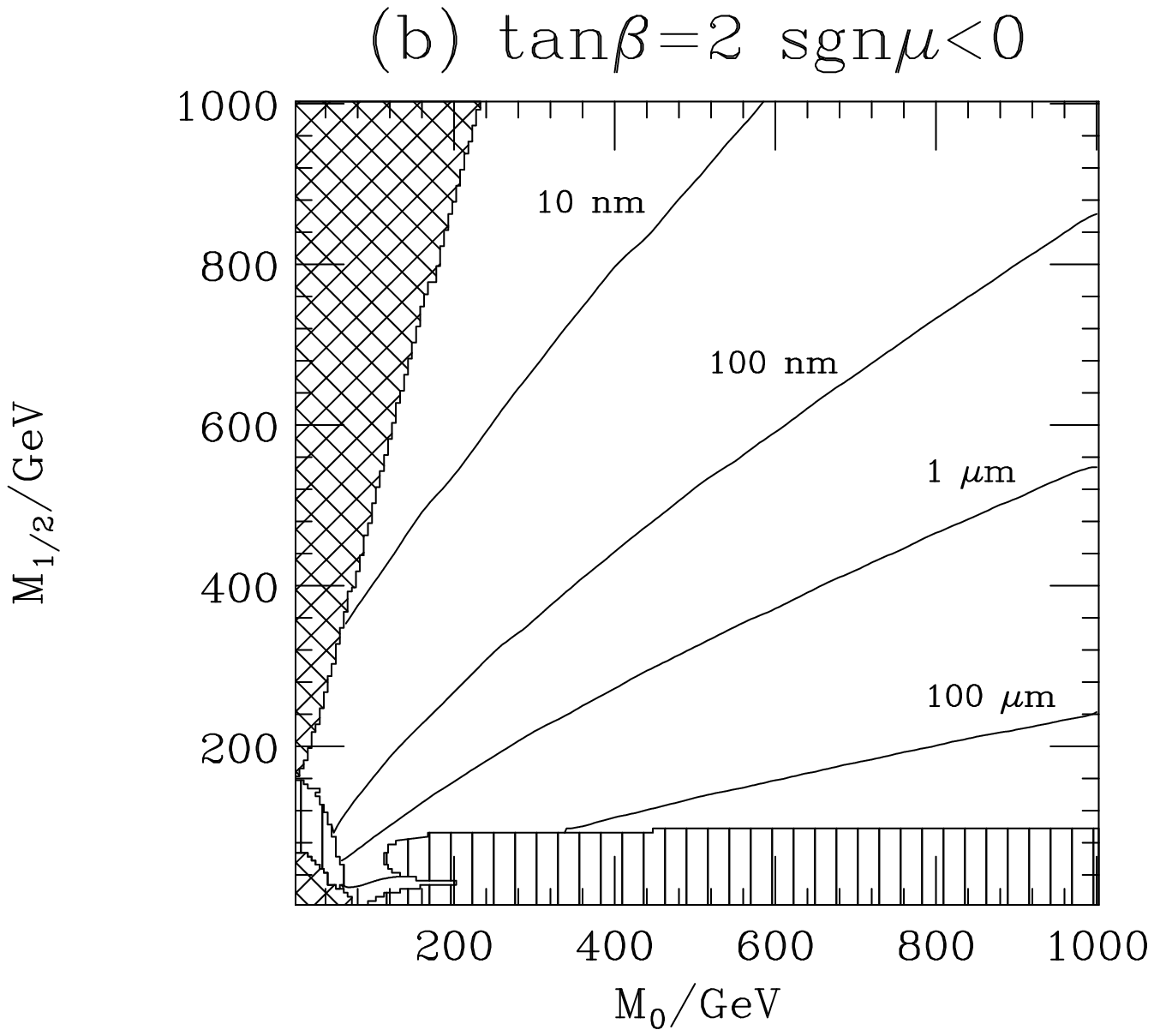}\\
\vskip 7mm
\includegraphics[angle=90,width=0.48\textwidth]{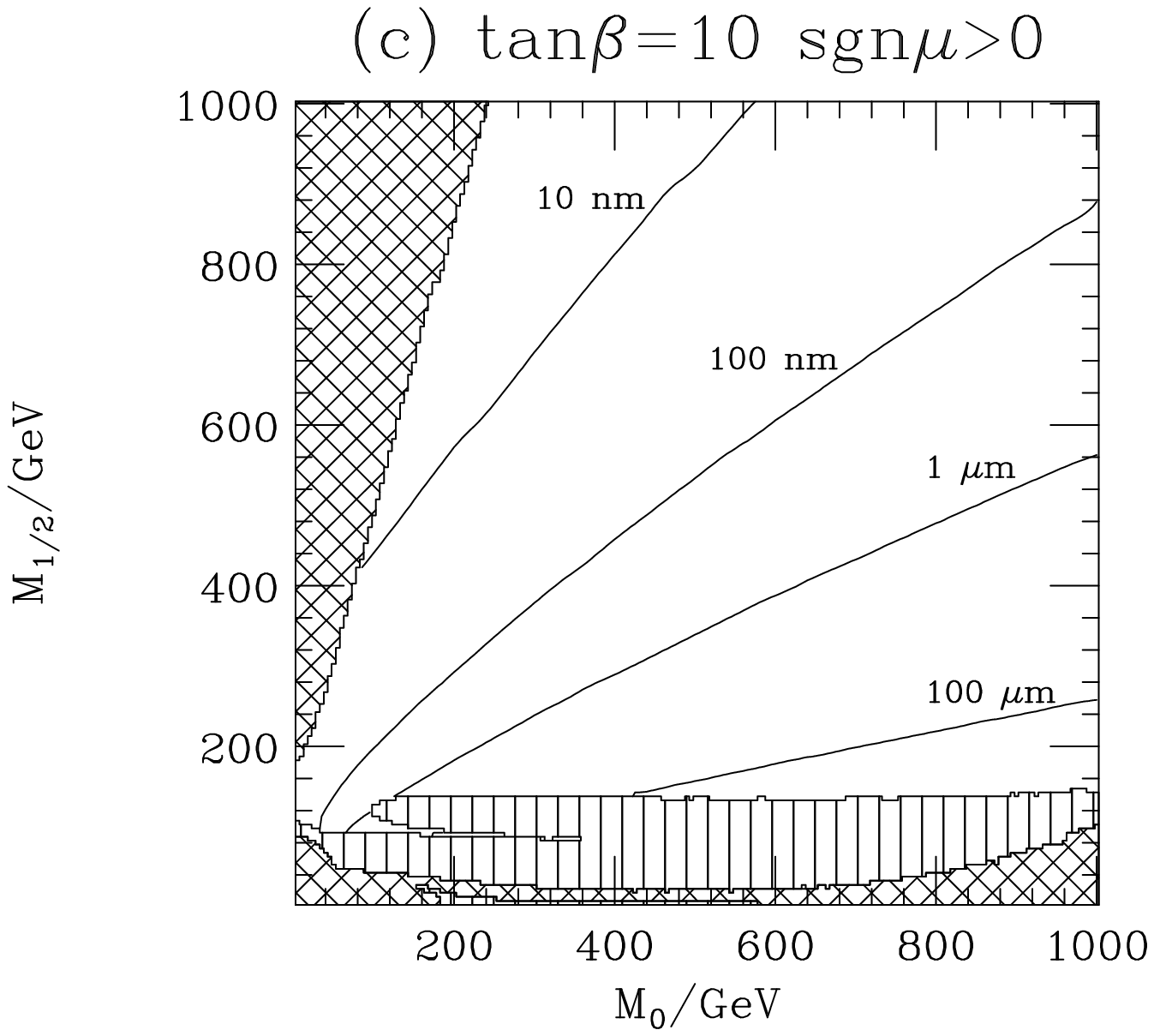}
\hfill
\includegraphics[angle=90,width=0.48\textwidth]{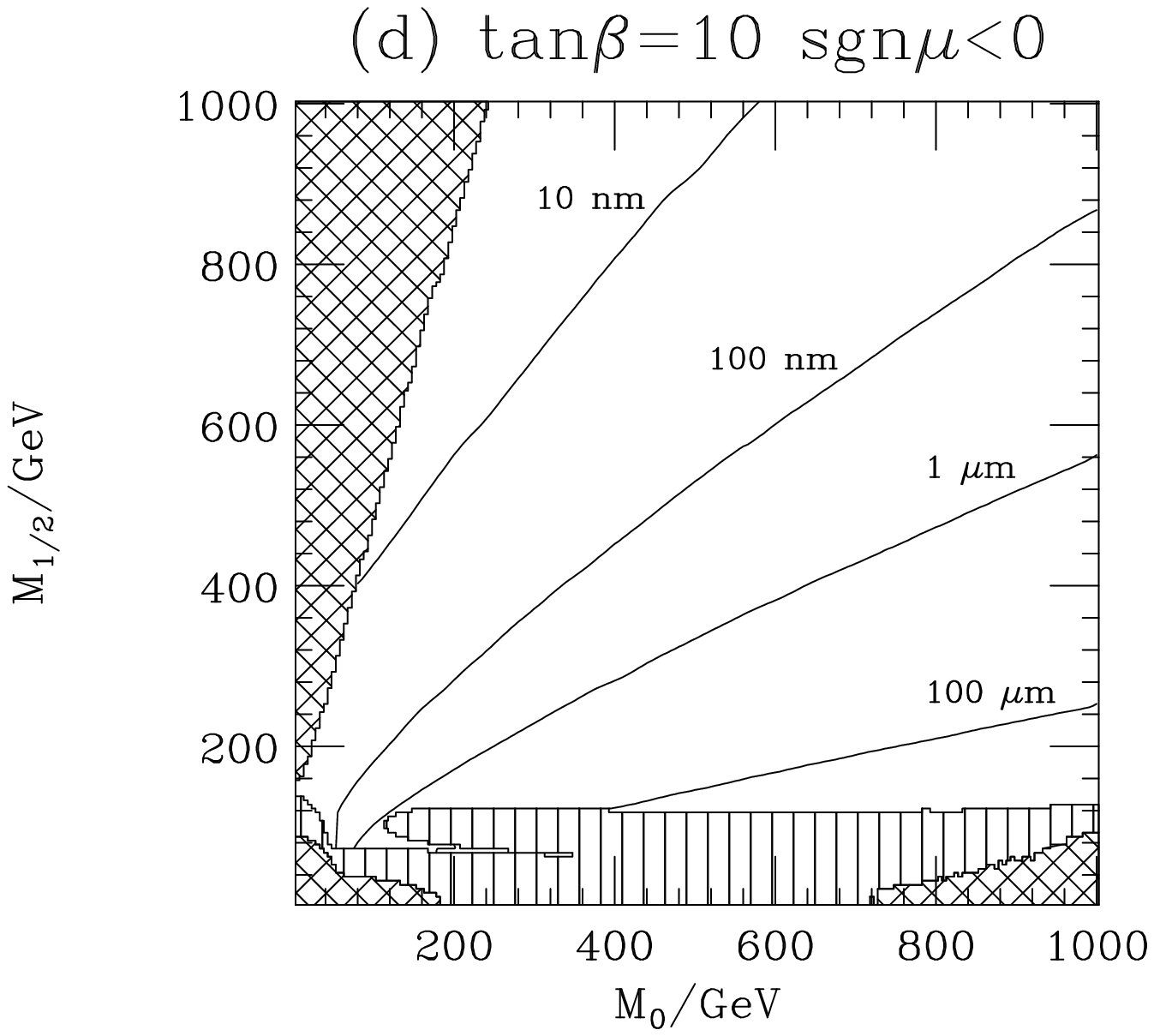}\\
\caption{Contours showing the decay length of a neutralino produced in the 
	 decay of an on-mass-shell slepton  in the 
	 $M_0$, $M_{1/2}$ plane for $A_0=0\, \mr{\gev}$,
	 ${\lam'}_{211}=10^{-2}$
	 and different values of $\tan\beta$ and  $\sgn\mu$. The
	 striped and hatched regions are the same as in 
	 Fig.\,\ref{fig:SUSYmass}.} 
\label{fig:SUSYlength}
\end{center}
\end{figure}

  The signature we are considering requires the neutralino to decay
  inside
  the detector. In practice, if the neutralino decays more than a few
  centimeters from the primary interaction point a different
  analysis including the displaced vertices would be necessary.
  The neutralino decay length is shown in Fig.\,\ref{fig:SUSYlength}
   and is small for all the currently allowed values
  of the SUGRA parameters. There will however be a lower limit on the \rpv\  
  couplings which can be probed using this process as the decay 
  length~$\sim1/{\lam'}^2_{211}$ \cite{Dawson:1985vr}.

%
%

%
%
\section{Backgrounds}
\label{sec:backgrounds}
\subsection{Standard Model Backgrounds}

  The dominant Standard Model backgrounds to like-sign dilepton production
come from:
\begin{itemize}
\item 	Gauge boson pair production, \ie production of WZ or ZZ 
followed by leptonic decays with some of the
leptons not being detected.

\item 	$\mr{t\bar{t}}$ production. Either the t or $\mr{\bar{t}}$ 
        decays semi-leptonically, giving one lepton. The second top
        decays hadronically. A second lepton with the same charge can
        be produced in a semi-leptonic decay of the bottom hadron
        formed in the decay of the second top, \ie
\barr
 \mr{t}	&\ra&  \mr{W^+ b} \ra \mr{\mu^{+}\bar{\nu_{\mu}} b},\nonumber \\
 \mr{\bar{t}}	&\ra& \mr{W^{-}\bar{b}}	\ra \mr{q\bar{q}\bar{b}},\quad 
 \mr{\bar{b}} \ra \mr{\mu^{+}\bar{\nu_{\mu}}\bar{c}}.
\earr

\item 	$\mathrm{b\bar{b}}$ production. If either of these quarks 
        hadronizes to form a $\mr{B^0_{d,s}}$ meson this can mix to
        give a $\mr{\bar{B}^0_{d,s}}$.  This means that if both the
        bottom hadrons decay semi-leptonically the leptons will have
        the same charge as they are both coming from either b or
        $\mr{\bar{b}}$ decays.

\item 	Single top production. A single top quark can be produced together 
        with a $\mr{\bar{b}}$ quark by either an $s$-  or $t$-channel W
        exchange. This can give one charged lepton from the top
        decay, and a second lepton with the same charge from the decay
        of the meson formed after the b quark hadronizes.

 \item Non-physics backgrounds. There are two major sources: (i) from
 misidentifying the charge of a lepton, \eg in Drell-Yan production, and 
 (ii) from incorrectly identifying an isolated hadron as a lepton. This
 means that there is a major source of background from W production
 with an additional jet faking a lepton.
\end{itemize}

  These processes have been extensively studied
  \cite{Baer:1992ef:Baer:1992xs,Jarlskog:1900dv,Barger:1985qa,Barnett:1993ea,
  Dreiner:1994ba:Guchait:1995zk,Armstrong:1994it,Baer:1994zt,Baer:1996va,
  Abdullin:1998nv,
  Matchev:1999nb,Matchev:1999yn,Nachtman:1999ua,Baer:1999bq} as they
  are also the major backgrounds to the production of like-sign
  dileptons in the MSSM. The first studies of like-sign dilepton
  production at the LHC \cite{Dreiner:1994ba:Guchait:1995zk} only
  considered the background from heavy quark production, \ie
  $\mr{t\bar{t}}$ and $\mr{b\bar{b}}$ production. More recent studies
  for both the LHC
  \cite{Armstrong:1994it,Baer:1994zt,Baer:1996va,Abdullin:1998nv} and
  Run II of the Tevatron
  \cite{Matchev:1999nb,Matchev:1999yn,Nachtman:1999ua,Baer:1999bq}
  have also considered the background from gauge boson pair
  production. In addition the Tevatron studies
  \cite{Matchev:1999nb,Matchev:1999yn,Nachtman:1999ua,Baer:1999bq}
  have included the non-physics backgrounds.  We have considered all
  the physics backgrounds, from both heavy quark production and gauge
  boson pair production, but have neglected the non-physics
  backgrounds which would require a full simulation of the
  detector. As we discuss below, since our signal is different, the
  optimal cuts are also different in our case.

  In these studies a number of different cuts have been used to suppress the
  backgrounds. These cuts can be split into two groups, firstly those
  cuts which are designed to reduce the background from
  heavy quark production:
\begin{itemize}
\item A cut on the $p_T$ of the leptons requiring 
	\beq
		p_T^{\mr{lepton}}>p^{\mr{CUT}}_T.
	\eeq
	The values of $p^{\mr{CUT}}_T$ for Tevatron studies have been between
	5 and $20\, \mr{\gev}$. Higher values, between 20 and
	$50\, \mr{\gev}$,
  	have usually
 	been used in LHC simulations.
\item A cut requiring that the leptons are isolated, \ie imposing a 
      cut on the transverse energy, 
      $E^{IC}_T$, in a cone about the direction of the lepton such that
	\beq
		E^{IC}_T < E_0.
		\label{eqn:etcut}
	\eeq
	$E_0$ has usually been taken to be less than $5\, \mr{\gev}$
      for Tevatron
      simulations and between 5 and $10\, \mr{\gev}$ for LHC studies.
      The radius of the cone is usually taken to be 
	\beq
  		\Delta R = \sqrt{\Delta\phi^2+\Delta\eta^2} < 0.4,
	\eeq
      where $\Delta\phi$ is the azimuthal angle and $\Delta\eta$ 
      the pseudo-rapidity of the particles with respect to the lepton.
      
\end{itemize}
  It was shown in \cite{Dreiner:1994ba:Guchait:1995zk} that these cuts
  can reduce the background from heavy quark production by several orders
  of magnitude. Any high $p_T$ lepton from a bottom hadron decay must come
  from a high $p_T$ hadron. This is due
  to the small mass of the bottom hadron relative
  to the lepton $p_T$ which means the lepton will be travelling in
  the same direction as the other decay products 
  \cite{Mondal:1994vi:Godbole:1983yb:Barger:1983qr:Roy:1987mg}. Hence the
  isolation and $p_T$ cuts remove the majority of these events.

  The analyses of 
  \cite{Baer:1996va,Abdullin:1998nv,Matchev:1999nb,Matchev:1999yn,
        Baer:1999bq,Nachtman:1999ua}
  then imposed further cuts to reduce the backgrounds from gauge boson pair
  production, which is the major contribution to the SM background after 
  the imposition of the isolation and $p_T$ cuts:
\begin{itemize}
\item A cut on the invariant mass, $m_{\ell^+\ell^-} $,
	 of any pair of opposite sign same 
      flavour (OSSF)
      leptons to remove those leptons coming from Z decays, \ie
	\beq
		|M_{\mr{Z}}-m_{\ell^+\ell^-}| < m^{\mr{CUT}}_{\ell^+\ell^-},
	\eeq	
      was used in the analyses of 
	\cite{Matchev:1999nb,Nachtman:1999ua,Baer:1999bq}.
\item Instead of a cut on the mass of OSSF lepton pairs some analyses
      considered a veto on the presence of an OSSF lepton pair in the event.
\item In \cite{Baer:1999bq,Matchev:1999yn} a cut on the transverse mass was
      imposed to reject leptons which come from the decays of W bosons. 
      The transverse mass, $M_T$, of a lepton neutrino pair is given by
	\beq
	M^2_T = 2|p_{T_{\ell}}||p_{T_\nu}|(1-\cos\Delta\phi_{\ell\nu}),
	\label{eqn:MTdef}
	\eeq
	where   $p_{T_{\ell}}$ is the transverse momentum of the charged
  	          	 lepton,
		$p_{T_\nu}$ is the transverse momentum of the neutrino 
			 (assumed to be the total missing transverse momentum
			 in the event) and
	        $\Delta\phi_{\ell\nu}$ is the azimuthal angle between
			 the lepton and the neutrino, 
			\ie the missing momentum in the event.
        This cut is applied to both of the like-sign leptons in the event
	to reject events in which either of them came
        from the decay of a W boson.
        A cut removing events with \mbox{$60\ \mr{GeV}<M_T<85\ \mr{GeV}$} was
	used in \cite{Baer:1999bq}
	to reduce the background from WW and WZ production.
\item For the MSSM signatures considered in 
      \cite{Matchev:1999nb,Matchev:1999yn,Baer:1999bq,Nachtman:1999ua}
      there is missing transverse energy, \met, due to the LSP escaping
      from the detector. This allowed them to impose a cut on the
      \met, $\not\!\!E_T > E_T^{\mr{CUT}}$, to reduce the background.
\end{itemize}

  There are however differences between the MSSM signatures which were
  considered in 
  \cite{Matchev:1999nb,Matchev:1999yn,Baer:1999bq,Nachtman:1999ua}
  and the \rpv\  processes we are considering here. In particular as the
  LSP decays, there will be little missing transverse energy in the \rpv\  
  events. 
  This means that instead of a cut requiring the \met\  to be above
  some value we will consider a cut requiring the \met\  to be less than
  some value, \ie
	\beq
		\not\!\!E_T < E_T^{\mr{CUT}}.
	\eeq
  This cut will remove events from some of the possible resonant production
  mechanisms, \ie those channels where a neutrino is produced in either
  the slepton decay or the cascade decay of a chargino, or one of the
  heavier neutralinos, to the lightest neutralino.
  However it will not affect the decay of a charged slepton to a neutralino
  which is the dominant production mechanism over most of the SUSY parameter
  space.

  Similarly, the signal we are considering in general will not contain more
  than two leptons. Further leptons can only come from cascade decays
  following the production of either a chargino or one of the heavier
  neutralinos, or from semi-leptonic
  hadron decays. This means that instead of the cut on the invariant mass of
  OSSF lepton pairs we will only consider the effect of a veto on the 
  presence of OSSF pairs. This veto was considered in 
   \cite{Matchev:1999nb,Matchev:1999yn} but for the MSSM signal  
  considered there it removed more signal than background.

%
%
\subsection{SUSY backgrounds}

  So far we have neglected what may be the major source of background to this
  process, \ie supersymmetric particle pair production. If we only consider 
  small \rpv\  couplings the dominant effect in sparticle pair production is 
  that the LSP produced at the end of the cascade decays of the other SUSY
  particles will decay. For large \rpv\  
  couplings the cascade decay chains can also
  be affected by the heavier SUSY particles decaying via \rpv\  modes, which
  we will not consider here.\footnote{These additional decays are included
  in HERWIG6.1.}
  The LSP will decay giving a quark-antiquark pair and
  either a charged lepton or a
  neutrino. There will usually be two LSPs in each event, one from the 
  decay chain of each of the sparticles produced in the hard collision. This
  means that they can both decay to give leptons with the same charge. 
  Leptons
  can also be produced in the cascade decays. These processes will therefore
  be a major background to like-sign dilepton production via resonant slepton
  production.

  The cuts which were intended to reduce the Standard Model background will
  also significantly reduce the background from sparticle pair production.
  However we will need to impose additional cuts to suppress this background.
  In the signal events there will be at least two high $p_T$ jets from the
  neutralino decay, there may be more jets from either initial-state QCD
  radiation or radiation from quarks produced in the neutralino decay.
  In the dominant production mechanism, \ie $\mr{\mut\ra\mu^-\cht^0_1}$,
  this will be the only source  of jets, however additional jets can be
  produced
  in the cascade SUSY decays if a chargino or one of the heavier neutralinos
  is produced.
  In the SUSY background there will be at least four high $p_T$ jets
  from the neutralino decays, plus other jets formed in the decays of the
  coloured sparticles which are predominantly formed in hadron--hadron
  collisions. This suggests two possible strategies for reducing 
  the sparticle 
  pair production background:
\begin{enumerate}
\item	A cut such that there are at most 2 or 3 jets (allowing for some QCD
 	radiation) above a given $p_T$. This will reduce the SUSY background
	which typically has more than four high $p_T$ jets.

\item	A cut such that there are exactly two jets, or only two or three
	jets above a given $p_T$. This will reduce the gauge boson pair
	background where typically the only jets come from initial-state
 	radiation, as well as the background from sparticle pair production.

\end{enumerate}

  In practice we would use a much higher momentum cut in the first case, as
  we only need to ensure that the cut is sufficiently high that most of the 
  sparticle pair production events give more than 2 or 3 jets above the cut.
  However with the second cut we need to ensure that the jets in the signal
  have sufficiently high $p_T$ to pass the cut as well. In practice we found
  that the first cut significantly reduced the sparticle pair production 
  background while
  having little effect on the signal, while the second cut dramatically
  reduced the signal as well. In the next section we will
  consider the effects of cut 1. on both the signal and background
  at Run II of the Tevatron and the LHC.

%
%
\section{Simulations}
\label{sec:results}
  
  HERWIG~6.1 \cite{Corcella:1999qn:Marchesini:1991ch} was used to
  simulate the
  signal and the backgrounds from sparticle pair, $\mr{t\bar{t}}$,
  $\mr{b\bar{b}}$ and single top production. HERWIG does not include gauge
  boson pair production in hadron--hadron collisions and we therefore used  
  PYTHIA~6.1 \cite{Sjostrand:1994yb} to simulate this background.
  The simulation of the signal includes all the non-\rpv\  decay modes
  given in Table~\ref{tab:decaymodes}. We used a cone algorithm 
  with $R=0.4$ radians for all the jet reconstructions. The algorithm is
  similar to that used by CDF, apart from using the midpoints 
  between two particles as a seed for the algorithm in addition to
  the particles themselves. The inclusion
  of the midpoints as seeds improves the 
  infra-red safety of the cone algorithm.

  Due to the large cross sections for some of the Standard Model backgrounds
  before any cuts we imposed parton-level cuts and forced certain decay
  modes in order to simulate a sufficient number of 
  events with the resources available. We designed these cuts in such a
  way that hopefully they are weaker than any final cut we apply, so that
  we do not lose any of the events that would pass the final cuts. 
  We imposed the following cuts
  for the various backgrounds:
\begin{itemize}
\item \underline{$\mr{b\bar{b}}$ production.}
	We forced the B hadrons produced by the
	hadronization to decay semi-leptonically.
	This neglects the production
        of leptons in charm decays which has a higher cross section but
        which we would expect to have a lower $p_T$
	and be less well isolated than those leptons produced in
	bottom decays. If there was only one 
  	$\mr{B^0_{d,s}}$ meson in the event this was forced to mix.
	When there was more than one $\mr{B^0_{d,s}}$ 
	meson then one of them was
	forced to mix and the others were forced not to mix. Similarly we
	imposed a parton-level cut on the transverse momentum of the initial
	b and $\mr{\bar{b}}$, 
	$p_T^{\mr{b},\mr{\bar{b}}}\geq p_T^{\mr{parton}}$.
 	This parton-level cut should not affect the
	background provided that we impose a 
	cut on the transverse momentum of
  	the leptons produced in the decay, 
	$p_T^{\mr{lepton}}\geq p_T^{\mr{parton}}$.

\item 	\underline{$\mr{t\bar{t}}$ production.}
        While not as large as the $\mr{b\bar{b}}$
	production cross section the cross section for $\mr{t\bar{t}}$ is
	large, particularly at the LHC. We improved the efficiency
	by forcing one of the top quarks in each
	event to decay semi-leptonically, again this neglects events in which
	there are leptons from charm decay. However we did not impose a cut
	on the $p_T$ of the top quarks as due 
	to the large top quark mass even
	relatively low $p_T$ top quarks can give high $p_T$ leptons.

\item 	\underline{Single top production.}
        While the cross section for this process is relatively small
        compared to the heavy quark pair production cross sections we
        still forced the top to decay semi-leptonically as above to
        reduce the number of events we need to simulate.

\item 	\underline{Gauge Boson Pair Production.}
	The cross sections for these processes
	are relatively small and it was not necessary to impose
        any parton-level
	cuts, or force particular decay modes.
\end{itemize}

  Where possible the results of the Monte Carlo simulations have been
  normalized by using next-to-leading-order cross sections for the various
  background processes. We used the next-to-leading-order calculation of 
  \cite{Campbell:1999ah} for gauge boson pair production.
  The $\mr{t\bar{t}}$ simulations were normalized using 
  the next-to-leading-order with next-to-leading-log resummation calculation
  from \cite{Bonciani:1998vc}.

  The calculation of a next-to-leading-order
  cross section for $\mr{b\bar{b}}$ production is more problematic
  due to the parton-level cuts we imposed on the simulated
  events. There are a range of possible options for applying the $p_T$ cut we
  imposed on the bottom quark at next-to-leading order. At leading order the
  transverse momenta of the quarks are identical and therefore the cut
  requires them both to have transverse momentum 
  \mbox{$p_T>p_T^{\mr{CUT}}$}. However at 
  next-to-leading order, due to gluon radiation, the transverse momenta of 
  the quarks are no longer equal. Therefore a cut on for example the $p_T$ 
  of the hardest quark, \mbox{$p_{T_1}>p_T^{\mr{CUT}}$}, together with a
  cut on the lower $p_T$ quark, \mbox{$p_{T_2}>p_T^{\mr{CUT}}-\delta$},
  with any positive value of \mbox{$\delta<p_T^{\mr{CUT}}$} is the same
  as the leading-order cut we applied. Given that we need a high
  transverse momentum bottom hadron to give a high $p_T$ lepton and 
  only events with two such high $p_T$ leptons will contribute to the
  background a cut requiring both bottom quarks to have 
  \mbox{$p_{T}>p_T^{\mr{CUT}}$}, \ie $\delta=0$, is most appropriate.
  However at this point the perturbation theory is
  unreliable \cite{Frixione:1997ks} and for the cuts we applied the
  next-to-leading-order cross section is smaller than the leading-order
   result. We therefore applied the cut 
  \mbox{$p_{T_1}>p_T^{\mr{CUT}}$} with no cut on the softer bottom 
  as this avoids the point 
  at which the perturbative expansion is unreliable, 
  \ie $\delta=p_T^{\mr{CUT}}$. We used the program 
  of \cite{Mangano:1992jk} to calculate the next-to-leading-order
  cross section with these cuts.

  All of the simulations and SUSY cross section calculations 
  used the latest MRS parton distribution set 
  \cite{Martin:1999ww:Martin:1998sq}, as did the calculation of the
  single top production cross section. The parton distribution sets used
  in the various next-to-leading-order cross sections are described in the
  relevant papers.

  We can now study the signal and background in more detail for both the
  Tevatron and the LHC. This is followed by a discussion of methods to 
  reconstruct the masses of both the lightest neutralino and the resonant
  slepton.

%
%
\subsection{Tevatron}
\label{sub:tevatron}

\begin{figure}
\begin{center}
\includegraphics[angle=90,width=0.48\textwidth]{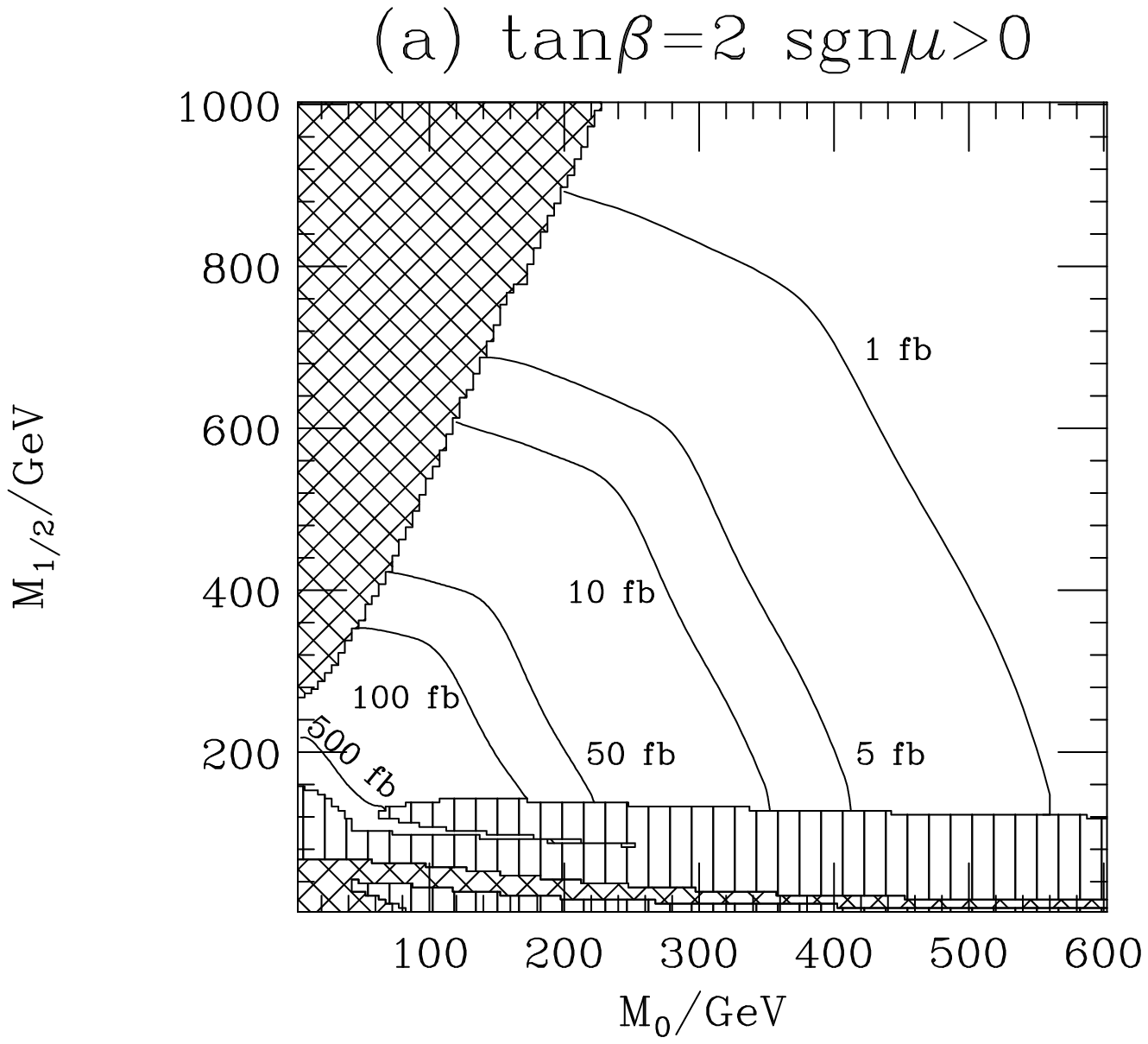}
\hfill
\includegraphics[angle=90,width=0.48\textwidth]{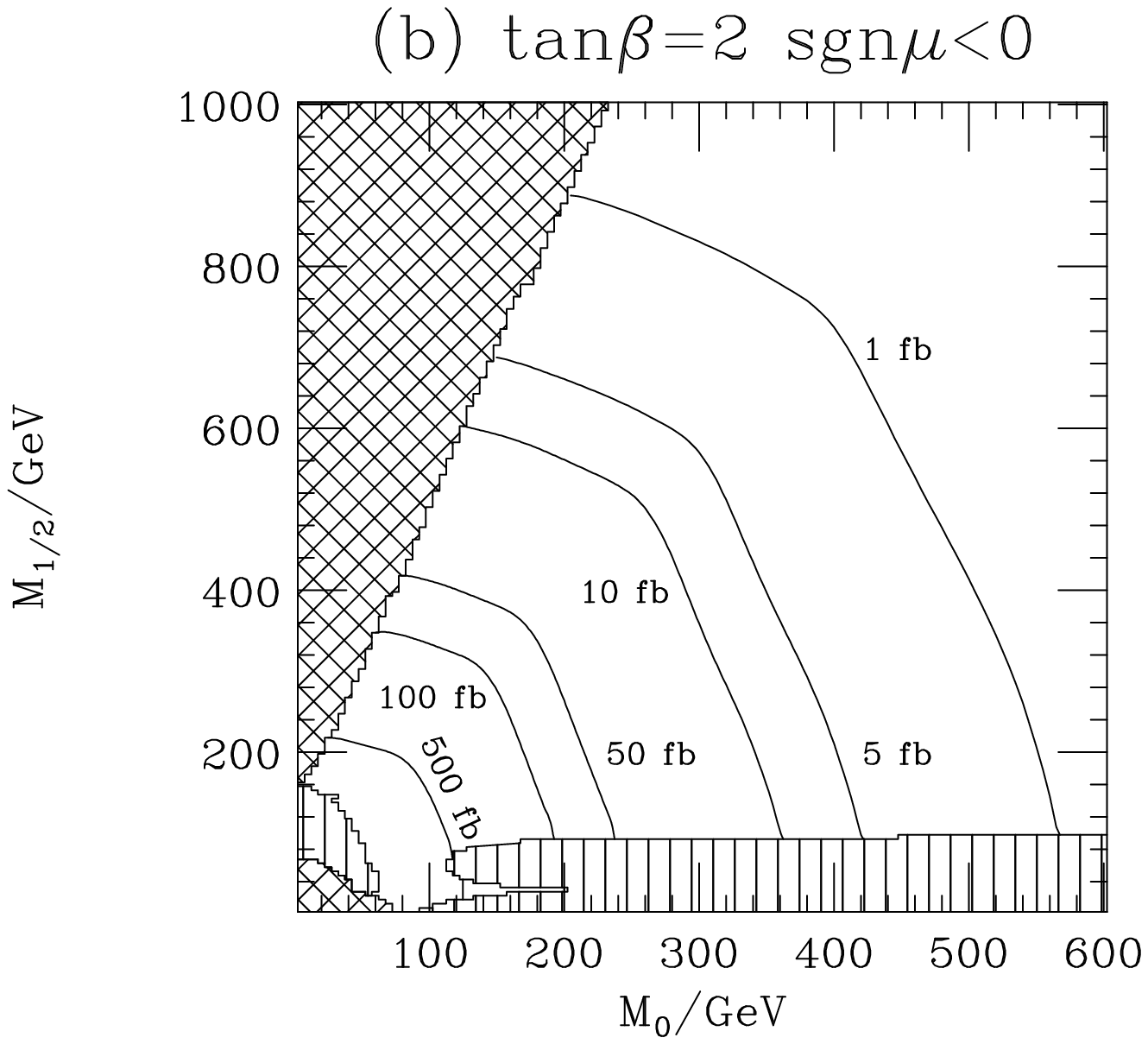}\\
\vskip 7mm
\includegraphics[angle=90,width=0.48\textwidth]{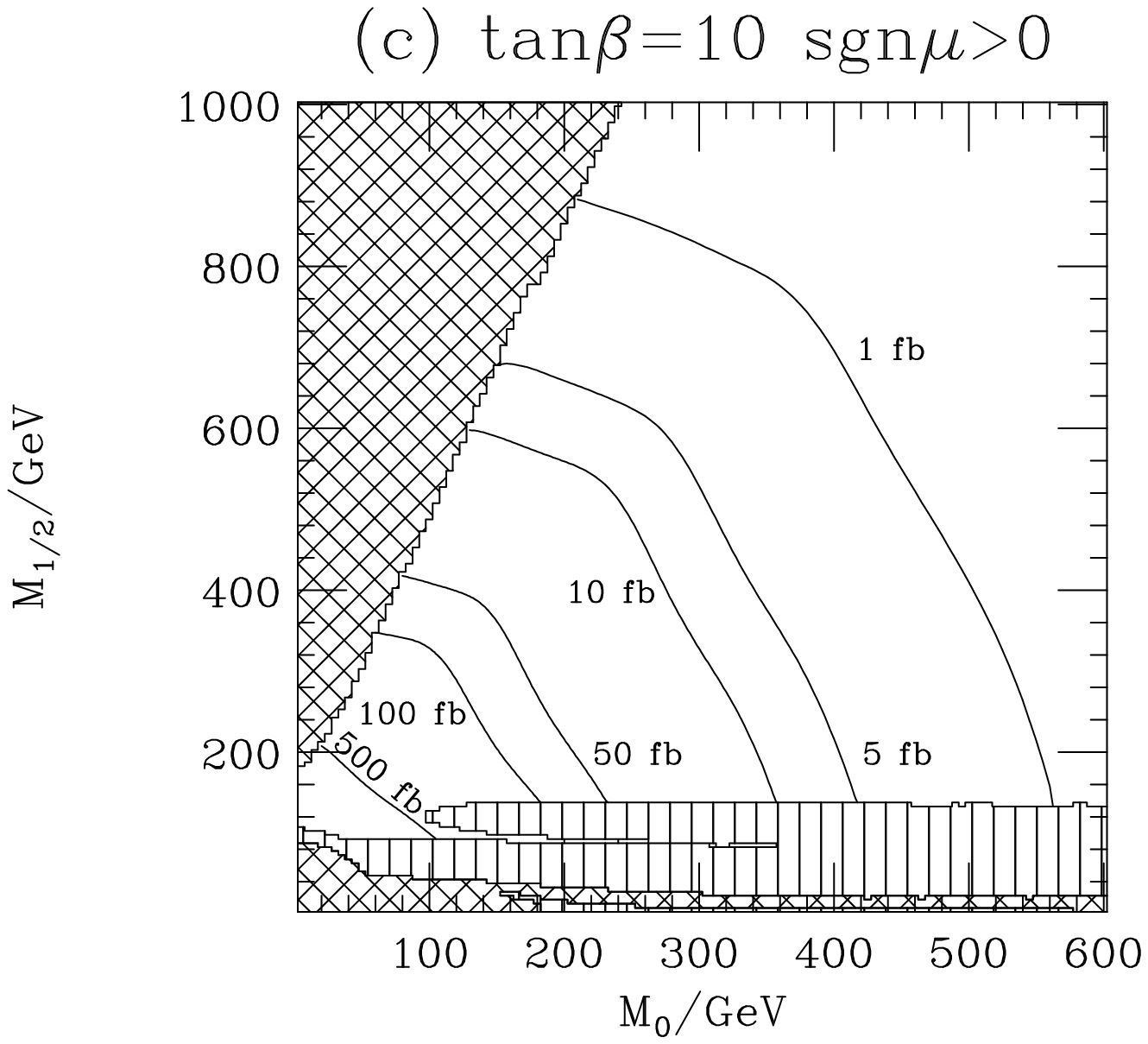}
\includegraphics[angle=90,width=0.48\textwidth]{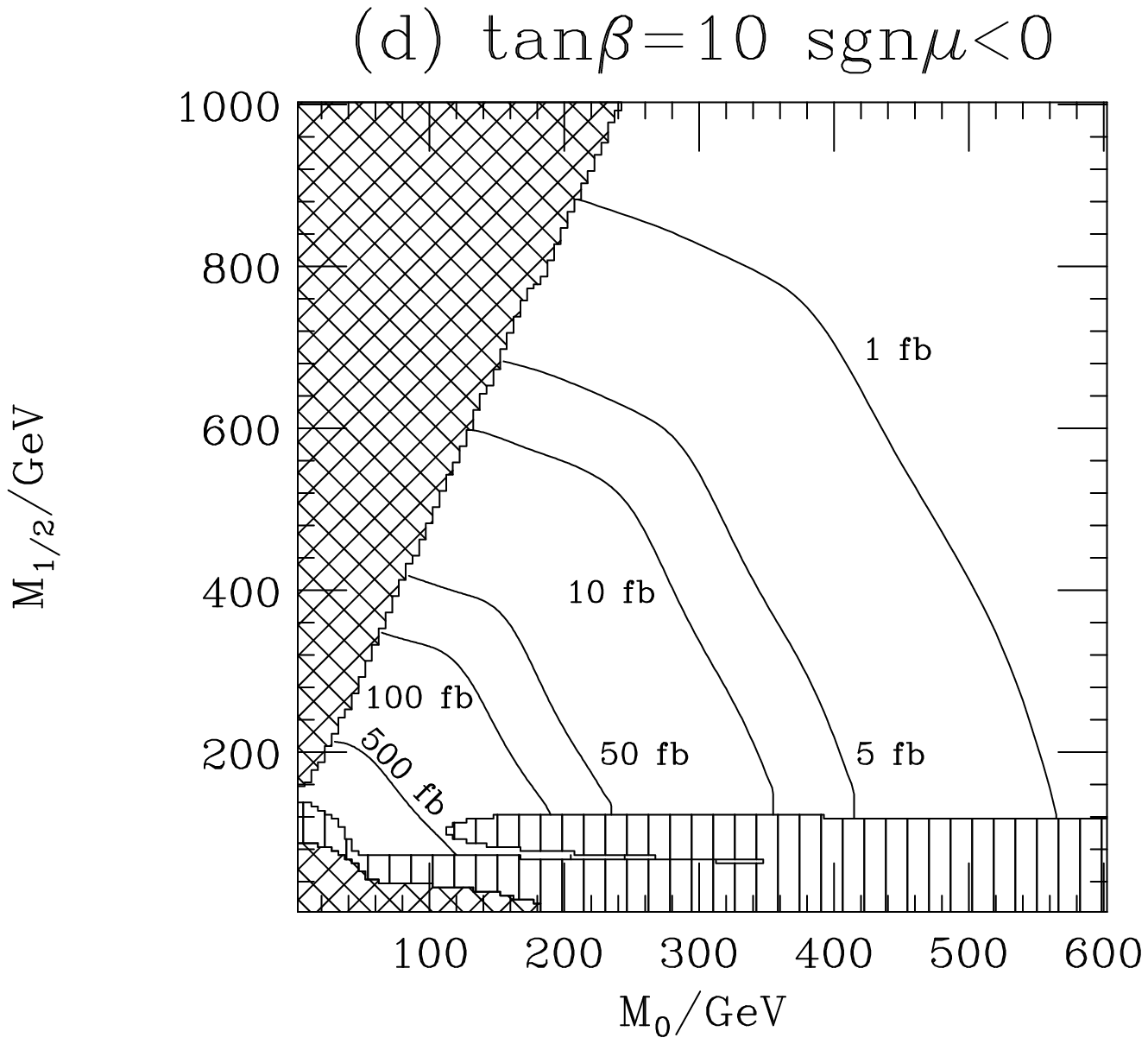}\\
\caption{Contours showing the cross section for the production of a 
	 neutralino and a charged lepton at Run II of the Tevatron
	 in the $M_0$, $M_{1/2}$ plane 
	 for $A_0=0\, \mr{\gev}$   and ${\lam'}_{211}=10^{-2}$ with
         different values
	 of $\tan\beta$ and $\sgn\mu$. The striped and hatched regions are
	 described in the caption of Fig.\,\ref{fig:SUSYmass}.} 
\label{fig:tevcross}
\end{center}
\end{figure}

\begin{figure}
\begin{center}
\includegraphics[angle=90,width=0.48\textwidth]{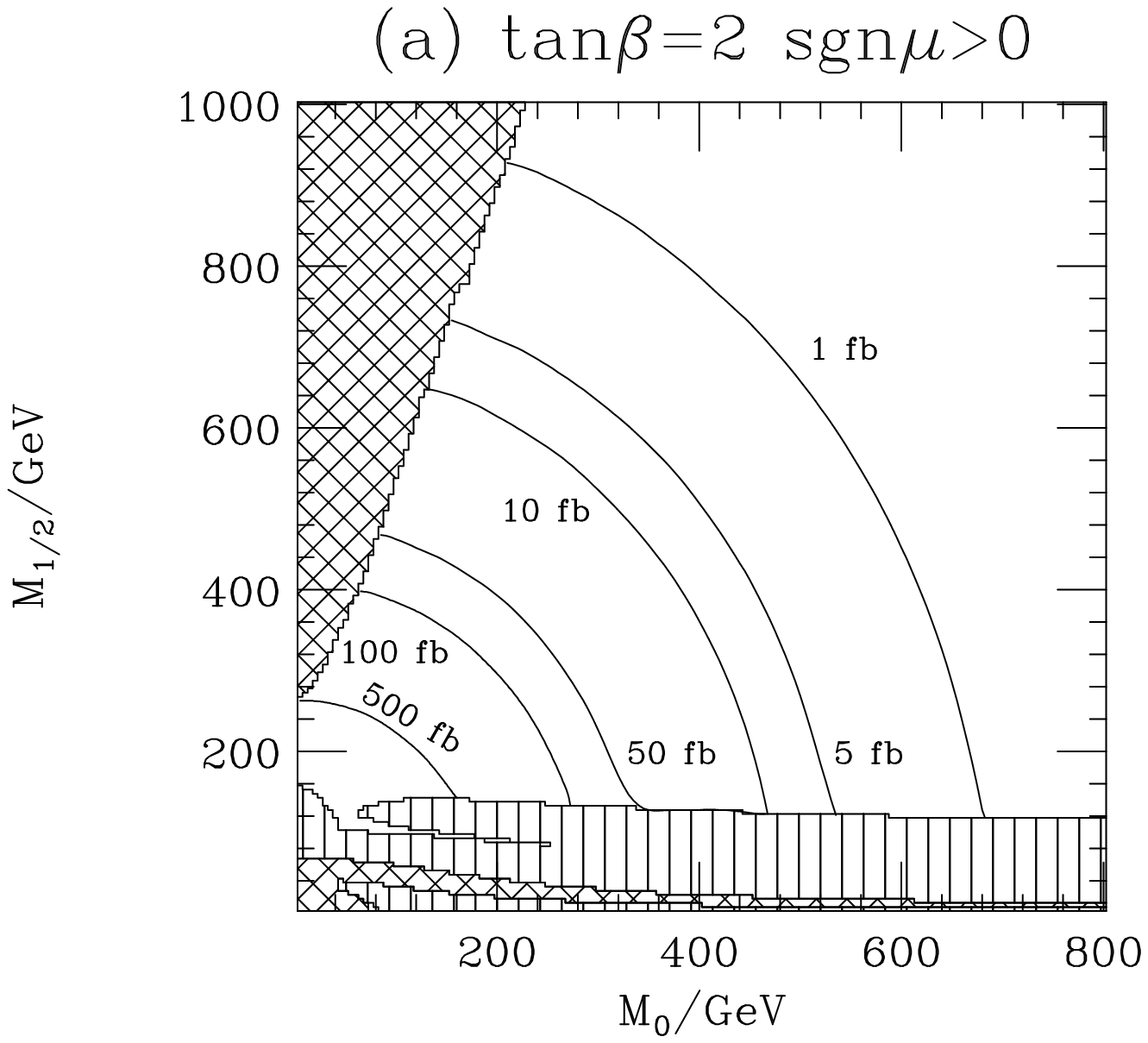}
\hfill
\includegraphics[angle=90,width=0.48\textwidth]{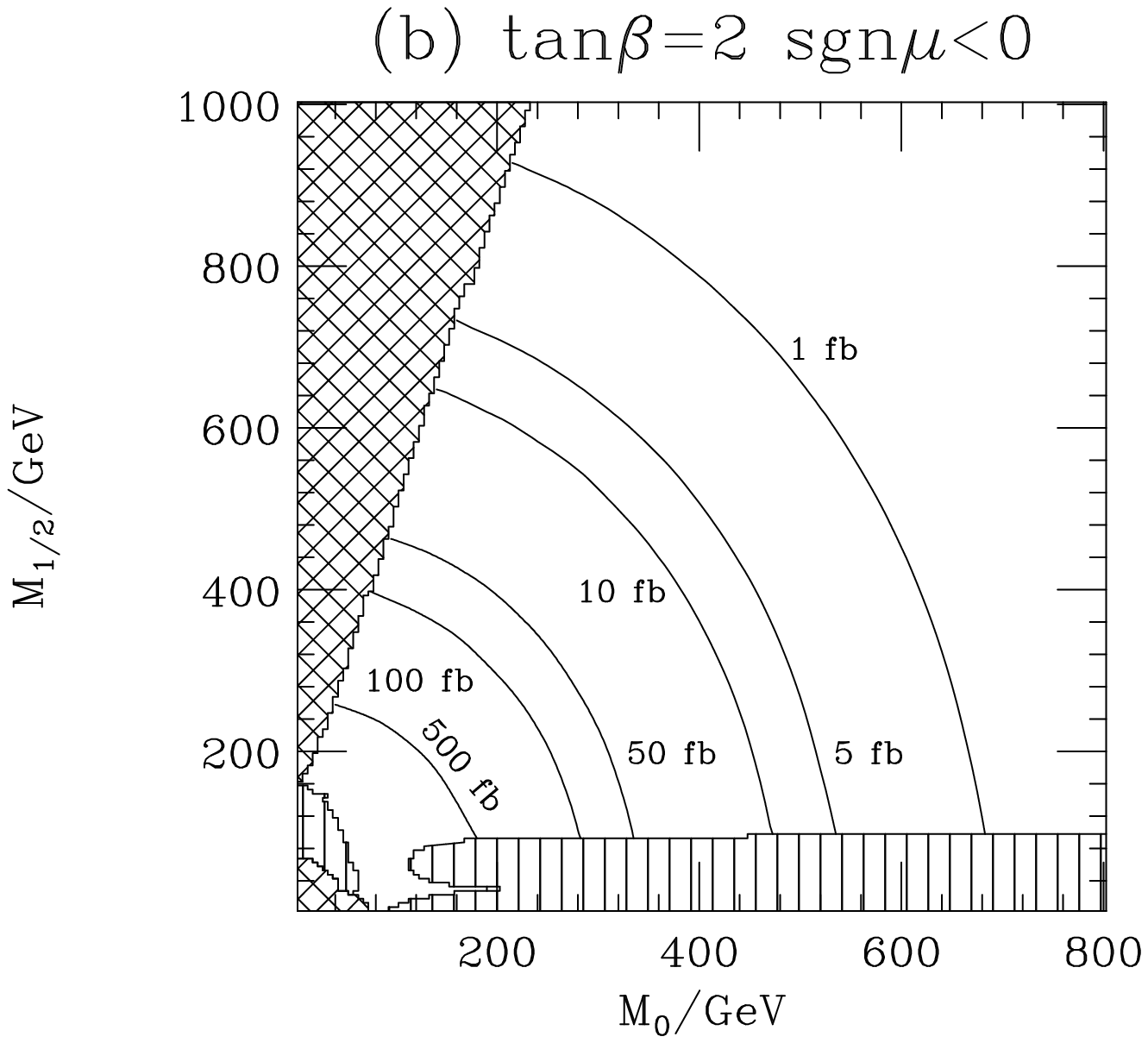}\\
\vskip 7mm
\includegraphics[angle=90,width=0.48\textwidth]{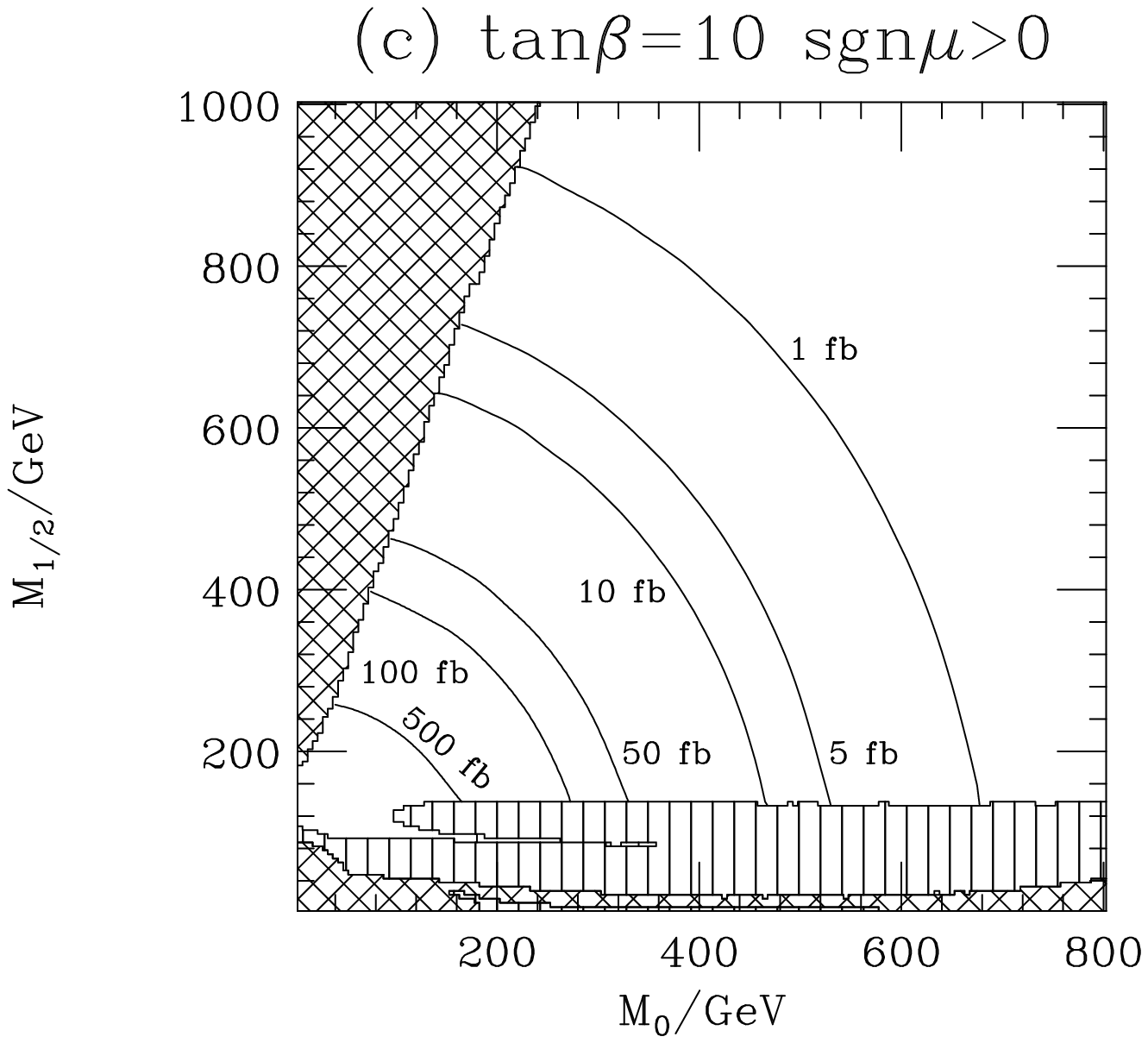}
\includegraphics[angle=90,width=0.48\textwidth]{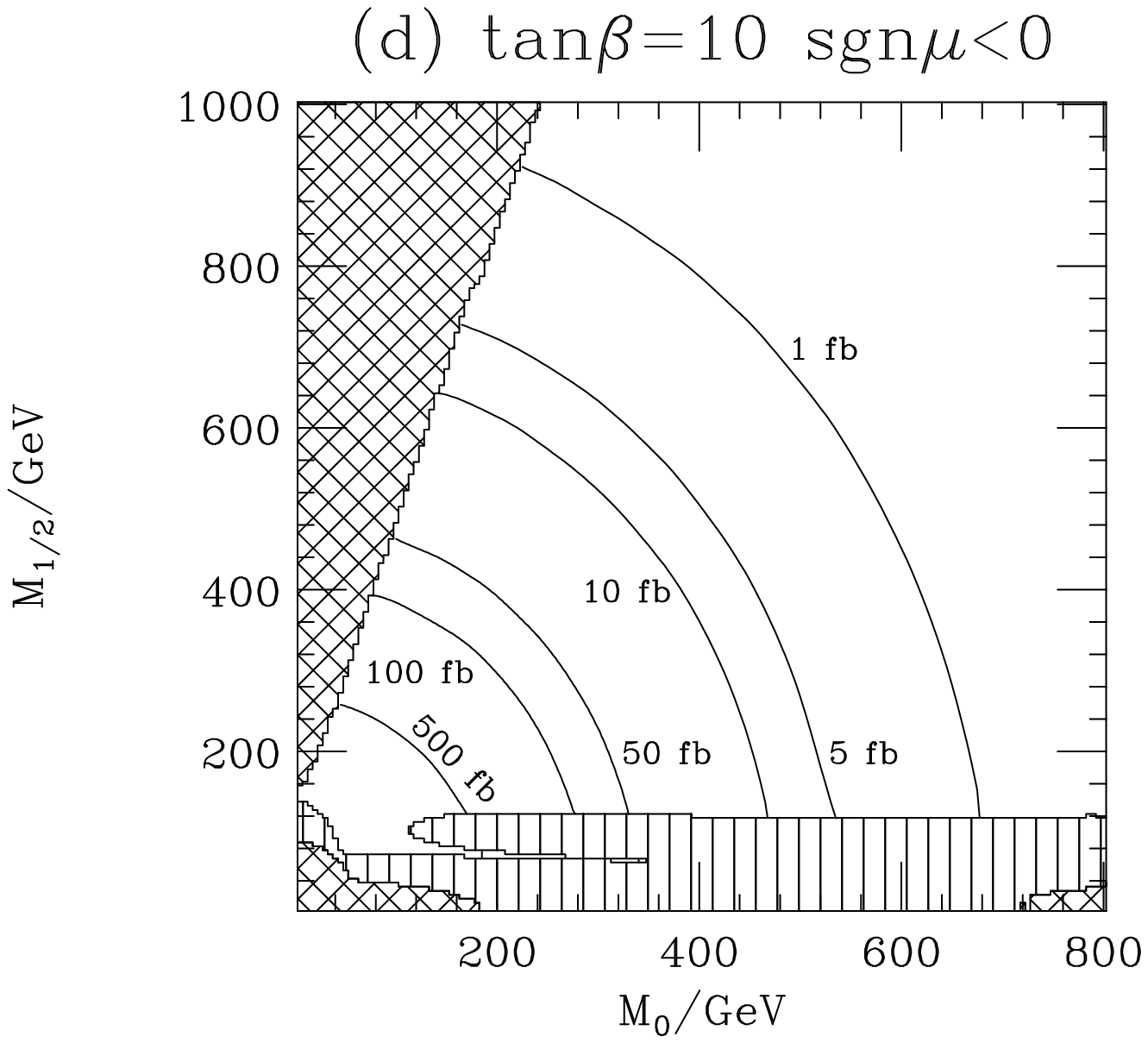}\\
\caption{Contours showing the cross section for the production of a slepton
	 followed by a supersymmetric gauge decay at
	 Run II of the Tevatron
	 in the $M_0$, $M_{1/2}$ plane 
	 for $A_0=0\, \mr{\gev}$   and ${\lam'}_{211}=10^{-2}$ with
         different values
	 of $\tan\beta$ and $\sgn\mu$. The striped and hatched regions are
	 described in the caption of Fig.\,\ref{fig:SUSYmass}.} 
\label{fig:tevcross2}
\end{center}
\end{figure}

  The cross section for the production of a neutralino and a charged
  lepton, which is the dominant production mechanism, is shown in
  Fig.\,\ref{fig:tevcross} in the $M_0$, $M_{1/2}$ plane with
  \mbox{$A_0=0\, \mr{\gev}$} and ${\lam'}_{211}=10^{-2}$ for two
  different values of $\tan\beta$ and both values of $\sgn\mu$. The
  total cross section for resonant slepton production followed by
  supersymmetric gauge decays is shown in Fig\,\ref{fig:tevcross2}.
  As can be seen the total cross section closely follows the slepton
  mass contours shown in Fig.\,\ref{fig:SUSYmass} whereas the
  neutralino lepton cross section falls-off more quickly at small
  $M_{1/2}$ where the charginos and heavier neutralinos can be
  produced.  This cross section must be multiplied by the the
  acceptance, \ie the fraction of signal events which pass the cuts,
  to give the number of observable events in the experiment.

  We will first discuss the cuts applied to reduce the various Standard Model
  backgrounds and then present the discovery potential at the Tevatron if we
  only consider these backgrounds. This is followed by a discussion of the
  additional cuts needed to reduce the background from sparticle pair
  production.

%
%

%
%

\subsubsection{Standard Model Backgrounds}

%
%
\begin{figure}
\begin{center}
\includegraphics[angle=90,width=0.48\textwidth]{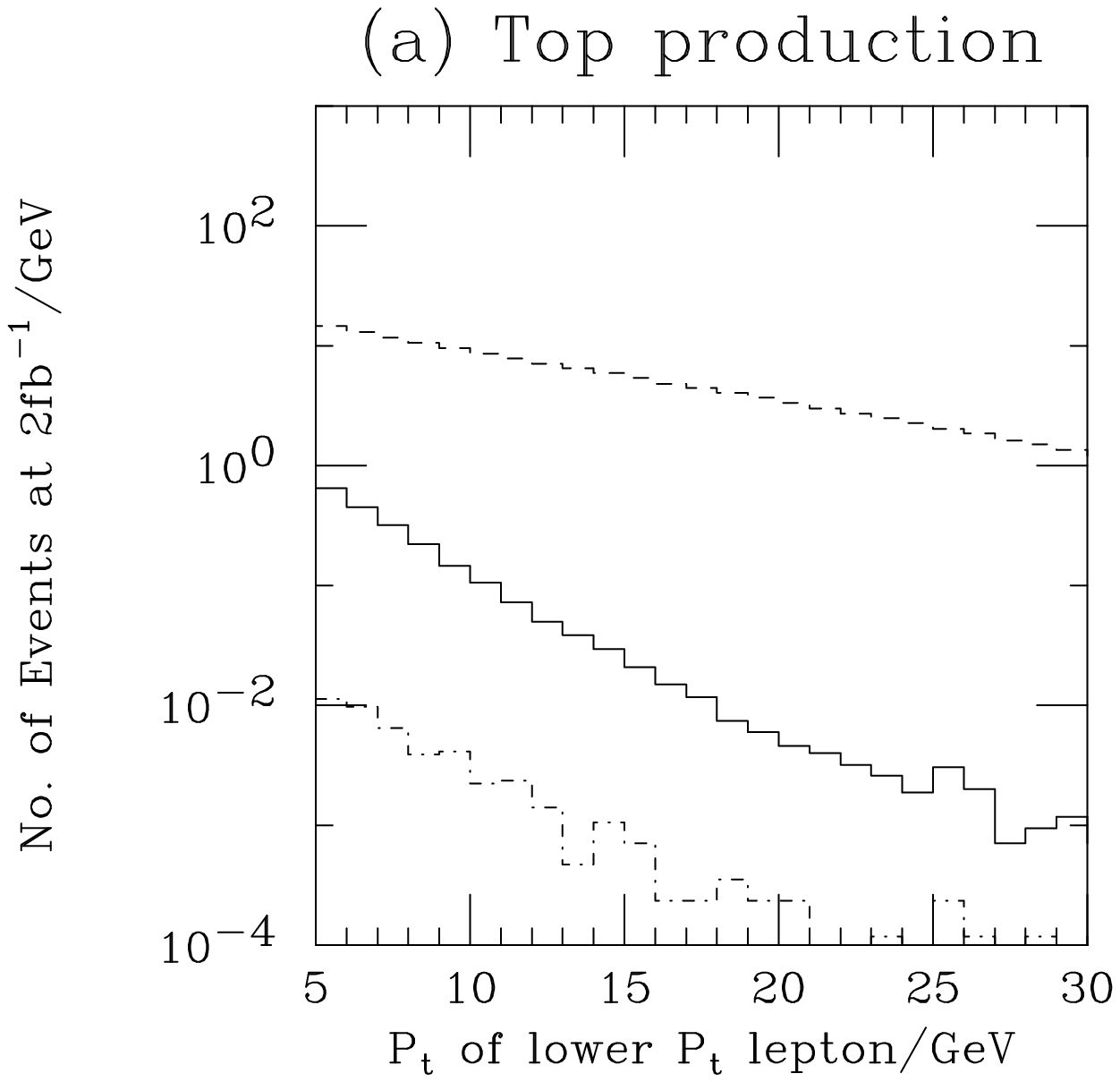}
\hfill
\includegraphics[angle=90,width=0.48\textwidth]{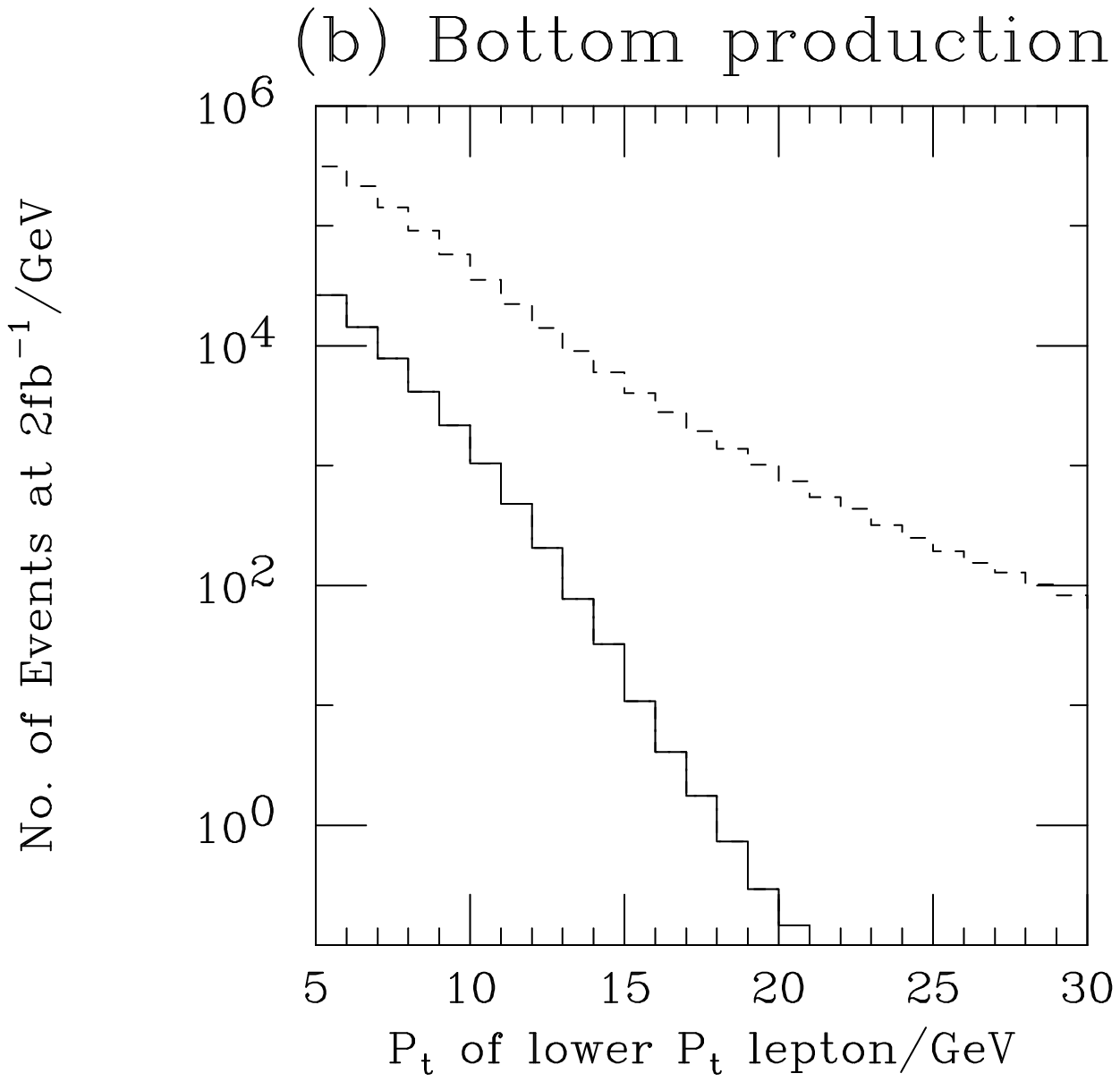}\\
\caption{Effect of the isolation cuts on the $\mr{t\bar{t}}$ and 
	 $\mr{b\bar{b}}$ backgrounds at Run II of
	the Tevatron. The dashed line gives the background before any cuts,
	the solid line shows the effect of the isolation cut described in the
	text. The dot-dash line gives the effect of all the cuts, including
	the cut on the number of jets, for the $\mr{b\bar{b}}$ background 
 	this is  indistinguishable from the solid line.
	The distributions have been normalized
 	to $2\  \mr{fb}^{-1}$ integrated luminosity. As a parton-level cut
	of $20\, \mr{\gev}$  was used in simulating
        the $\mr{b\bar{b}}$ background
	the results below $20\, \mr{\gev}$  for the
        lepton $p_T$ do not correspond to
 	the full number of background events.} 
\label{fig:tevheavyiso}
\end{center}
\end{figure}

  We have applied the following cuts to
  reduce the Standard Model backgrounds:
\begin{enumerate}

\item A cut requiring all the leptons to be in the central region of the
      detector, $|\eta|<2.0$.

\item	A cut on the transverse momentum of each of the like-sign leptons
	$p_T^{\mr{lepton}} \geq 20\, \mr{\gev}$,
	this is the lowest cut we could apply given our parton level cut of
	$p_T^{\mr{parton}}=20\, \mr{\gev}$,
        for the $\mr{b\bar{b}}$ background.
	
\item 	An isolation cut on the like-sign leptons so that the
	transverse energy in a cone of radius,
	$R = \sqrt{\Delta\phi^2+\Delta\eta^2} = 0.4$, about the direction of
	the lepton is less than $5\, \mr{\gev}$.

\item   We reject events with \mbox{$60\, \mr{\gev} < M_T < 85\, \mr{\gev}$}
	($c.f.$ Eqn.\,\ref{eqn:MTdef}.)
	This cut is applied to both of the like-sign leptons.

\item   A veto on the presence of a lepton in the event with the same flavour
        but opposite charge as either of the leptons in the like-sign
        pair if the lepton has  $p_T>10\, \mr{\gev}$  
	and passes the same isolation cut as the like-sign leptons.

\item   A cut on the missing transverse energy,
	 $\not\!\!\!E_T<20\, \mr{\gev}$.
	In our analysis we have assumed that the missing transverse energy 
	is solely due to the momenta of the neutrinos produced. 
\end{enumerate}

%
%
\begin{figure}
\begin{center}
\includegraphics[angle=90,width=0.48\textwidth]{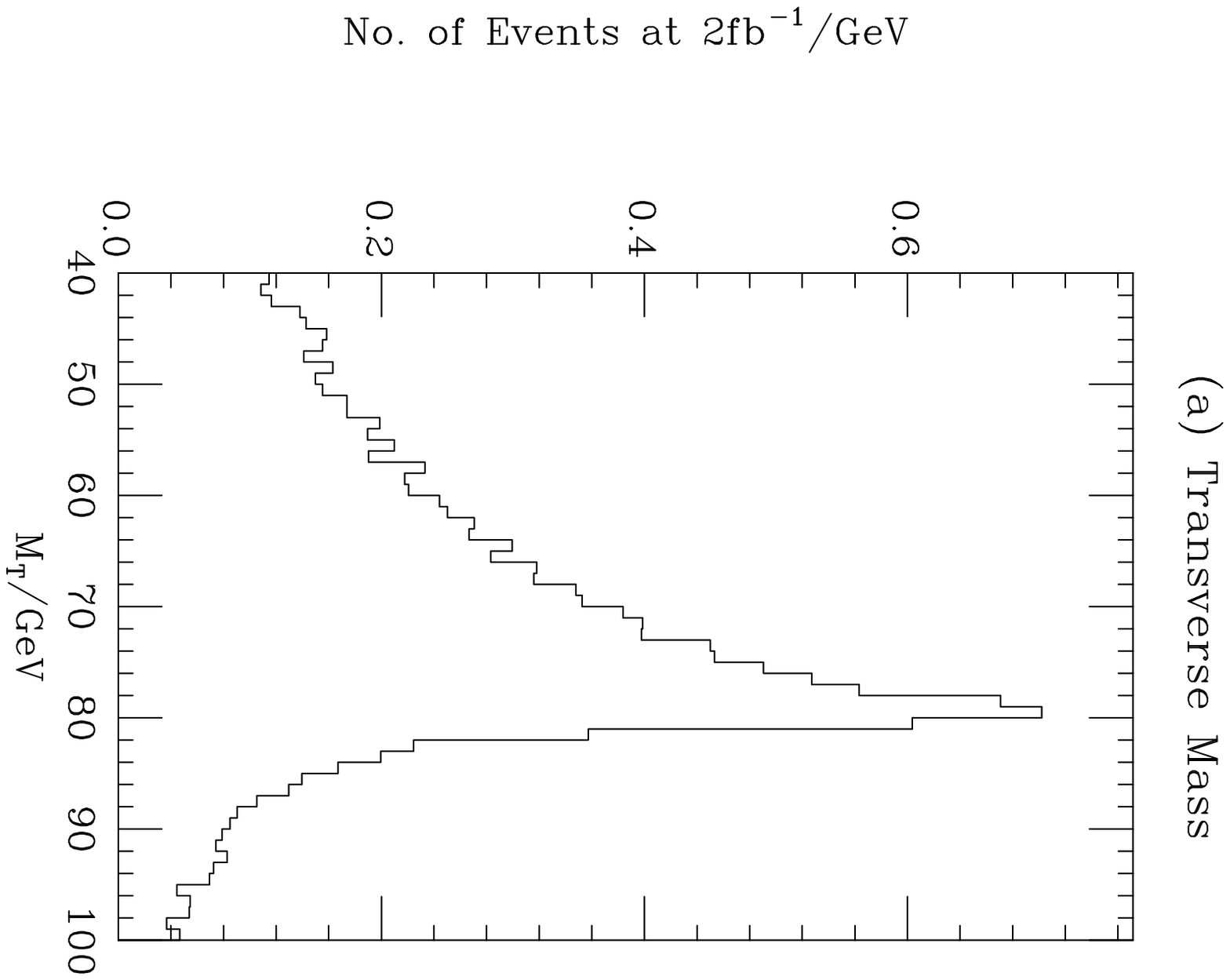}
\hfill
\includegraphics[angle=90,width=0.48\textwidth]{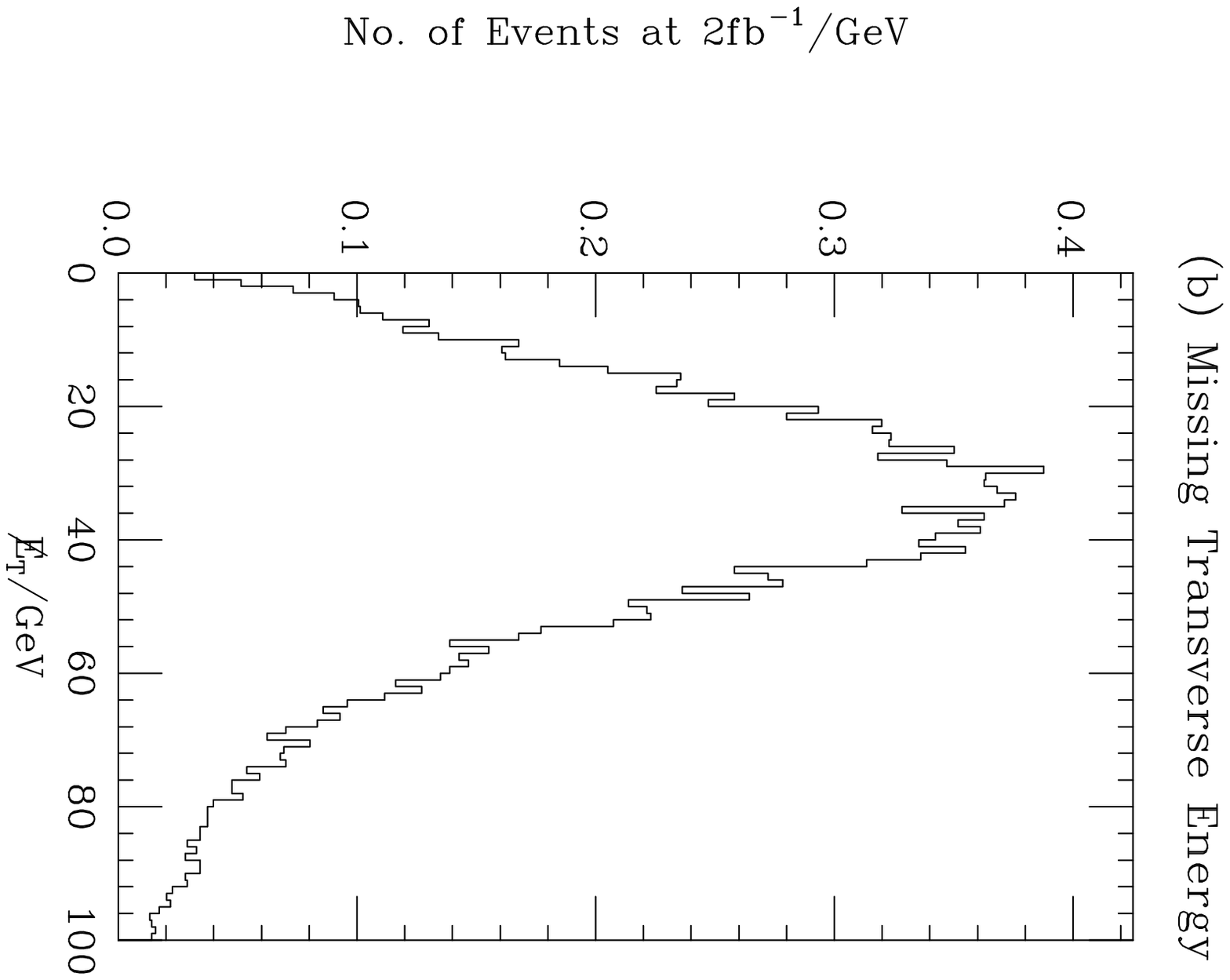}\\
\caption{The transverse mass and missing transverse energy in WZ events
	 at Run II of the Tevatron. The distributions are normalized to
	$2\  \mr{fb}^{-1}$ luminosity.} 
\label{fig:tevemiss}
\end{center}
\begin{center}
\includegraphics[angle=90,width=0.48\textwidth]{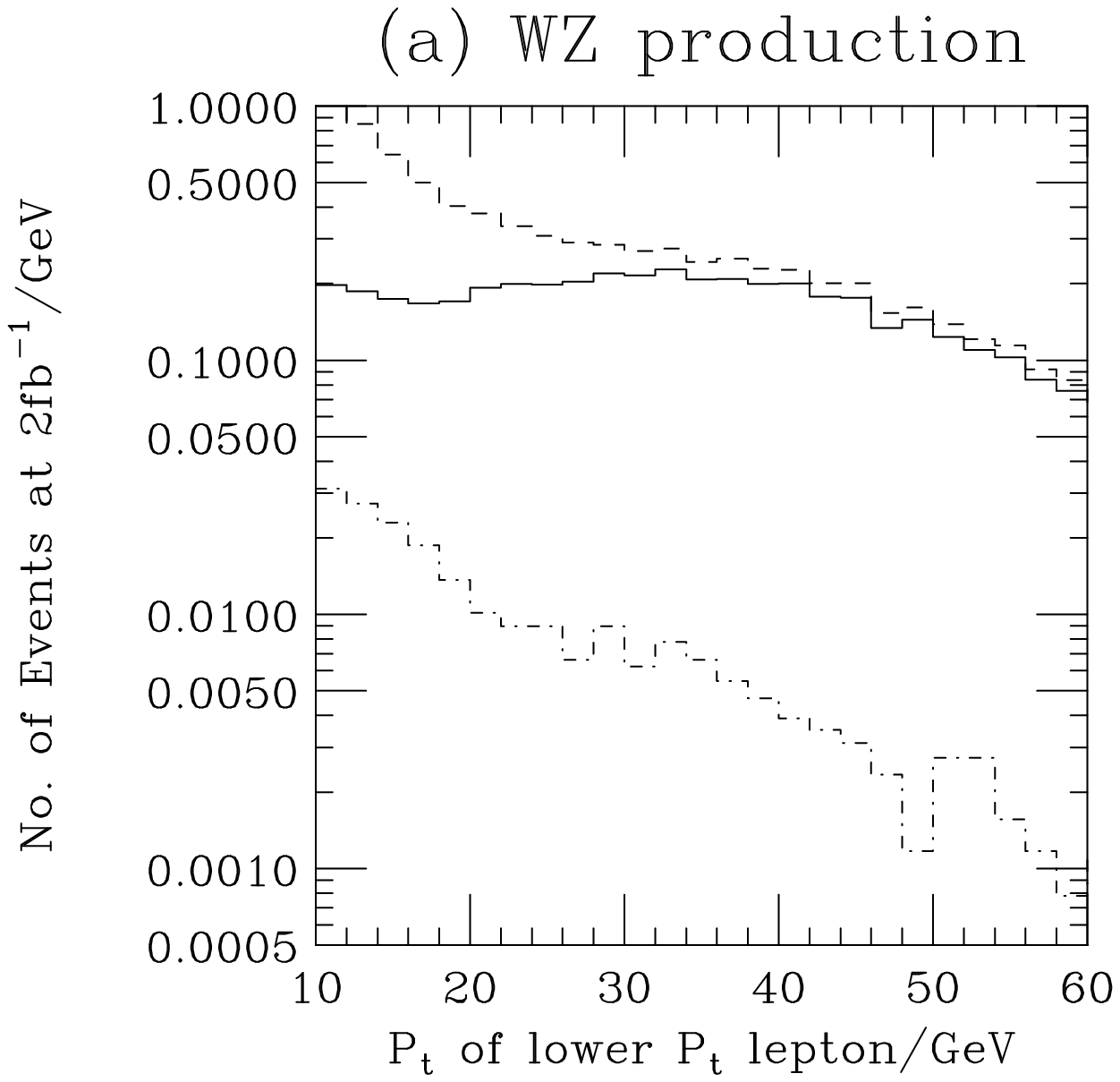}
\hfill
\includegraphics[angle=90,width=0.48\textwidth]{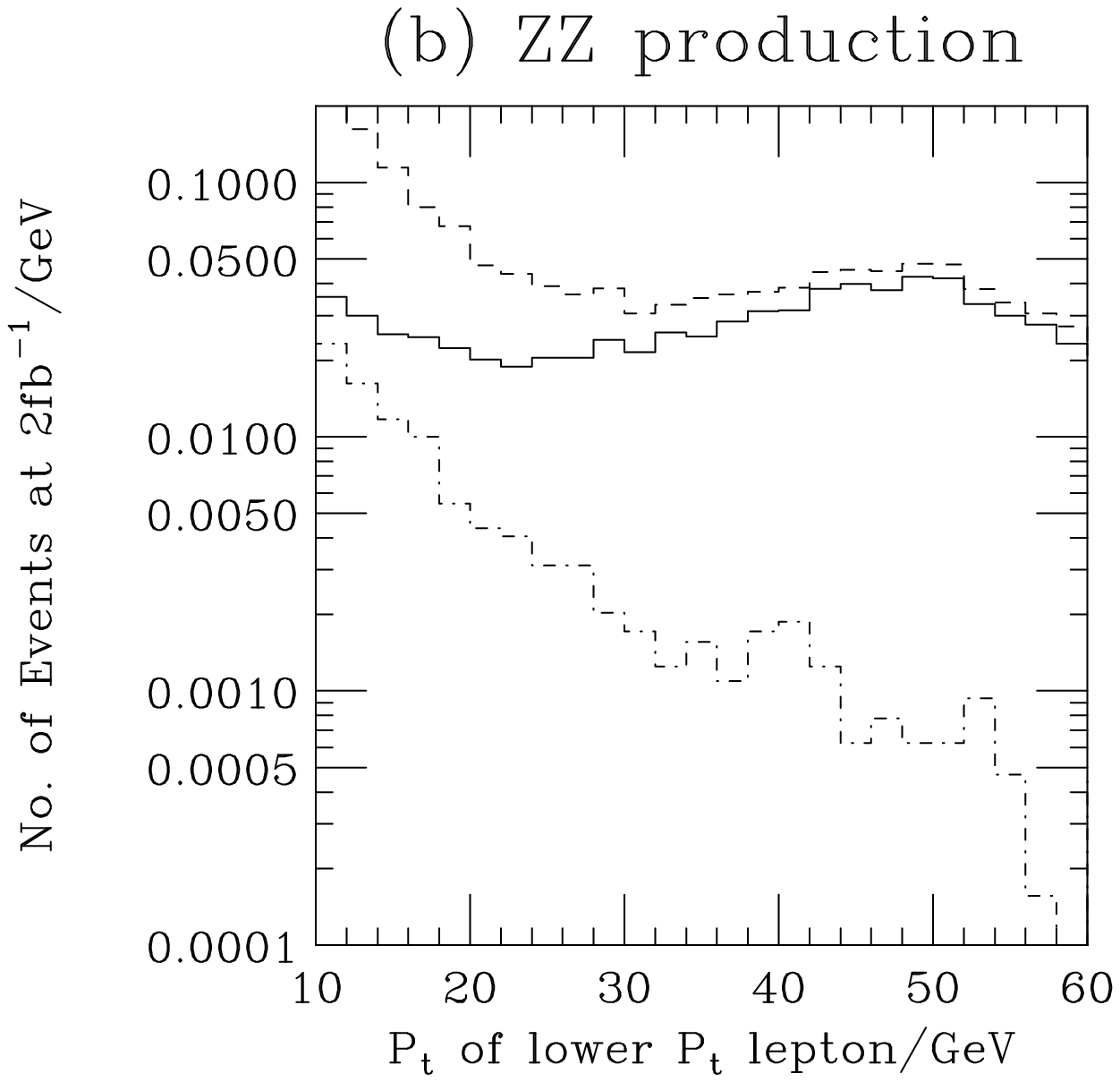}\\
\caption{Effect of the isolation cuts on the WZ and ZZ backgrounds at Run II
	of
	the Tevatron. The dashed line gives the background before any cuts,
	the solid line shows the effect of the isolation cut described in the
	text. The dot-dash line gives the effect of all the cuts, including
	the cut on the number of jets. The distributions are normalized to
	$2\  \mr{fb}^{-1}$ luminosity.} 
\label{fig:tevgaugeiso}
\end{center}
\end{figure}

%
%
\begin{table}
\begin{center}
\begin{tabular}{|c|c|c|c|c|}
\hline
	& \multicolumn{4}{c|}{Number of Events} \\
\cline{2-5}
 		& 	     	  &  		    & After isolation,&\\
Background      & After $p_T$ cut & After isolation & $p_T$, $M_T$, \met\  
cuts & 
		 After all cuts \\
 Process        &                 & and $p_T$ cuts  & and OSSF lepton & \\
	        & 		  &		    & veto.  & \\
\hline
WW 		& $0.23\pm0.02$			& $0.0\pm0.003$ 
	        & $0.0\pm0.003$	 		& $0.0\pm0.003$ \\
\hline		                
WZ 		& $9.96\pm0.09$			& $7.93\pm0.08$
		& $0.21\pm0.01$ 		& $0.21\pm0.01$ \\
\hline				                
ZZ 		& $2.05\pm0.03$			& $1.61\pm0.02$	
		& $0.069\pm0.005$ 		& $0.069\pm0.005$ \\
\hline		                
$\mr{t\bar{t}}$ & $34.1\pm1.6$ 			& $0.028\pm0.002$
		& $0.0032\pm0.0006$ 		& $0.0016\pm0.0004$ \\
\hline		                
$\mr{b\bar{b}}$ & $(3.4\pm1.3)\times10^3$	& $0.15\pm0.16$
		& $0.15\pm0.16$ 		& $0.15\pm0.16$ \\
\hline	                
Single Top 	& $1.77\pm0.01$ 		& $0.0014\pm0.0003$	
		& $0.0001\pm0.0001$ 		& $0.0001\pm0.0001$ \\
\hline		                
Total 	 	& $(3.4\pm1.3)\times10^3$	& $9.72\pm0.18$ 
		& $0.43\pm0.16$ 		& $0.43\pm0.16$ \\
\hline
\end{tabular}
\end{center}
\caption{Backgrounds to like-sign dilepton production at
         Run II of the Tevatron. The numbers of events
	are based on an integrated luminosity of $2\  \mr{fb}^{-1}$.
	We have calculated an error on the cross section by varying the
	scale between half and twice the hard scale, apart from the
	gauge boson pair production cross section where we do not have this
	information and the effect of varying the scale is expected
	to be small
	anyway. The error on the number of events is then the error on
	the cross section and the statistical error from the Monte
	Carlo simulation added in quadrature. If no events passed
	the cut the statistical error was taken to be the same
	as if one event had passed the cuts.}
\label{tab:tevback}
\end{table}

  The first two cuts are designed to reduce the background from heavy
  quark production, which is the major source of background before any
  cuts. As can be seen in Fig.\,\ref{fig:tevheavyiso} the cut on the
  transverse momentum, $p_T>20\, \mr{\gev}$,
  reduces the background by several
  orders of magnitude and the addition of the isolation cut reduces this
  background to less than one event at Run II of the Tevatron.

  The remaining cuts reduce the background from gauge boson pair
  production which dominates the Standard Model background after the
  imposition of the isolation and $p_T$ cuts. Fig.\,\ref{fig:tevemiss}a
  shows that the cut on the transverse mass,
  \ie removing the region \linebreak
  \mbox{$60\, \mr{\gev} < M_T < 85\, \mr{\gev}$}, for
  each of the like-sign leptons
  will reduce the background
  from WZ production, which is the largest of the gauge boson pair
  production backgrounds. Similarly the cut on the missing transverse
  energy, $\not\!\!E_T<20\, \mr{\gev}$, 
  will significantly reduce the background from WZ production as
  can be seen in Fig.\,\ref{fig:tevemiss}b. The effect of these cuts is
  shown in Fig.\,\ref{fig:tevgaugeiso}.  Our simulations do not
  include $\mr{W\gamma}$ production which was recently found to be
  a major source of background to like-sign
  dilepton production in the MSSM \cite{Matchev:1999yn}. However, we
  would expect this to be less important here due to the different cuts
  we have applied. In particular, in the analysis of \cite{Matchev:1999yn} a
  cut on the invariant mass of OSSF lepton pairs was imposed to reduce the
  background from Z production, rather than the veto on the presence of OSSF
  leptons which we have used. The veto and missing transverse energy cut
  will reduce the number of events from
  $\mr{W\gamma}$ production while the cut on the invariant mass will not
  suppress this background.

  The effect of all these cuts on the background is given in 
  Table~\ref{tab:tevback}. While the dominant background is from WZ
  production,
  the dominant contribution to the error comes from $\mr{b\bar{b}}$
  production. 
  This can only be reduced with a significantly more elaborate simulation.

  We also need to calculate the acceptance of these cuts for the signal.
  To estimate the acceptance of the cuts we simulated 
  twenty thousand events at one hundred points in the $M_0$, $M_{1/2}$ plane.
  The acceptance was then interpolated between the points and multiplied by
  the cross section to give the number of
  signal events passing the cuts.
  This can then be used to find the discovery potential
  by comparing the number of signal events with a $5\sigma$
  statistical fluctuation of the background. 

\begin{figure}
\begin{center}
\includegraphics[angle=90,width=0.48\textwidth]{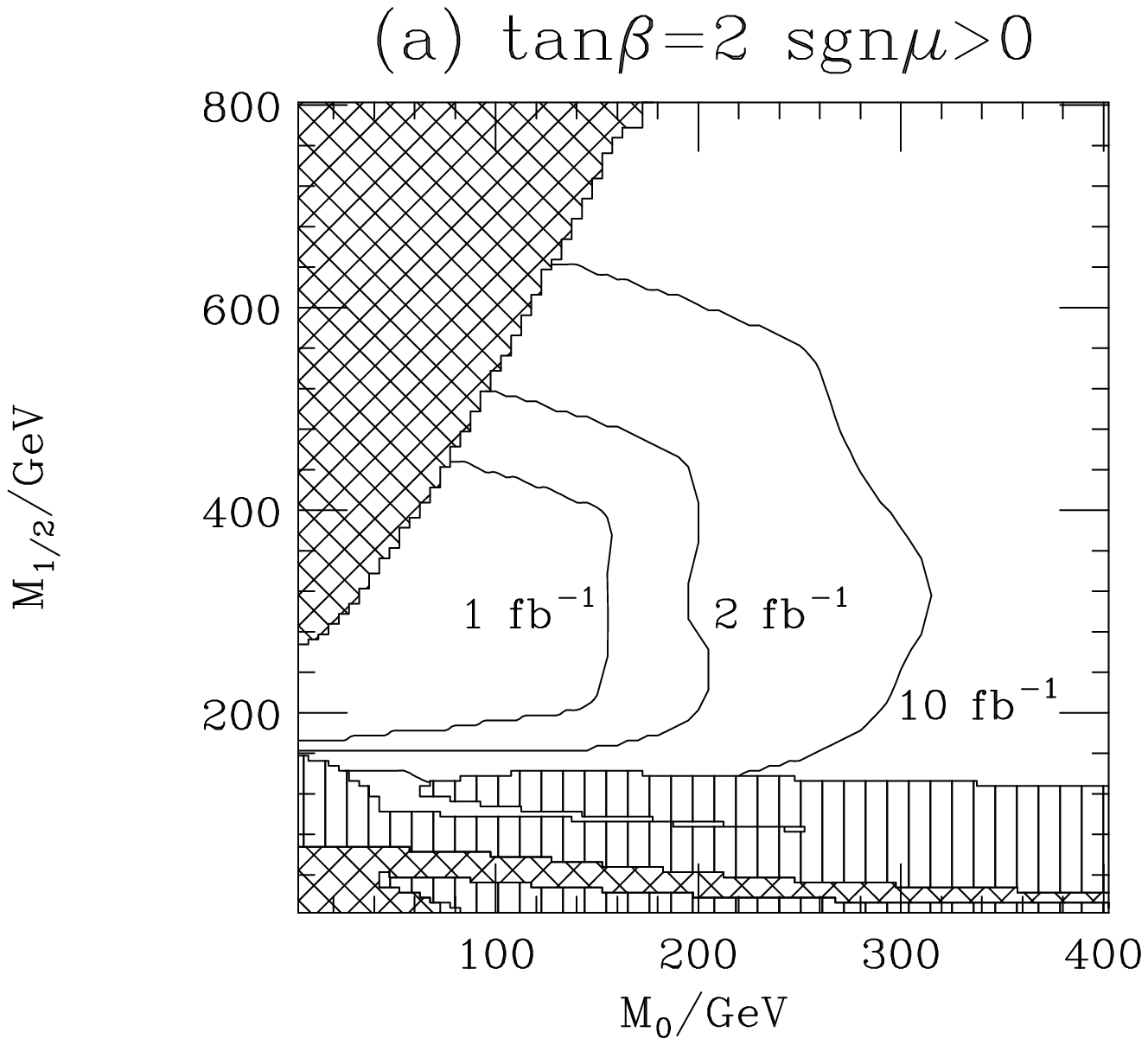}
\hfill
\includegraphics[angle=90,width=0.48\textwidth]{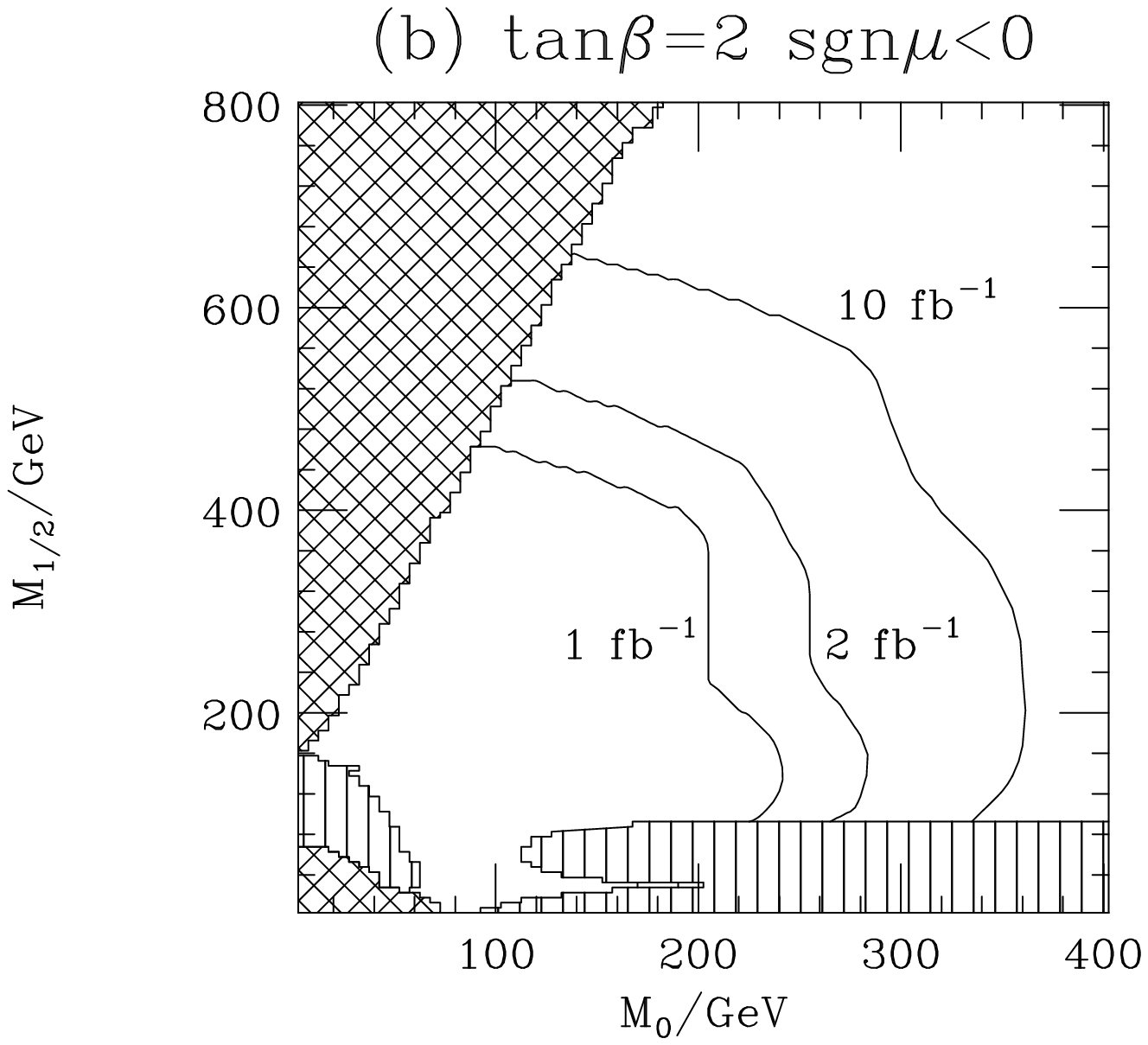}\\
\vskip 7mm
\includegraphics[angle=90,width=0.48\textwidth]{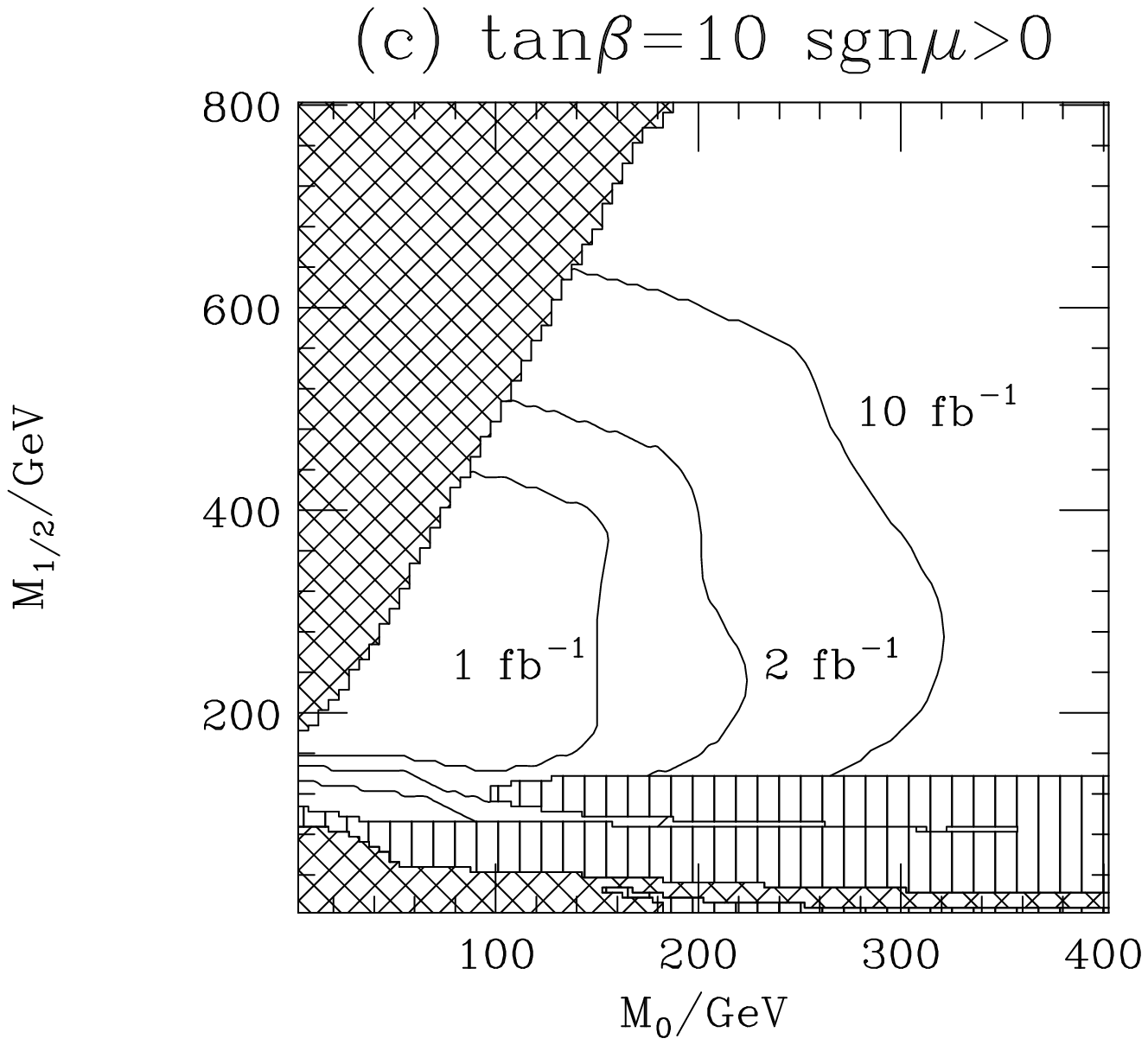}
\hfill
\includegraphics[angle=90,width=0.48\textwidth]{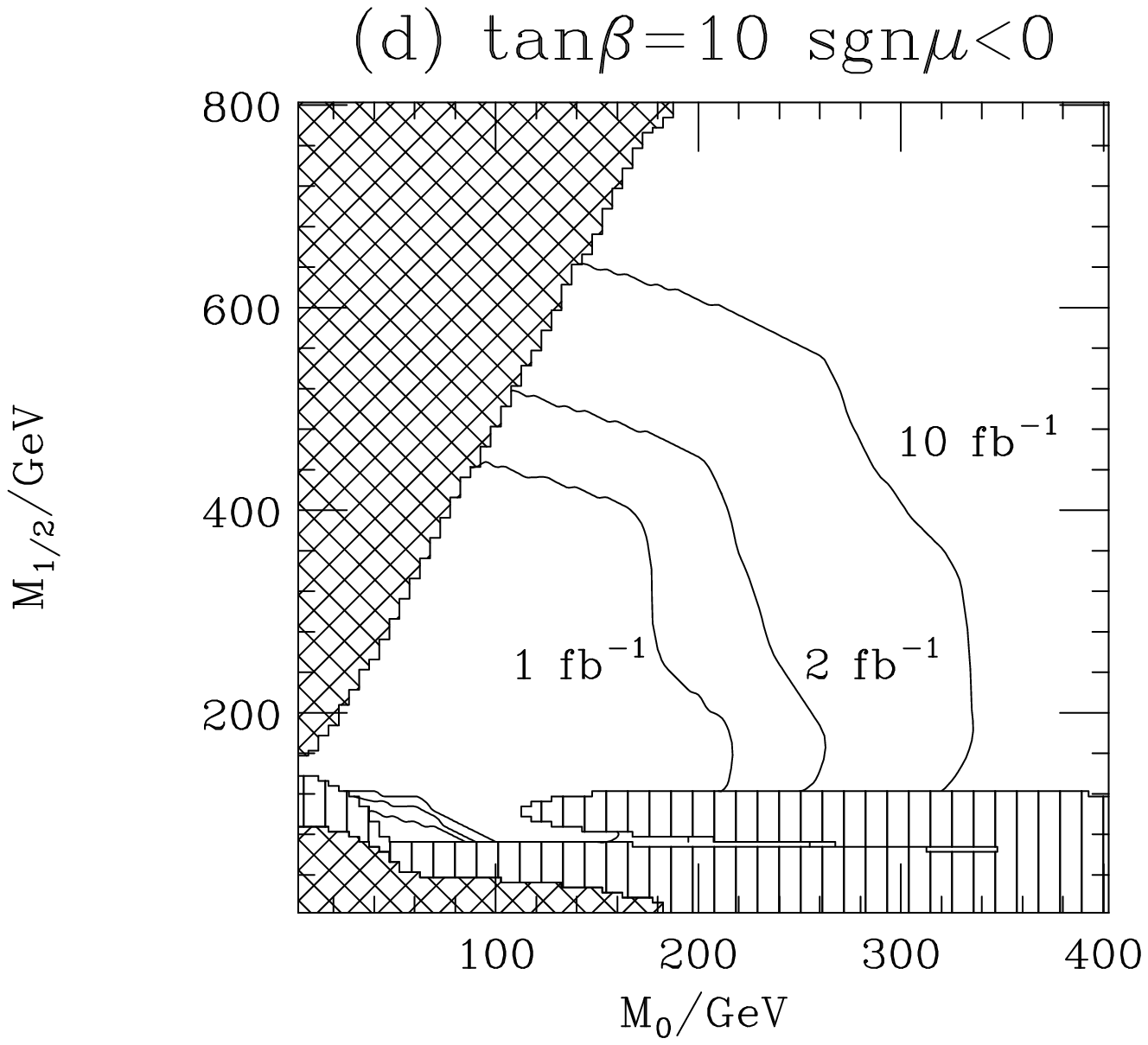}\\
\caption{Contours showing the discovery potential of the Tevatron in the
	 $M_0$,
	 $M_{1/2}$ plane for ${\lam'}_{211}=10^{-2}$ and $A_0=0\, \mr{\gev}$.
	 These are a $5\sigma$ excess of the signal above the
	 background. Here we have imposed the cuts on the isolation and $p_T$
  	 of the leptons, the transverse mass
	 and the missing transverse energy
	 described in the text, and a veto on the presence of OSSF leptons.
	 We have only considered the Standard Model background. The
         striped and hatched regions are
	 described in the caption of Fig.\,\ref{fig:SUSYmass}.} 
\label{fig:tevSMnojet}
\end{center}
\end{figure}

\begin{figure}
\begin{center}
\includegraphics[angle=90,width=0.48\textwidth]{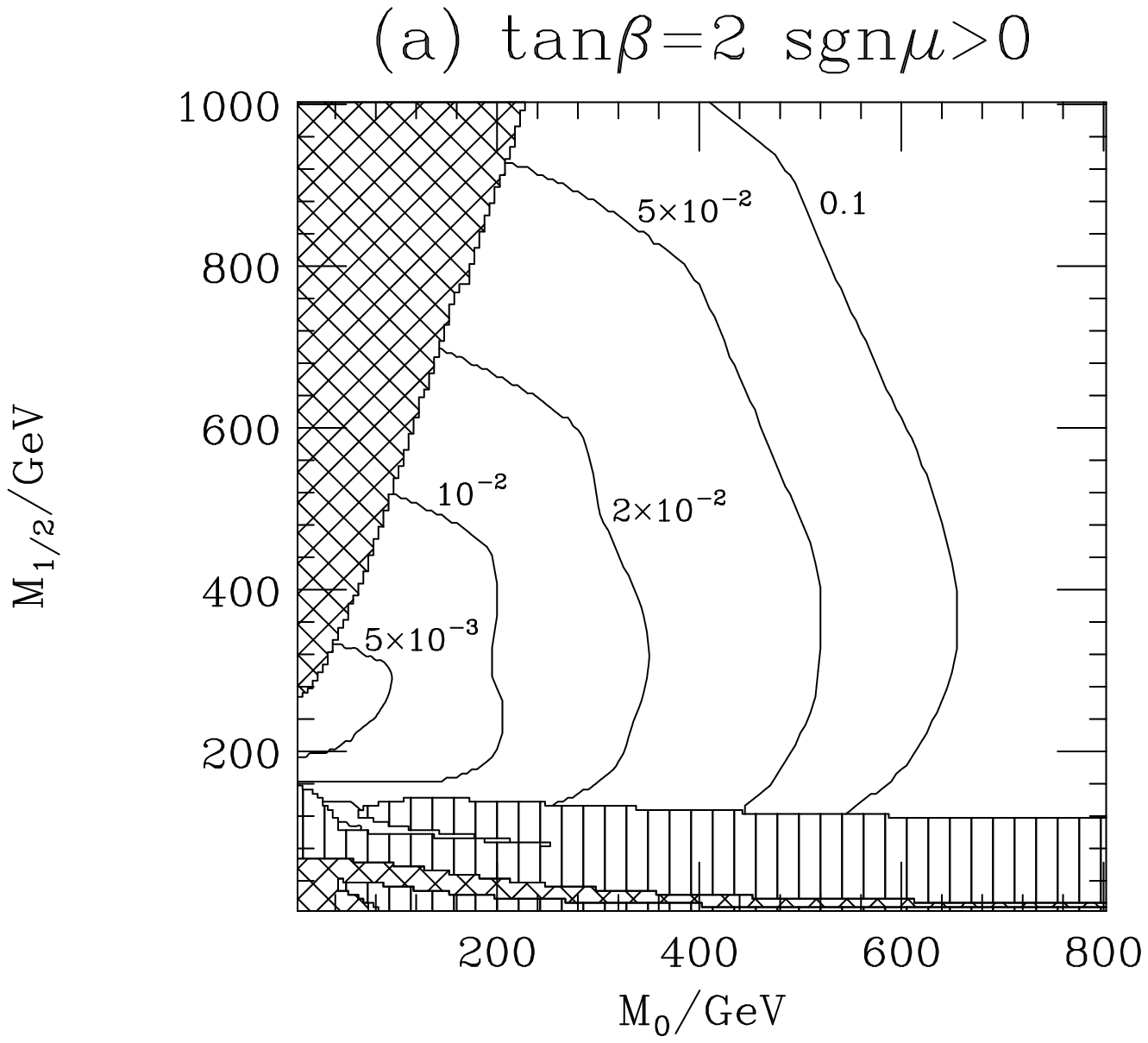}
\hfill
\includegraphics[angle=90,width=0.48\textwidth]{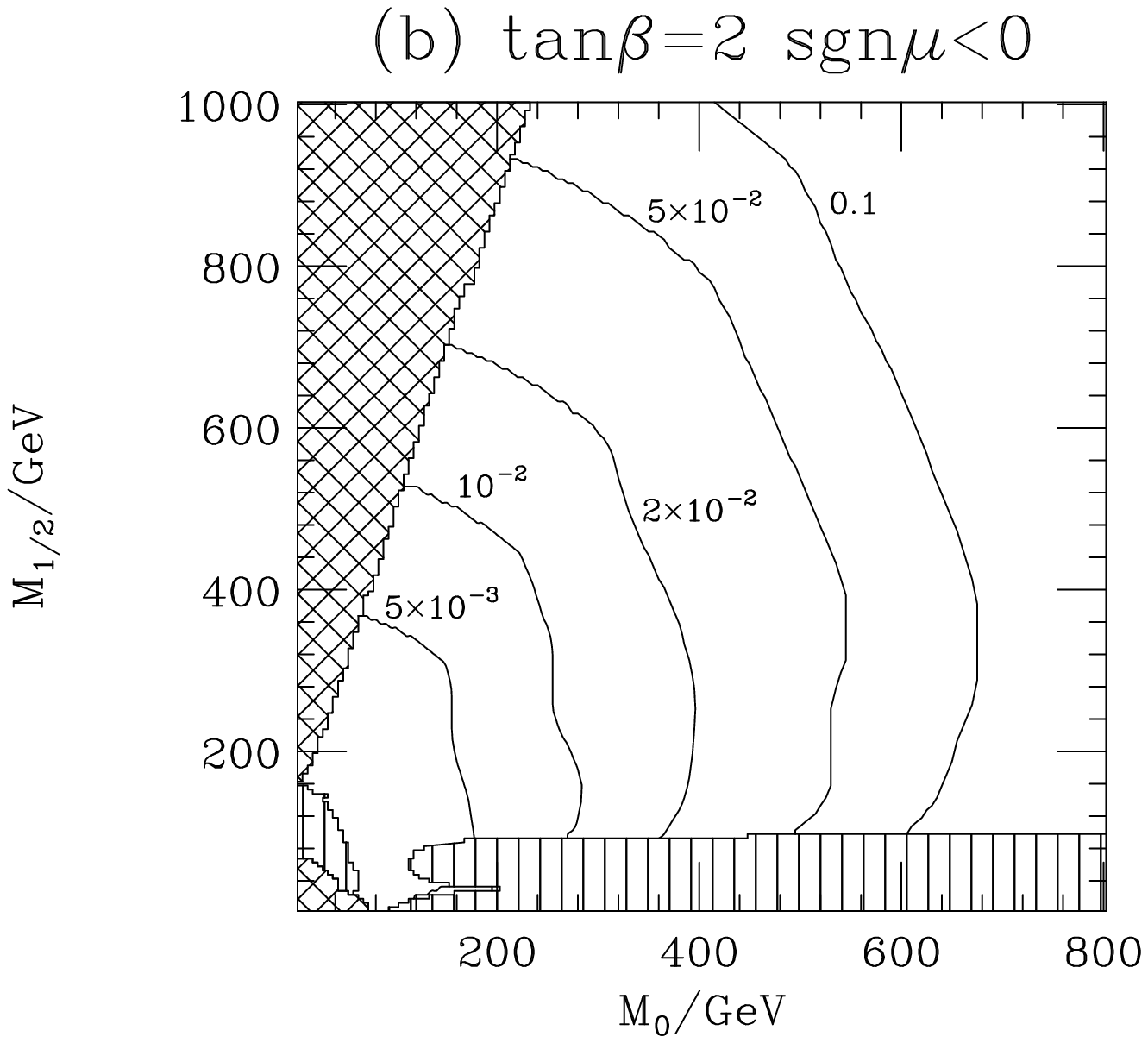}\\
\vskip 7mm
\includegraphics[angle=90,width=0.48\textwidth]{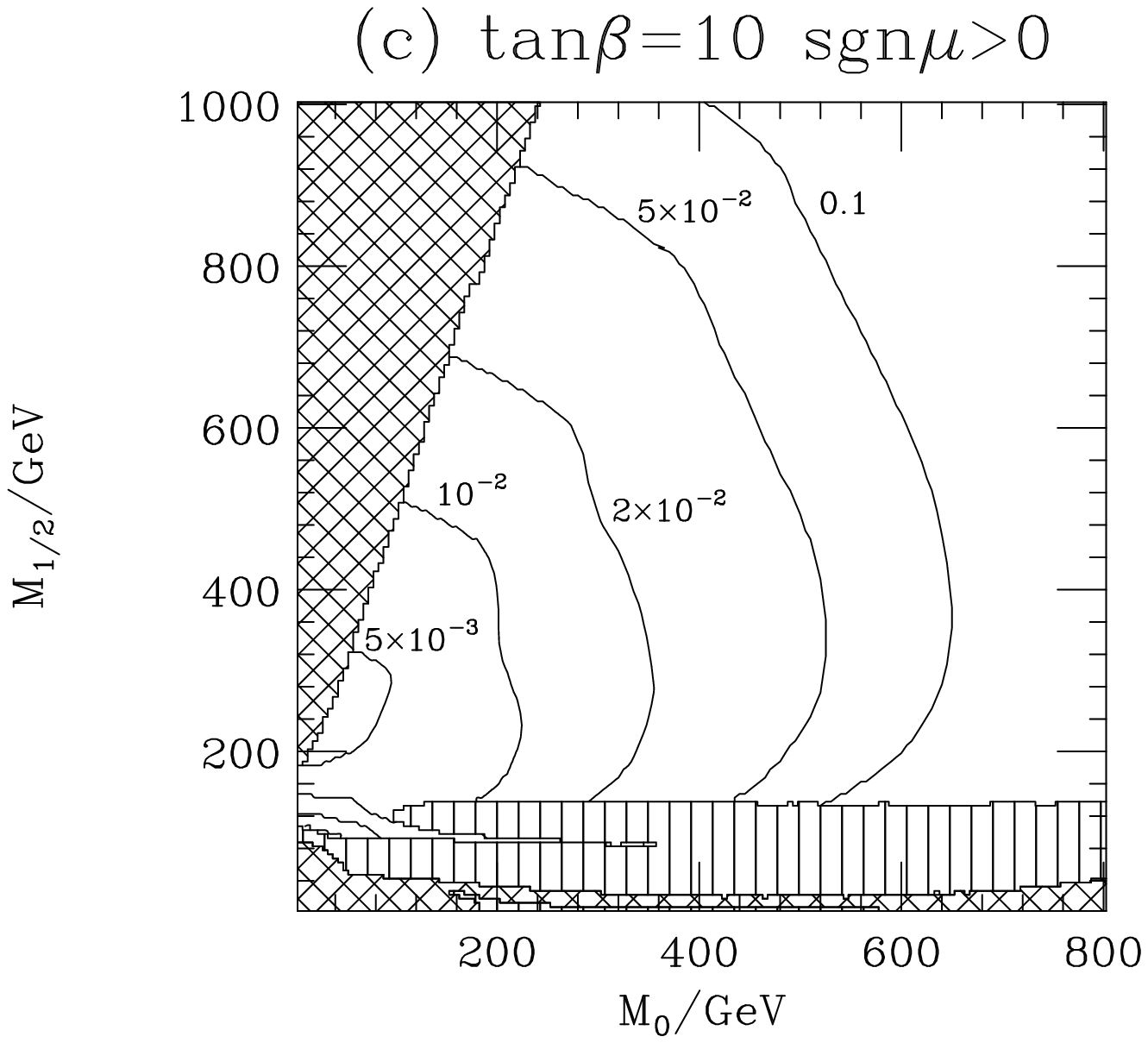}
\hfill
\includegraphics[angle=90,width=0.48\textwidth]{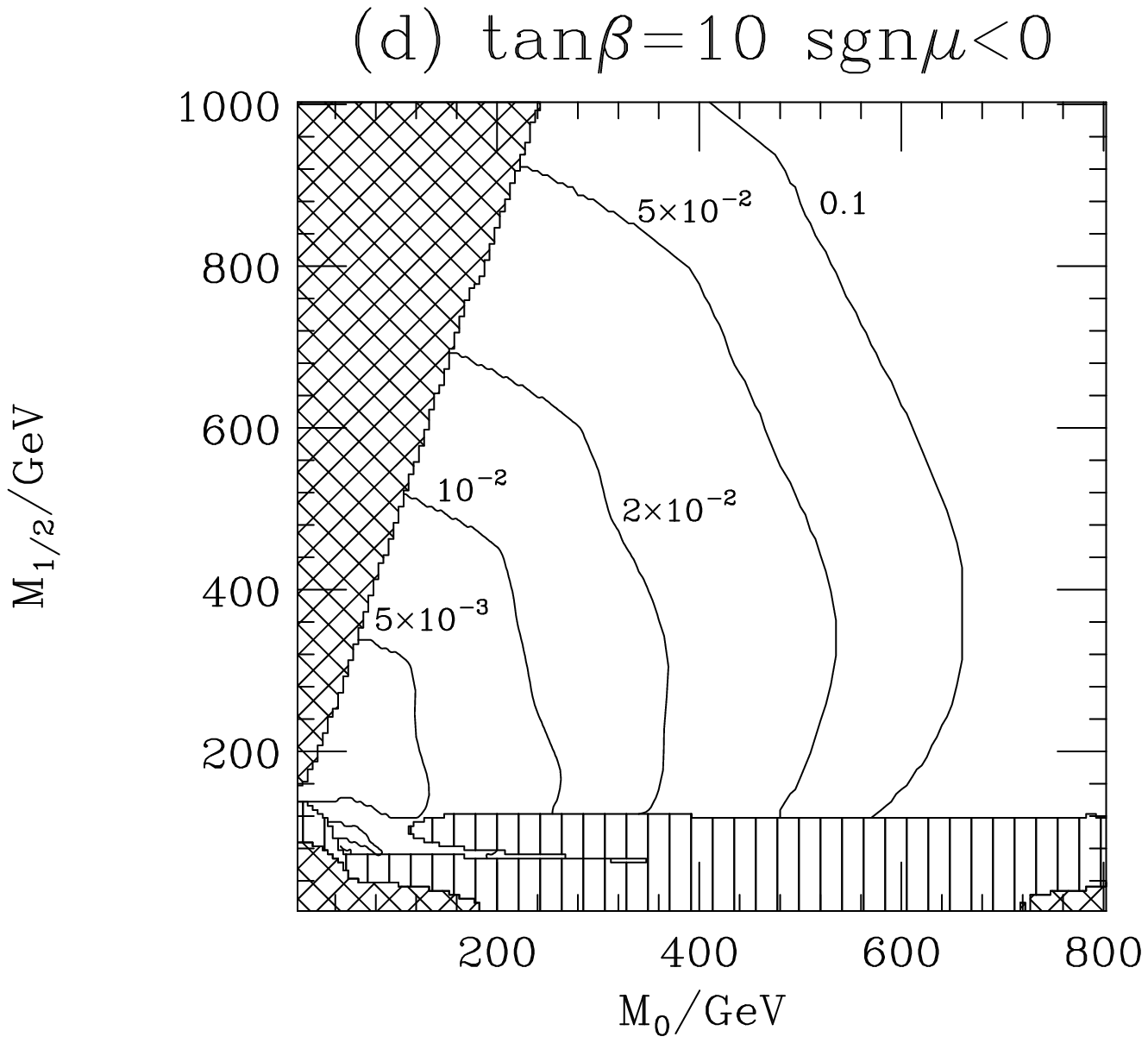}\\
\caption{Contours showing the discovery potential of the Tevatron in the
	 $M_0$,
	 $M_{1/2}$ plane for $A_0=0\, \mr{\gev}$   and 
         $2\ \mr{fb}^{-1}$ integrated luminosity for different values
	 of ${\lam'}_{211}$.
	 These are a $5\sigma$ excess of the signal above the
	 background. Here we have imposed the cuts on the isolation and $p_T$
  	 of the leptons, the transverse mass and 
	 the missing transverse energy
	 described in the text, and a veto on the presence of OSSF leptons.
	 We have only considered the Standard Model background. The
         striped and hatched regions are
	 described in the caption of Fig.\,\ref{fig:SUSYmass}.} 
\label{fig:tevSMnojetb}
\end{center}
\end{figure}

  Fig.\,\ref{fig:tevSMnojet} shows the discovery potential,
  for different integrated luminosities and a fixed
  value of the coupling ${\lam'}_{211}=10^{-2}$, if we only consider
  the Standard Model backgrounds and apply the cuts we described to suppress
  these backgrounds. Fig.\,\ref{fig:tevSMnojetb} shows the effect
  of varying the
  \rpv\  coupling for $2\  \mr{fb}^{-1}$ integrated luminosity with the
  same assumptions.

  We have taken a conservative approach where the background
  is taken to be one standard deviation above the central value. Due to the 
  small number of events we must use Poisson statistics, this means that for
  the Standard Model background given in Table~\ref{tab:tevback}, 7 events
  corresponds to the same probability as a  $5 \sigma$ statistical
  fluctuation for a
  Gaussian distribution. Here we have used 0.59 events as a conservative
  estimate of the background, \ie a $1\sigma$ fluctuation above our central
  value.

  For small couplings there are regions, for low $M_{1/2}$, which cannot
  be observed even for small smuon masses. For larger couplings
  however we can probe masses of up to $430\,(500)\, \mr{\gev}$ 
  for a coupling
  ${\lam'}_{211}=0.05$ with 2\,(10)~$\mr{fb}^{-1}$ integrated luminosity.
  Masses of up to $520\,(600)\,\mr{\gev}$  can be observed for a coupling of
  ${\lam'}_{211}=0.1$ with 2\,(10)~$\mr{fb}^{-1}$ integrated luminosity.

  We have neglected the non-physics background. This mainly comes from fake 
  leptons in W production. The cuts we have applied to reduce the gauge boson
  pair production backgrounds, in particular the cuts on the missing
  transverse
  energy and the transverse mass, will significantly reduce this
  background. It was noted in \cite{Matchev:1999nb} that the cross
  section
  falls extremely quickly with the $p_T$ of the fake lepton, and hence the
  large $p_T$ cut we have imposed will suppress this background.
  A proper treatment of the non-physics background 
  requires a simulation of the detector. This is
  beyond the scope of this paper.

%
%

%
%

  In Figs.\,\ref{fig:tevSMnojet} and \ref{fig:tevSMnojetb} the background
  from sparticle pair production is neglected. This is
  reasonable in an initial search where presumably an experiment would be
  looking for an excess of like-sign dilepton pairs, rather than worrying
  about
  precisely which model was giving the excess. If such an 
  excess were observed,
  it would then be necessary to establish which physical processes were
  producing
  the excess. In the \rpv\  MSSM there are two possible mechanisms which
  could produce such an excess: either resonant sparticle production; or
  sparticle pair production followed by the decay of the LSP.
  We will now consider additional cuts which will suppress the
  background to resonant slepton production 
  from sparticle pair production and hopefully allow these two 
  scenarios to be distinguished.

\subsubsection{SUSY backgrounds}

  We have seen that by imposing cuts on the transverse momentum and isolation
  of the like-sign dileptons, the missing transverse energy, the transverse
  mass and the presence of OSSF leptons the Standard Model backgrounds  
  can be significantly reduced. However a significant
  background from sparticle pair production still remains.
  We therefore imposed the following additional cut to 
  reduce this background:
\begin{itemize}
\item 	Vetoing all events when there are more than two jets each with
	$p_T>20\, \mr{\gev}$. 
\end{itemize}
  While this cut slightly reduces the signal
  it also dramatically reduces the
  background from sparticle pair production. We performed a scan of the
  SUGRA parameter space at the four values of $\tan\beta$ and $\sgn\mu$
  considered in Section~\ref{sec:signal}. We generated fifty thousand
  events at each of one hundred points in the $M_0$, $M_{1/2}$ plane at
  each value of $\tan\beta$ and $\sgn\mu$, and then interpolated between
  these points as for the signal process. This allowed us to estimate an
  acceptance for the cuts which we multiplied by the sparticle pair
  production cross section to give a number of
  background events.

\begin{figure}
\begin{center}
\includegraphics[angle=90,width=0.48\textwidth]{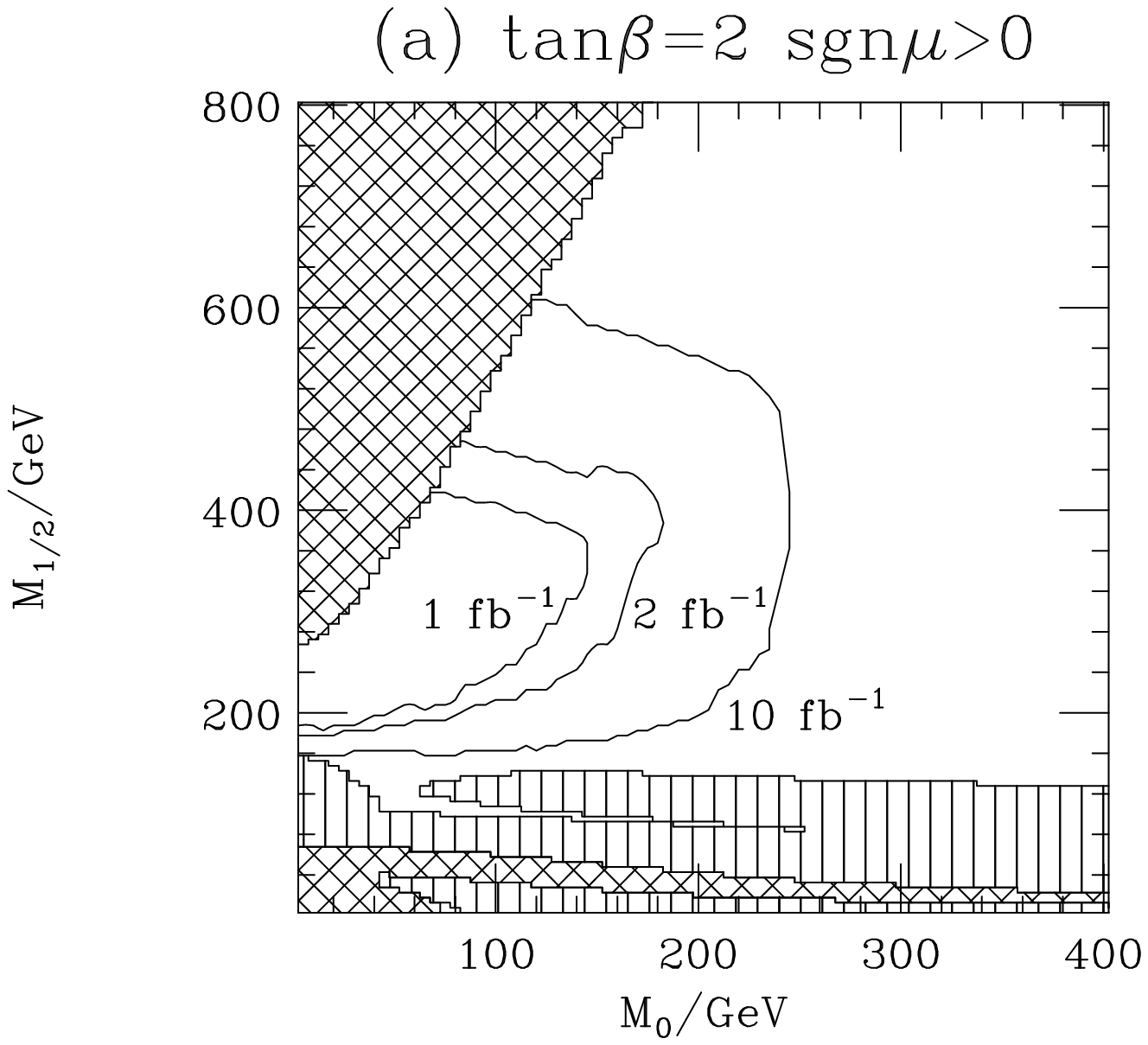}
\hfill
\includegraphics[angle=90,width=0.48\textwidth]{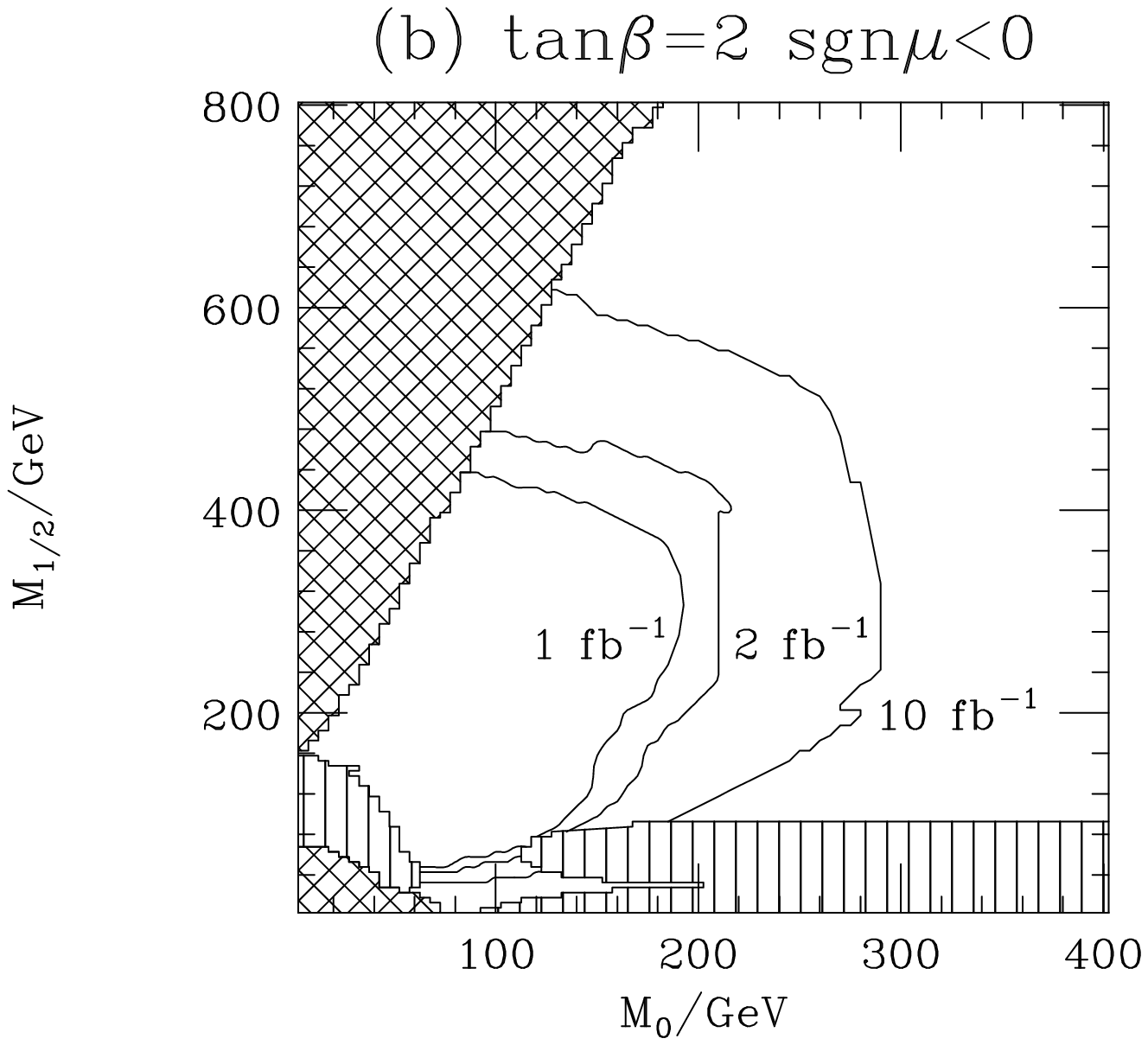}\\
\vskip 7mm
\includegraphics[angle=90,width=0.48\textwidth]{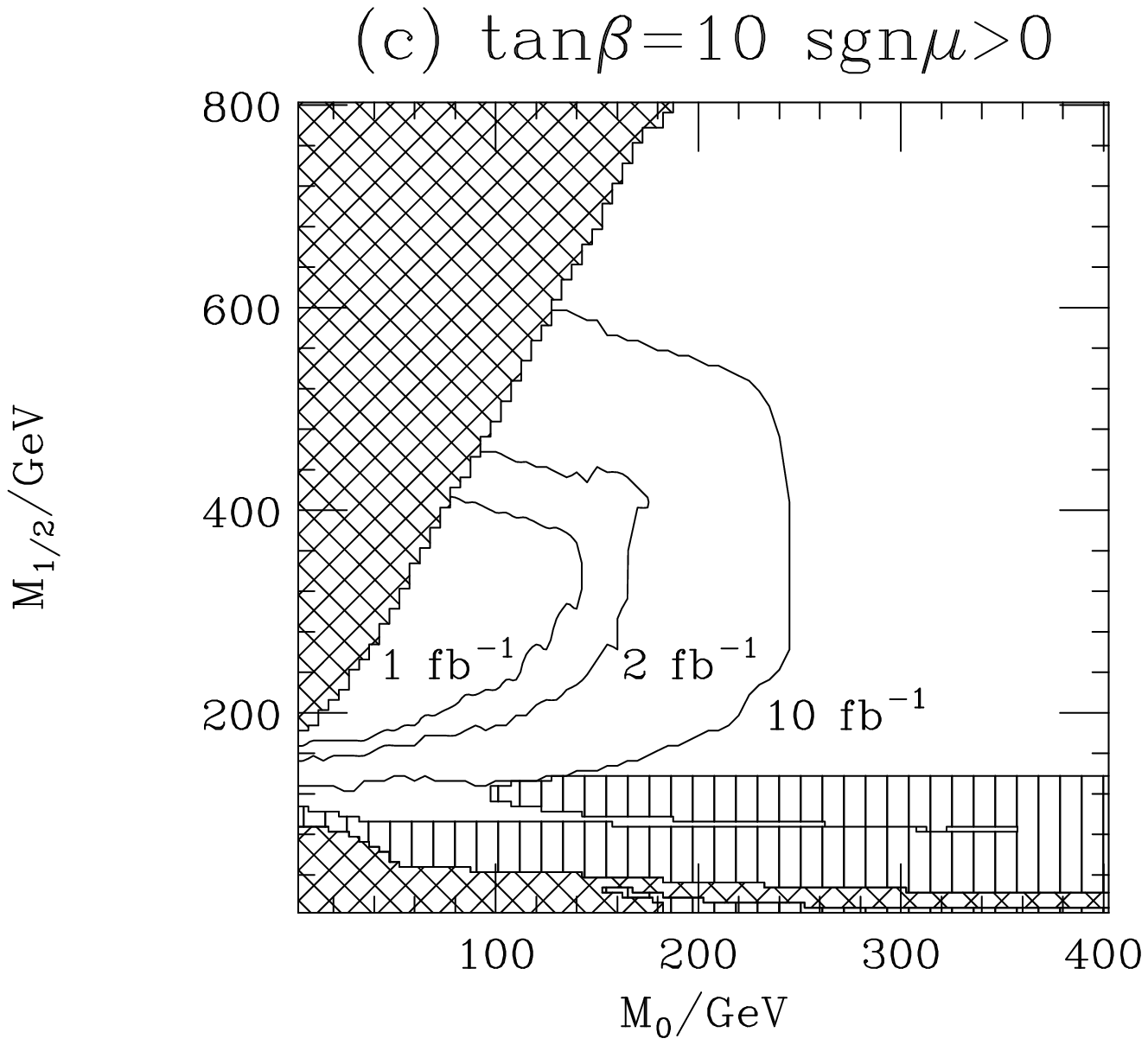}
\hfill
\includegraphics[angle=90,width=0.48\textwidth]{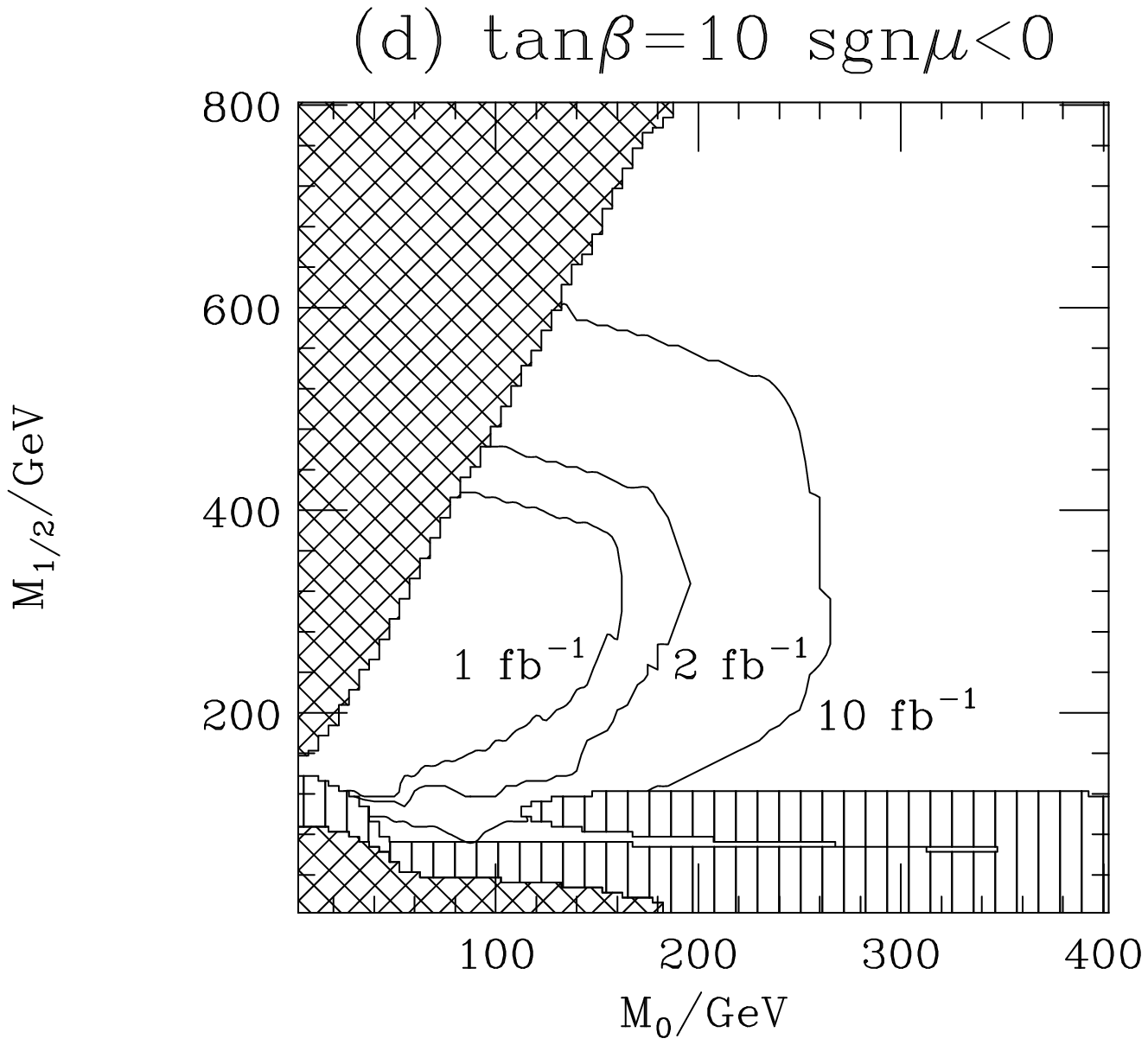}\\
\caption{Contours showing the discovery potential of the Tevatron in the
	 $M_0$,
	 $M_{1/2}$ plane for ${\lam'}_{211}=10^{-2}$ and $A_0=0\, \mr{\gev}$.
	 These are a $5\sigma$ excess of the signal above the
	 background. Here in addition to the cuts on the isolation and $p_T$
  	 of the leptons, the transverse mass
	 and the missing transverse energy
	 described in the text, and a veto on the presence of OSSF leptons
	 we have imposed a cut on the presence of more than two jets. This
	 includes the sparticle pair production background as well as the
	 Standard Model backgrounds. Again the striped and hatched regions
	 are as described in the caption of Fig.\,\ref{fig:SUSYmass}.}
\label{fig:tevSUSYjet}
\end{center}
\end{figure}

\begin{figure}
\begin{center}
\includegraphics[angle=90,width=0.48\textwidth]{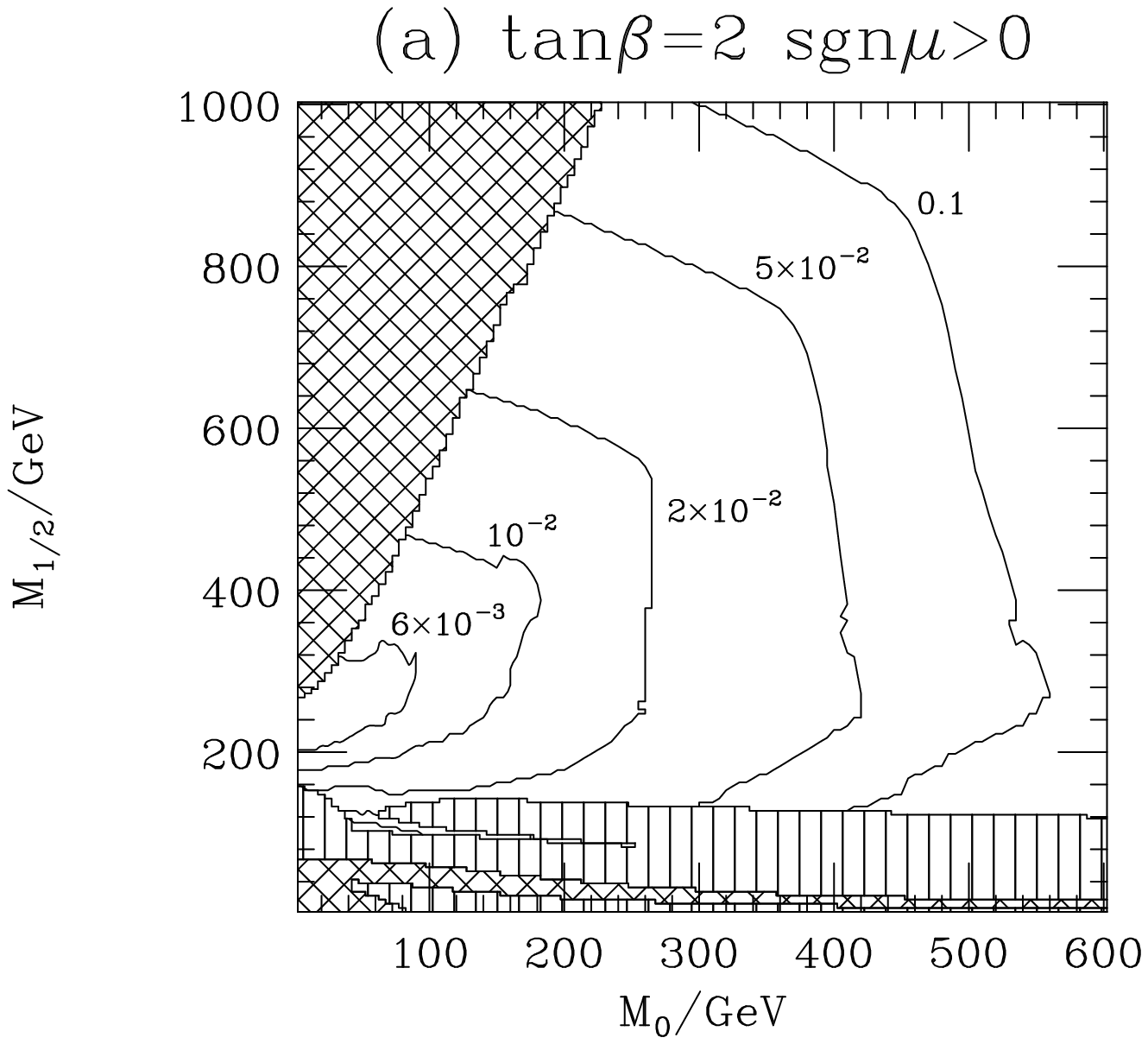}
\hfill						
\includegraphics[angle=90,width=0.48\textwidth]{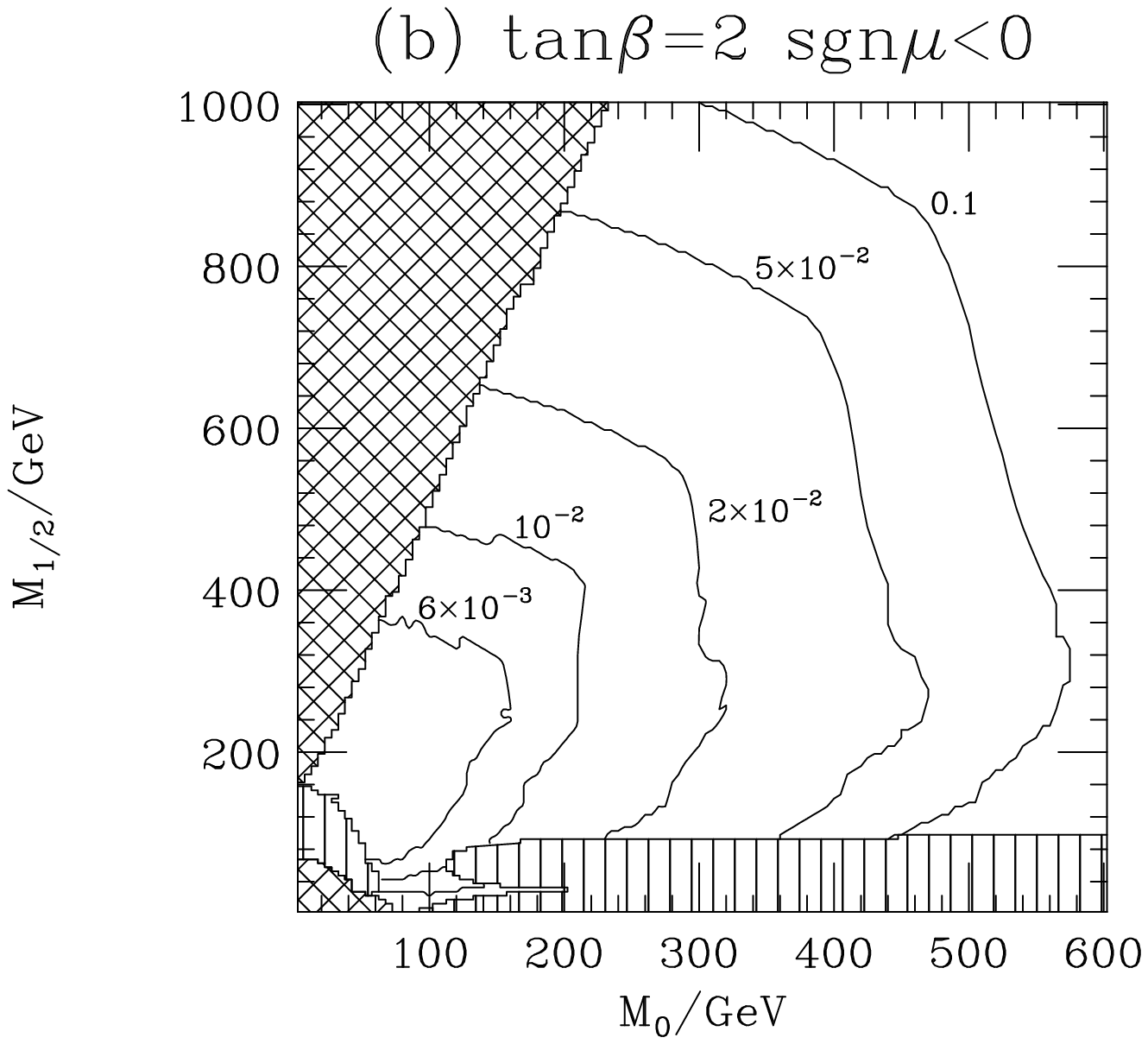}\\
\vskip 7mm					
\includegraphics[angle=90,width=0.48\textwidth]{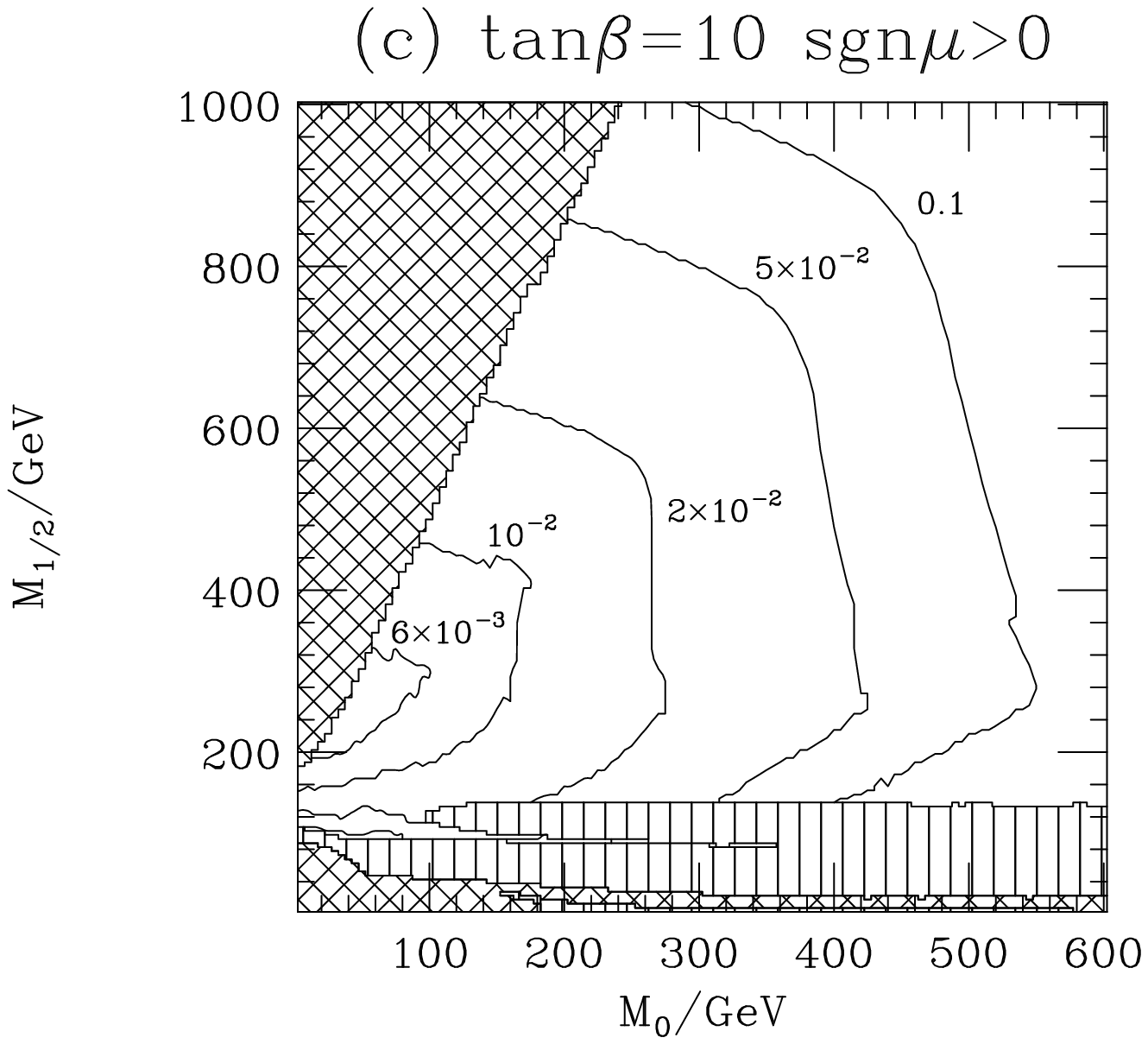}
\hfill						
\includegraphics[angle=90,width=0.48\textwidth]{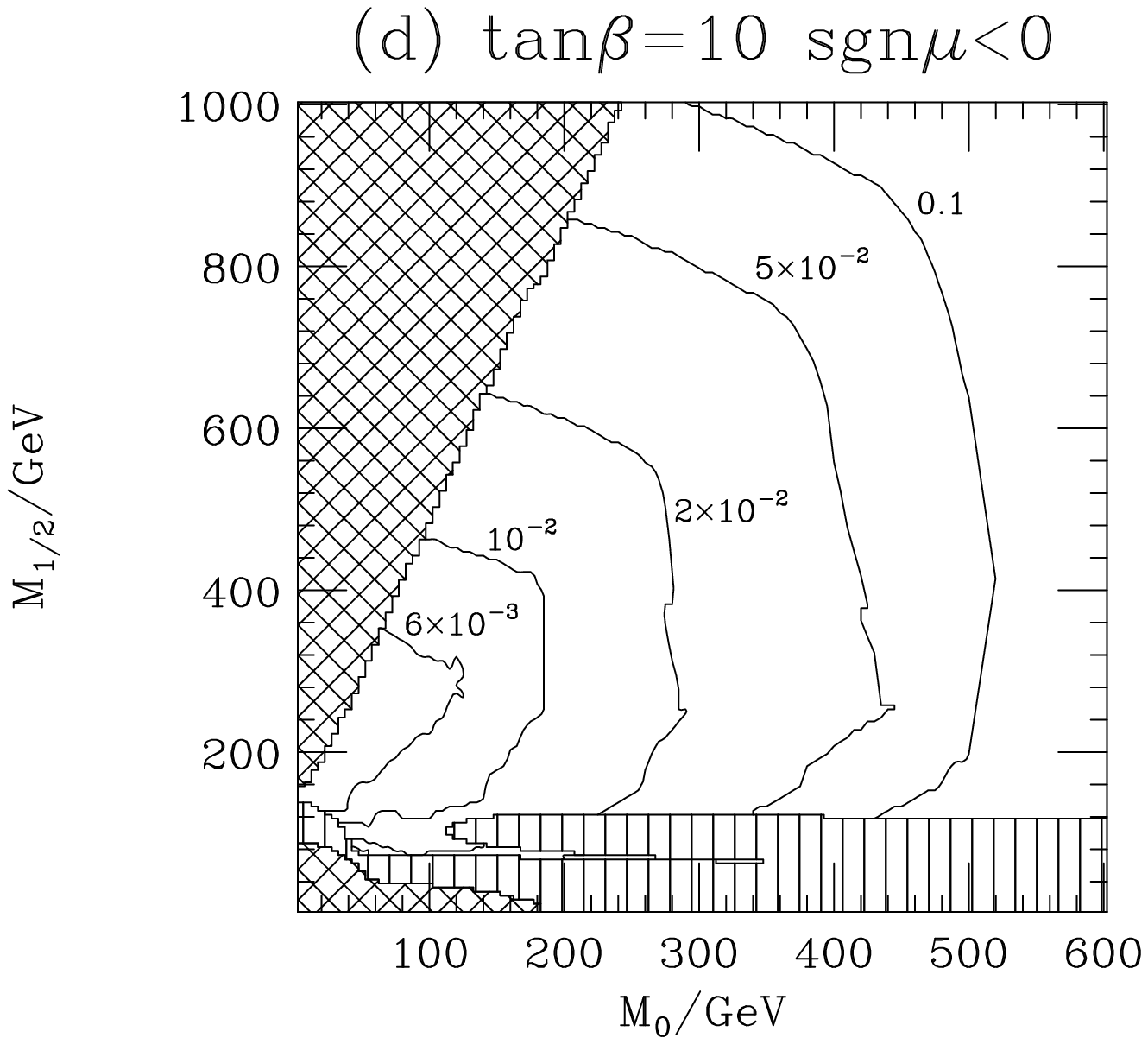}\\
\caption{Contours showing the discovery potential of the Tevatron in the
	 $M_0$,
	 $M_{1/2}$ plane for $A_0=0\, \mr{\gev}$  and an
	 integrated luminosity of
  	 $2\  \mr{fb}^{-1}$ for different values of ${\lam'}_{211}$.
	 These are a $5\sigma$ excess of the signal above the
	 background. Here in addition to the cuts on the isolation and $p_T$
  	 of the leptons, the transverse mass
	 and the missing transverse energy
	 described in the text, and a veto on the presence of OSSF leptons
	 we have imposed a cut on the presence of more than two jets. This
	 includes the sparticle pair production background as well as the
	 Standard Model backgrounds. Again the striped and hatched regions
	 are as described in the caption of Fig.\,\ref{fig:SUSYmass}.}
\label{fig:tevSUSYjetb}
\end{center}
\end{figure}

  The effect of all the cuts on the total background, \ie the Standard Model
  background and the sparticle pair production background is shown in 
  Fig.\,\ref{fig:tevSUSYjet} for different integrated luminosities with 
  ${\lam'}_{211}=10^{-2}$ and in Fig.\,\ref{fig:tevSUSYjetb} for an
  integrated luminosity of $2\  \mr{fb}^{-1}$ with different values 
  of ${\lam'}_{211}$.

  As can be seen, the effect of including the sparticle pair production
  background is to reduce the $5\sigma$ discovery regions. These regions are
  reduced for two reasons: for large $M_{1/2}$ the additional cut removes 
  more signal events and hence reduces the statistical significance of the
  signal;  at
  small values of $M_{1/2}$ there is a large background from sparticle pair
  production, relative to the SM background, which
  also reduces the statistical
  significance of the signal. However even for this relatively small value of
  the coupling there are large regions of parameter space in
  which a signal is
  visible above the background. The ratio of signal to background is still
  larger than one for most of the region where the signal is detectable above
  the background. For $\sgn\mu>0$
  there is only a very
  small region at low $M_{1/2}$ where $S/B$ drops below one and even here
  $S/B>0.5$. However for $\sgn\mu<0$ there are regions of 
  low $S/B$ for small values of $M_{1/2}$.
  The discovery range for these \rpv\  processes extends to
  larger values of 
  $M_{1/2}$ than the $5\sigma$ discovery curve for sparticle pair production
  as only one sparticle is produced which requires a much lower parton-parton
  centre-of-mass energy than sparticle pair production. 

  Again even for small smuon masses with low values of the \rpv\  Yukawa 
  coupling there are regions where a signal of resonant slepton production
  is not visible above the background. However for large couplings the
  signal in these regions is visible above the background. For a coupling
  of ${\lam'}_{211}=0.05$ a smuon mass of $310\,(330)\, \mr{\gev}$
  is visible above
  the background with 2\,(10)~$\mr{fb}^{-1}$ integrated luminosity, 
  and for a coupling of ${\lam'}_{211}=0.1$
  a smuon mass of $400\,(430)\,\mr{\gev}$ 
  is visible above the background with 2\,(10)~$\mr{fb}^{-1}$ integrated
  luminosity.

%
%

%
%
%
%
\subsection{LHC}
\label{sub:LHC}
 
%
%
\begin{figure}
\begin{center}
\includegraphics[angle=90,width=0.48\textwidth]{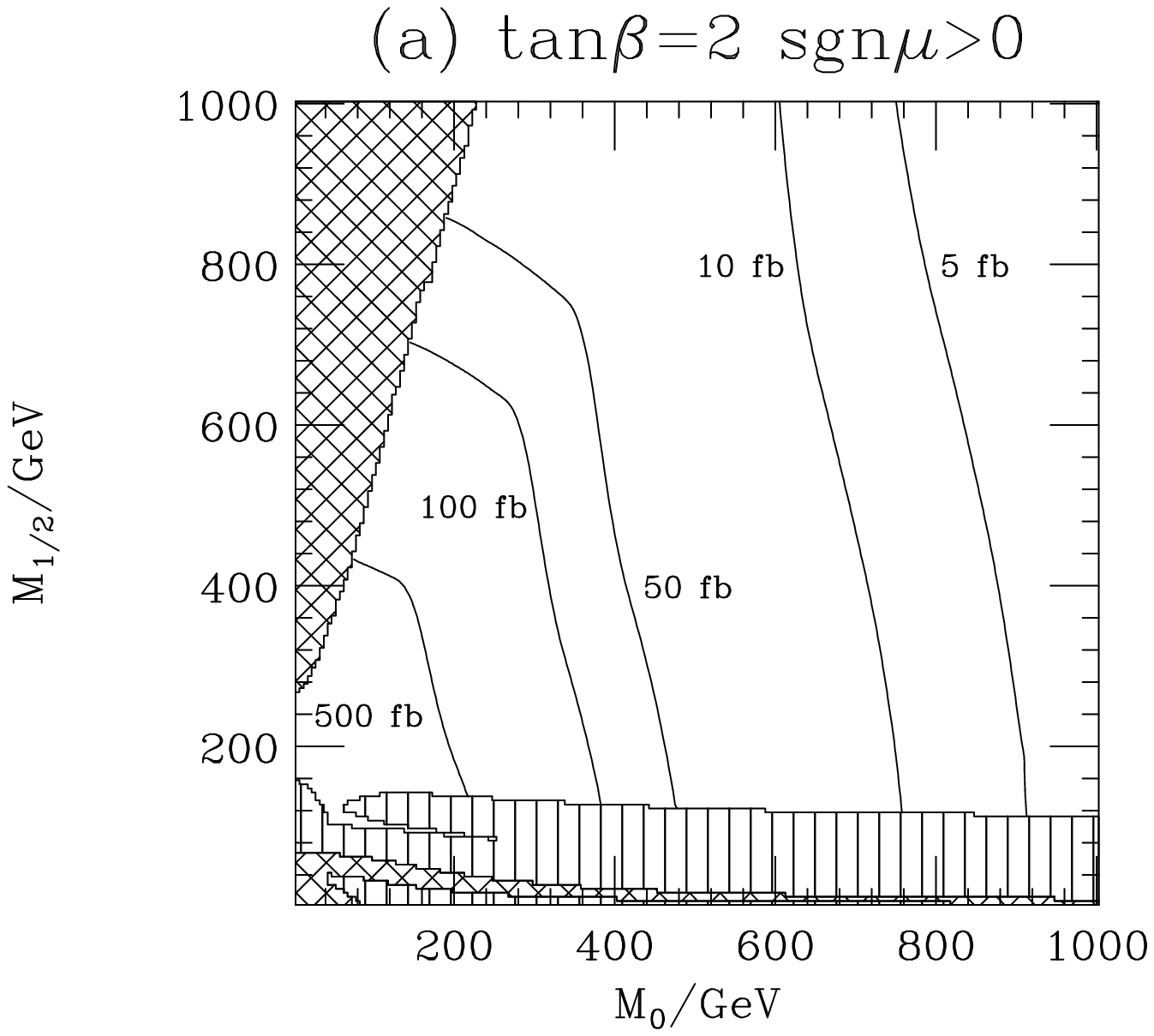}
\hfill
\includegraphics[angle=90,width=0.48\textwidth]{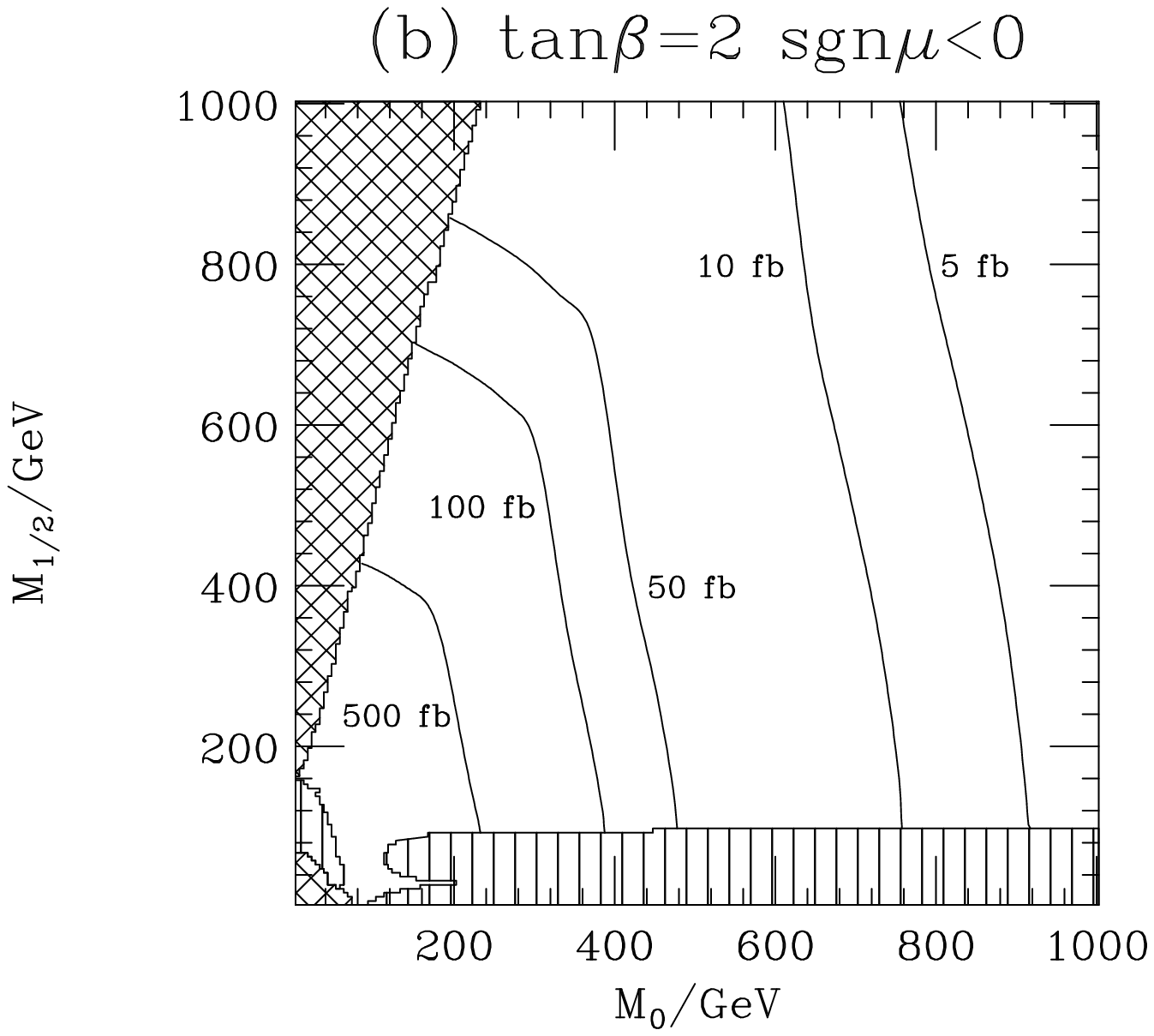}\\
\vskip 7mm
\includegraphics[angle=90,width=0.48\textwidth]{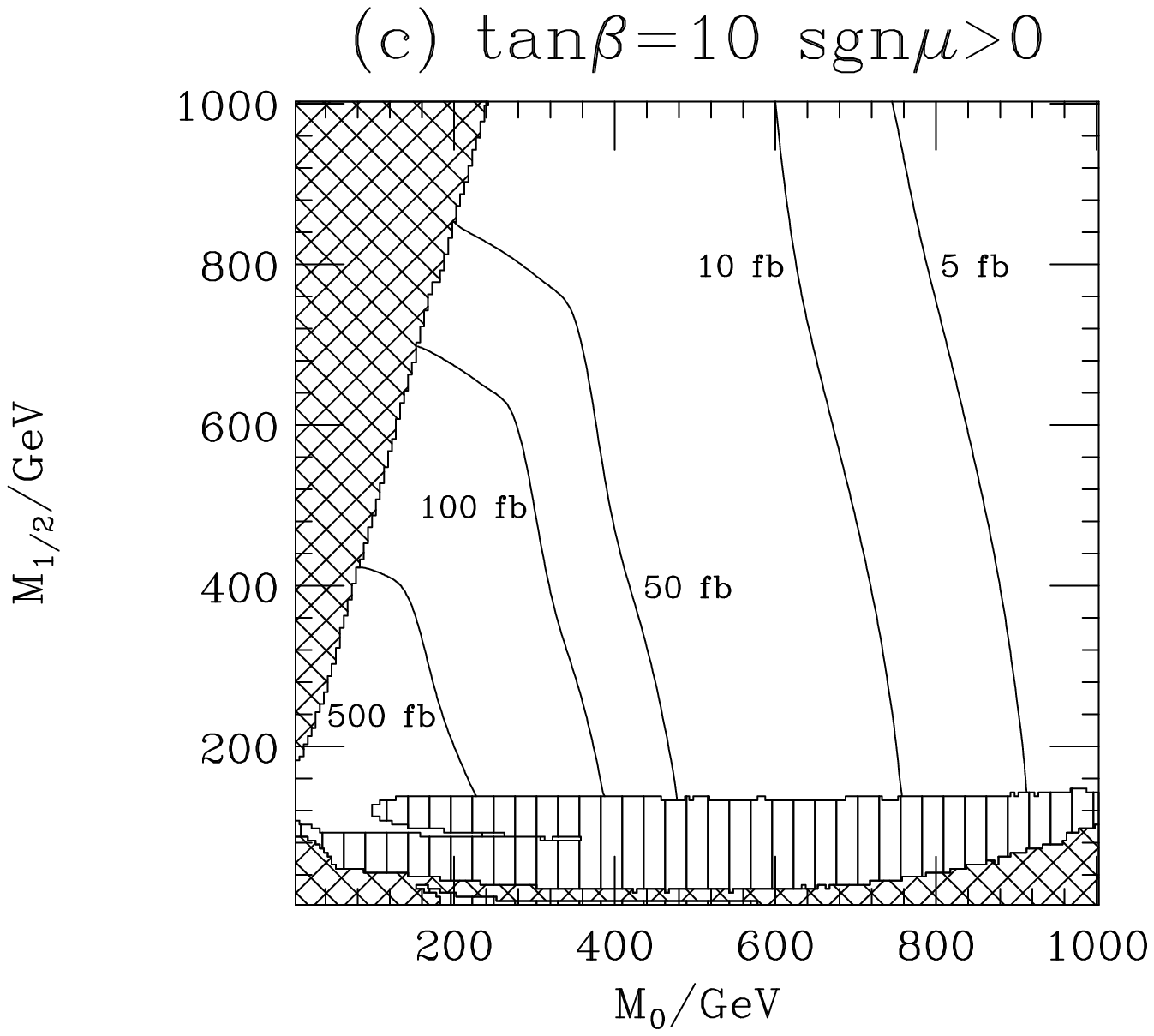}
\hfill
\includegraphics[angle=90,width=0.48\textwidth]{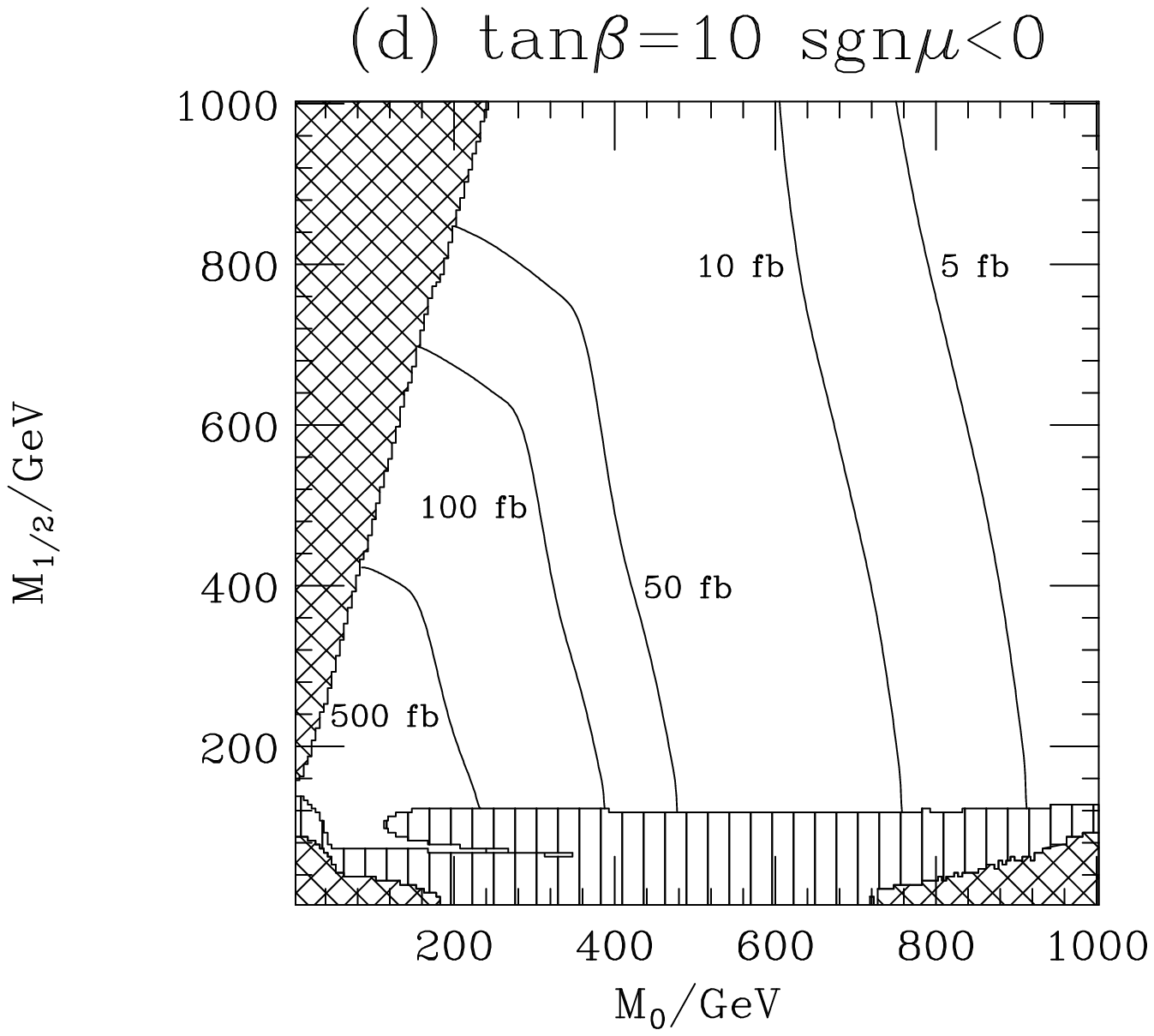}\\
\caption{Contours showing the cross section for the
	 production of a neutralino
	 and a charged lepton at the LHC in the $M_0$, $M_{1/2}$ plane 
	 for $A_0=0\, \mr{\gev}$
 	 and ${\lam'}_{211}=10^{-2}$ with different values
	 of $\tan\beta$ and $\sgn\mu$. The striped and hatched regions are
	 described in the caption of Fig.\,\ref{fig:SUSYmass}.} 
\label{fig:LHCcross}
\end{center}
\end{figure}

%
%
\begin{figure}
\begin{center}
\includegraphics[angle=90,width=0.48\textwidth]{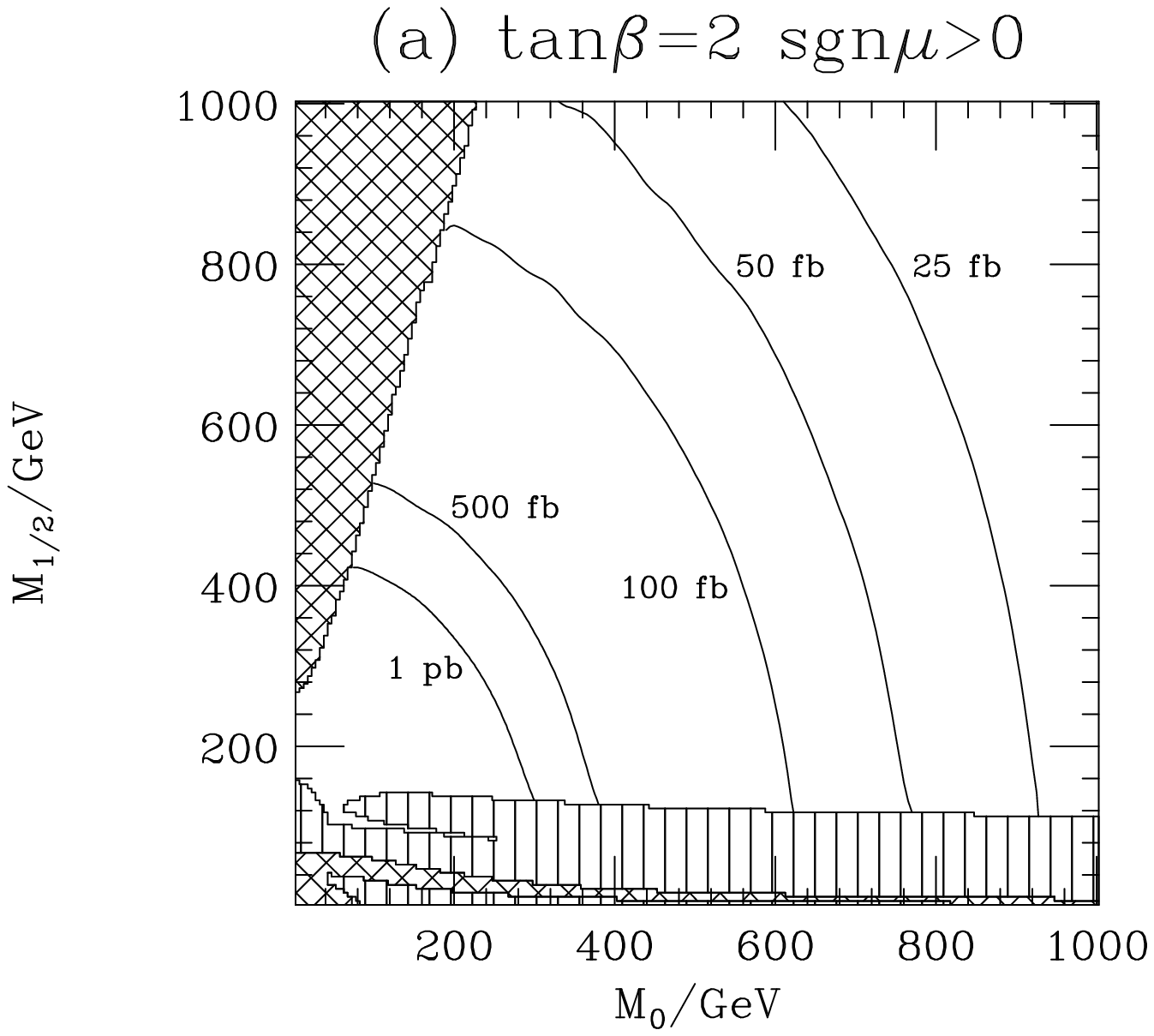}
\hfill
\includegraphics[angle=90,width=0.48\textwidth]{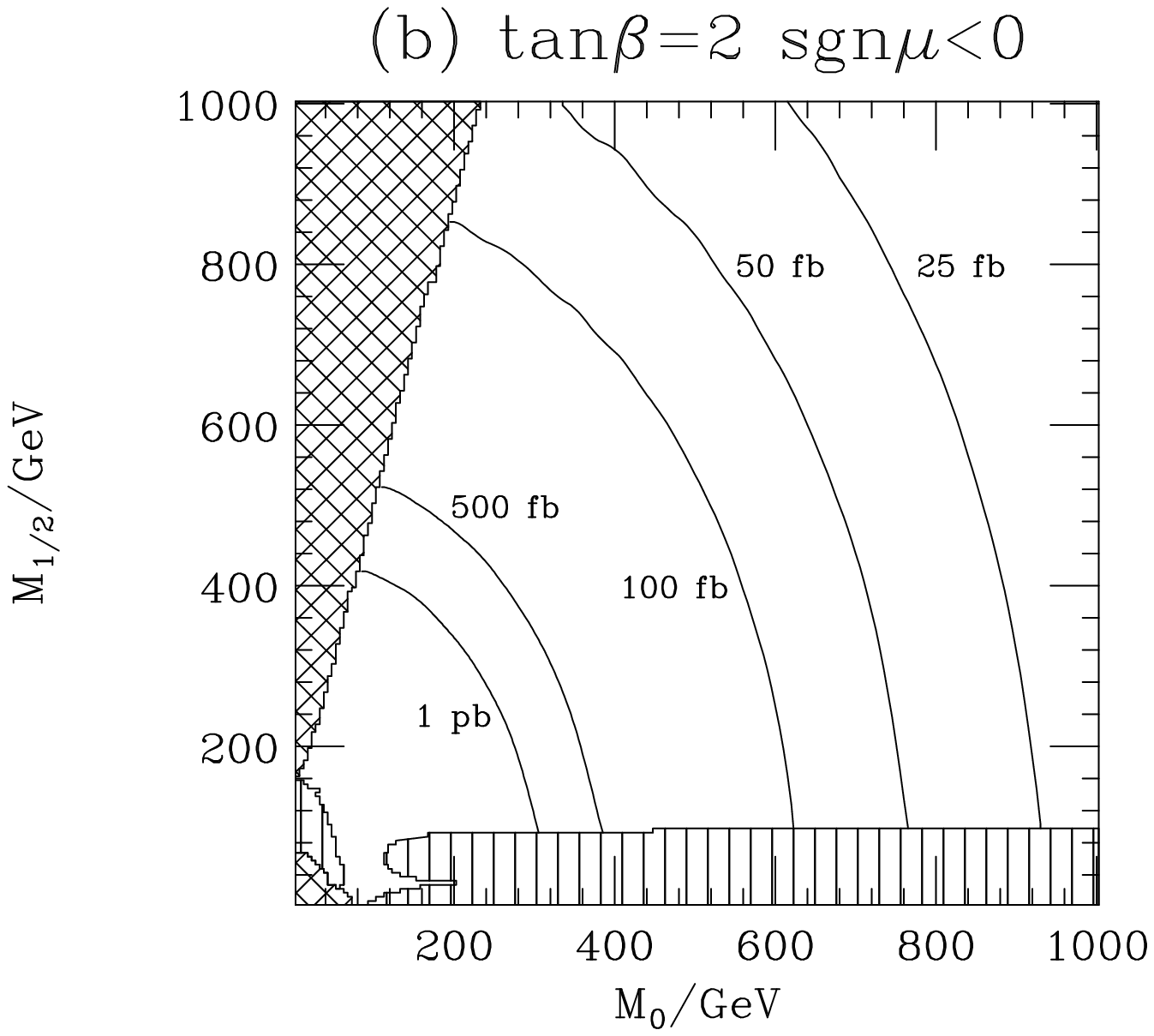}\\
\vskip 7mm
\includegraphics[angle=90,width=0.48\textwidth]{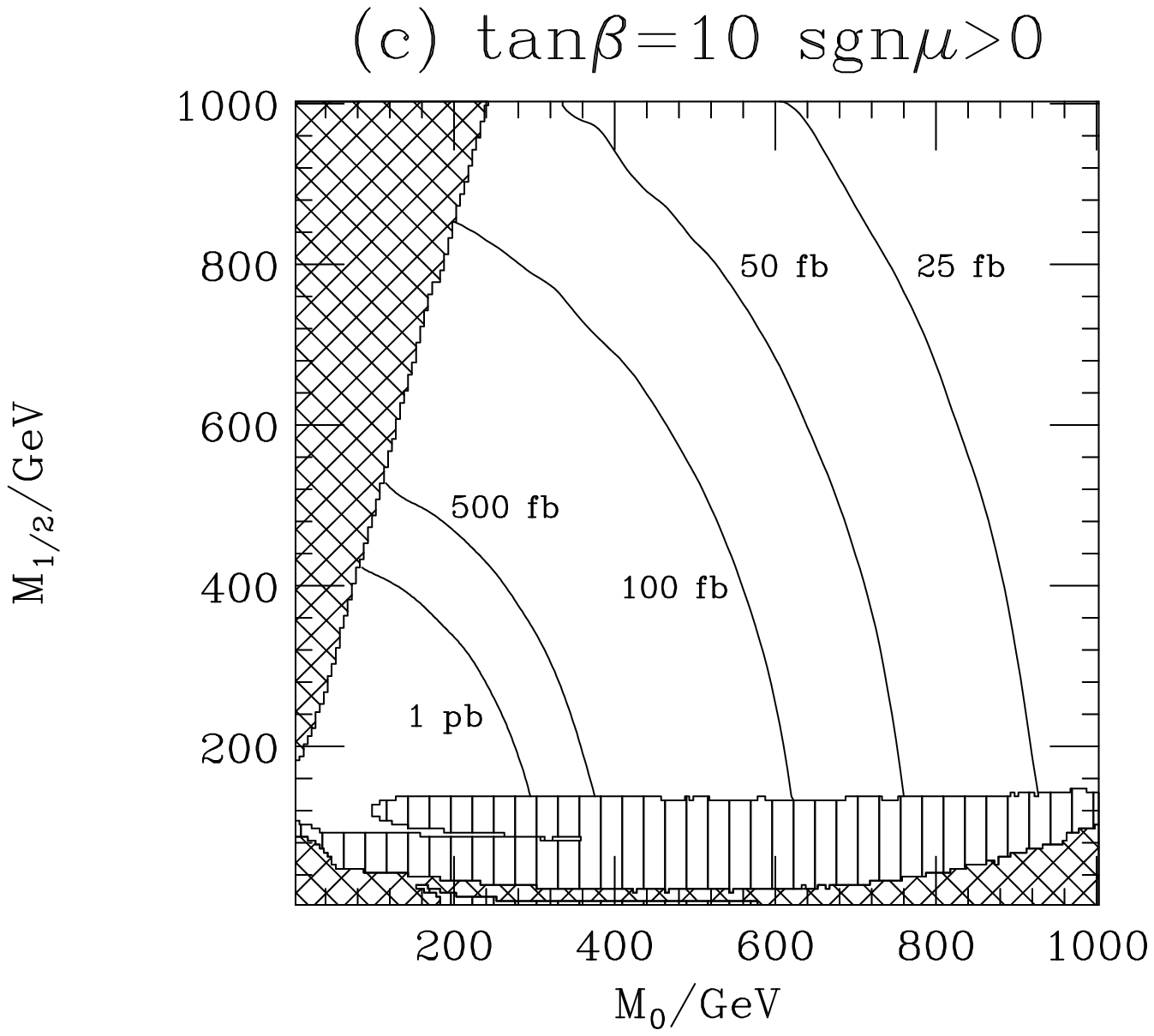}
\hfill
\includegraphics[angle=90,width=0.48\textwidth]{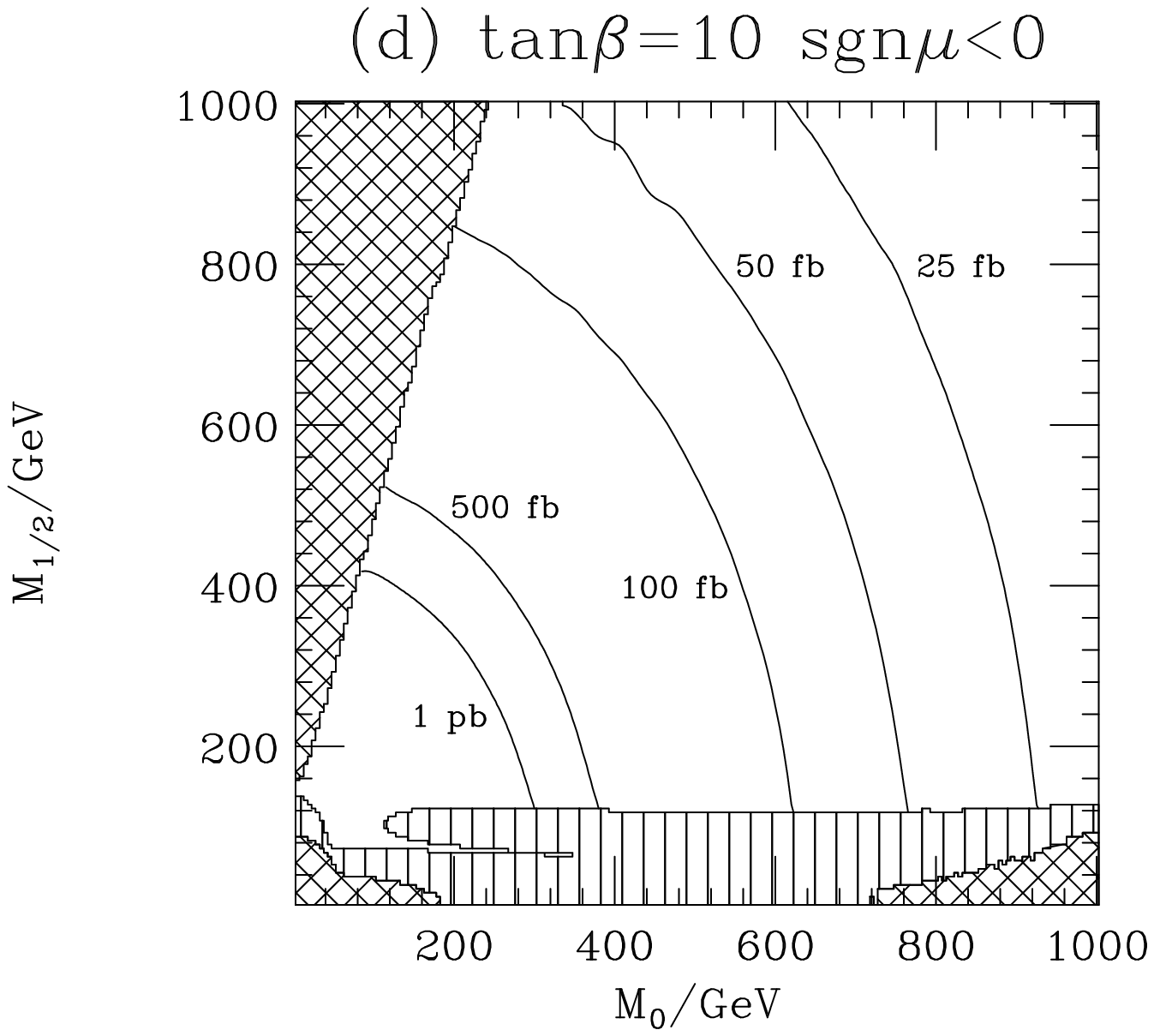}\\
\caption{Contours showing the cross section for resonant slepton 
	 production followed by a supersymmetric gauge decay at the
	 LHC in the $M_0$, $M_{1/2}$ plane 
	 for $A_0=0\, \mr{\gev}$  and 
	 ${\lam'}_{211}=10^{-2}$ with different values
	 of $\tan\beta$ and $\sgn\mu$. The striped and hatched regions are
	 described in the caption of Fig.\,\ref{fig:SUSYmass}.} 
\label{fig:LHCcross2}
\end{center}
\end{figure}
  The cross section for the production of a charged lepton and a neutralino,
  which is again the dominant production mechanism,
  at the LHC
  is shown in Fig.\,\ref{fig:LHCcross} in the  $M_0$, $M_{1/2}$ plane 
  with $A_0=0\, \mr{\gev}$ and ${\lam'}_{211}=10^{-2}$
  for two different values
  of $\tan\beta$ and both values of $\sgn\mu$. The total cross section for
  resonant slepton production followed by a supersymmetric 
  gauge decay is shown in
  Fig.\,\ref{fig:LHCcross2}. As for the Tevatron, the total resonant slepton
  cross section closely follows the slepton mass contours whereas the cross
  section for neutralino lepton production falls-off more quickly at small
  $M_{1/2}$ because the branching ratio for $\mut_L\ra\mu\cht^0_1$
  is reduced due to the production of charginos and the heavier neutralinos.
  We adopted the same procedure 
  described in Section~\ref{sub:tevatron} to
  estimate the acceptance of the cuts
  we have imposed. We will again first consider the cuts required to reduce
  the Standard Model backgrounds and then the additional cut used to suppress
  the sparticle pair production background.
   
%
%

\subsubsection{Standard Model Backgrounds}

\begin{figure}
\begin{center}
\includegraphics[angle=90,width=0.48\textwidth]{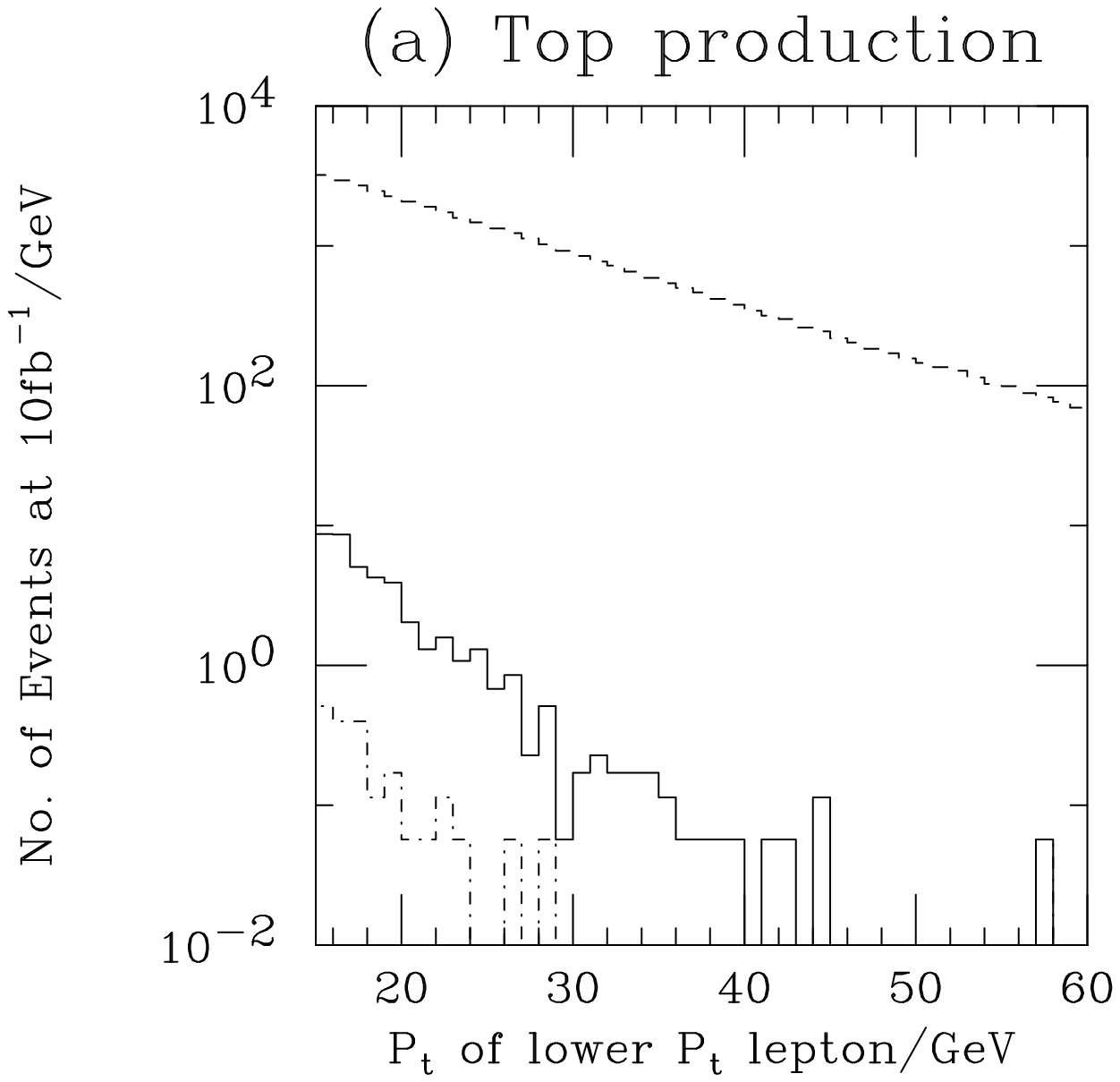}
\hfill
\includegraphics[angle=90,width=0.48\textwidth]{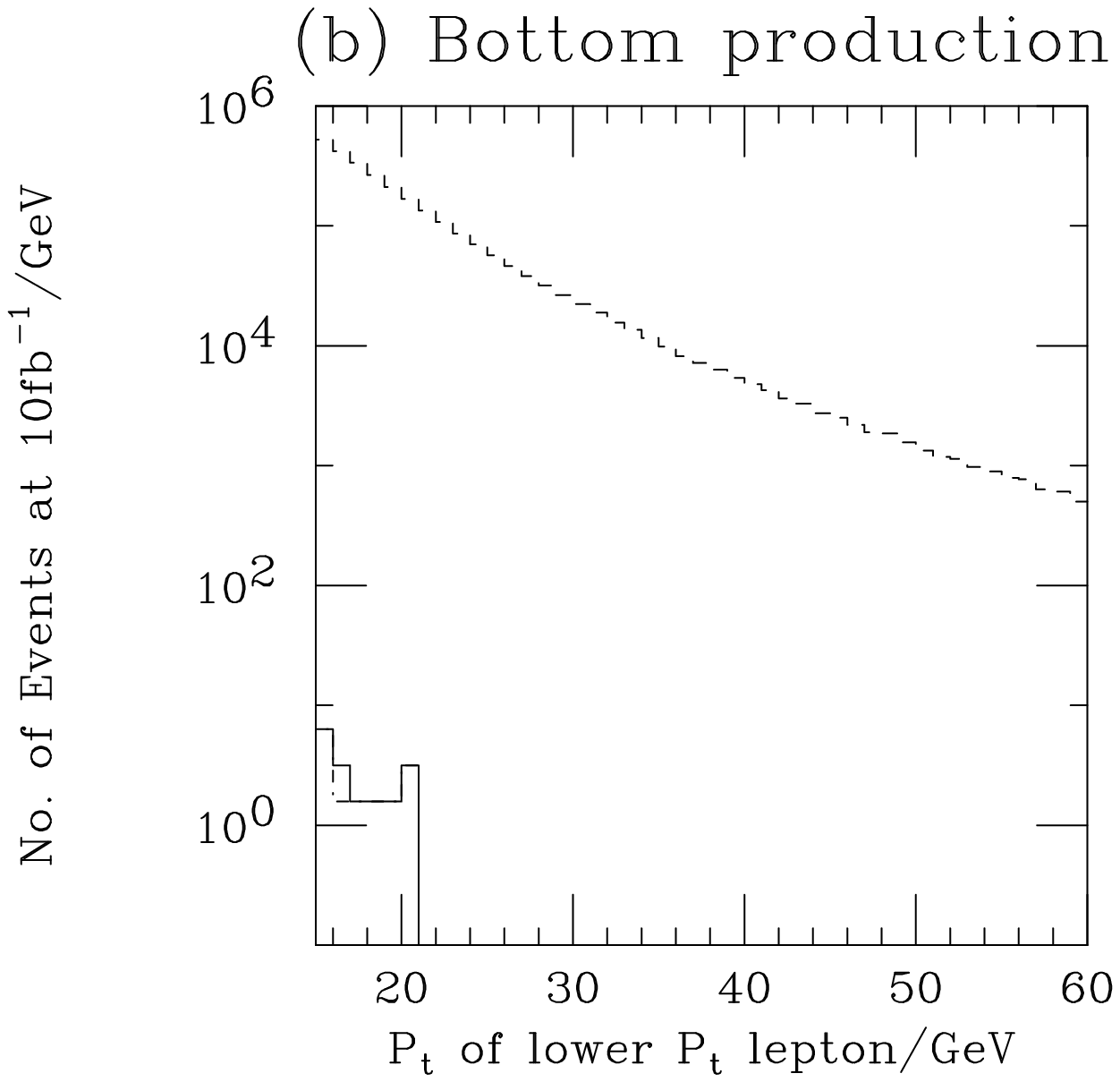}\\
\caption{Effect of the isolation cuts on the $\mr{t\bar{t}}$ and 
	 $\mr{b\bar{b}}$ backgrounds at 
	the LHC. The dashed line gives the background before any cuts,
	the solid line shows the effect of the isolation cut described in the
	text. The dot-dash line gives the effect of all the cuts, including
	the cut on the number of jets, for the $\mr{b\bar{b}}$ background 
 	this is almost  indistinguishable from the solid line.
	 As a parton-level cut
	of $40\, \mr{\gev}$  was used in simulating
	 the $\mr{b\bar{b}}$ background
	the results below $40\, \mr{\gev}$
	for the lepton $p_T$ do not correspond to
 	the full number of background events.
	The distributions have been normalized
 	to $10\  \mr{fb}^{-1}$ integrated luminosity.}
\label{fig:lhcheavyiso}
\end{center}
\end{figure}

  We applied the following cuts to reduce the Standard Model backgrounds:
\begin{enumerate}

\item A cut requiring all the leptons to be in the central region of the
      detector $|\eta|<2.0$.

\item	A cut on  the transverse momentum of the like-sign leptons
	$p_T^{\mr{lepton}} \geq 40\, \mr{\gev}$,
	this is the lowest cut we could apply given our parton-level cut of
	$p_T^{\mr{parton}}=40\, \mr{\gev}$,
	for the $\mr{b\bar{b}}$ background.
	
\item 	An isolation cut on the like-sign leptons so that the
	transverse energy in a cone of radius,
	$R = \sqrt{\Delta\phi^2+\Delta\eta^2} = 0.4$, about the direction of
	the lepton is less than $5\, \mr{\gev}$.

\item   We reject events with  $60\,\mr{\gev} < M_T < 85\,\mr{\gev}$ ($c.f.$ 
	Eqn.\,\ref{eqn:MTdef}.) This cut is
	applied to both of the like-sign leptons.

\item   A veto on the presence of a lepton in the event with the same flavour
        but opposite charge as either of the leptons in the like-sign
	pair if the lepton has  $p_T>10\, \mr{\gev}$  
	and passes the same isolation cut as the like-sign leptons.

\item   A cut on the missing transverse energy, $\not\!\!E_T<20\, \mr{\gev}$.
\end{enumerate}
  The first two cuts are designed to reduce the background from heavy
  quark, \ie $\mr{b\bar{b}}$ and $\mr{t\bar{t}}$ production which is again
  the major source of backgrounds before any cuts. However, as can be seen in
  Fig.\,\ref{fig:lhcheavyiso}, after the
  imposition of the $p_T$ and isolation cuts this background is significantly
  reduced. It remains the major source of the error on the background however
  due to the large cross section for $\mr{b\bar{b}}$ production which makes
  it impossible to simulate the full luminosity of the LHC with the resources
  available.

\begin{figure}
\begin{center}
\includegraphics[angle=90,width=0.48\textwidth]{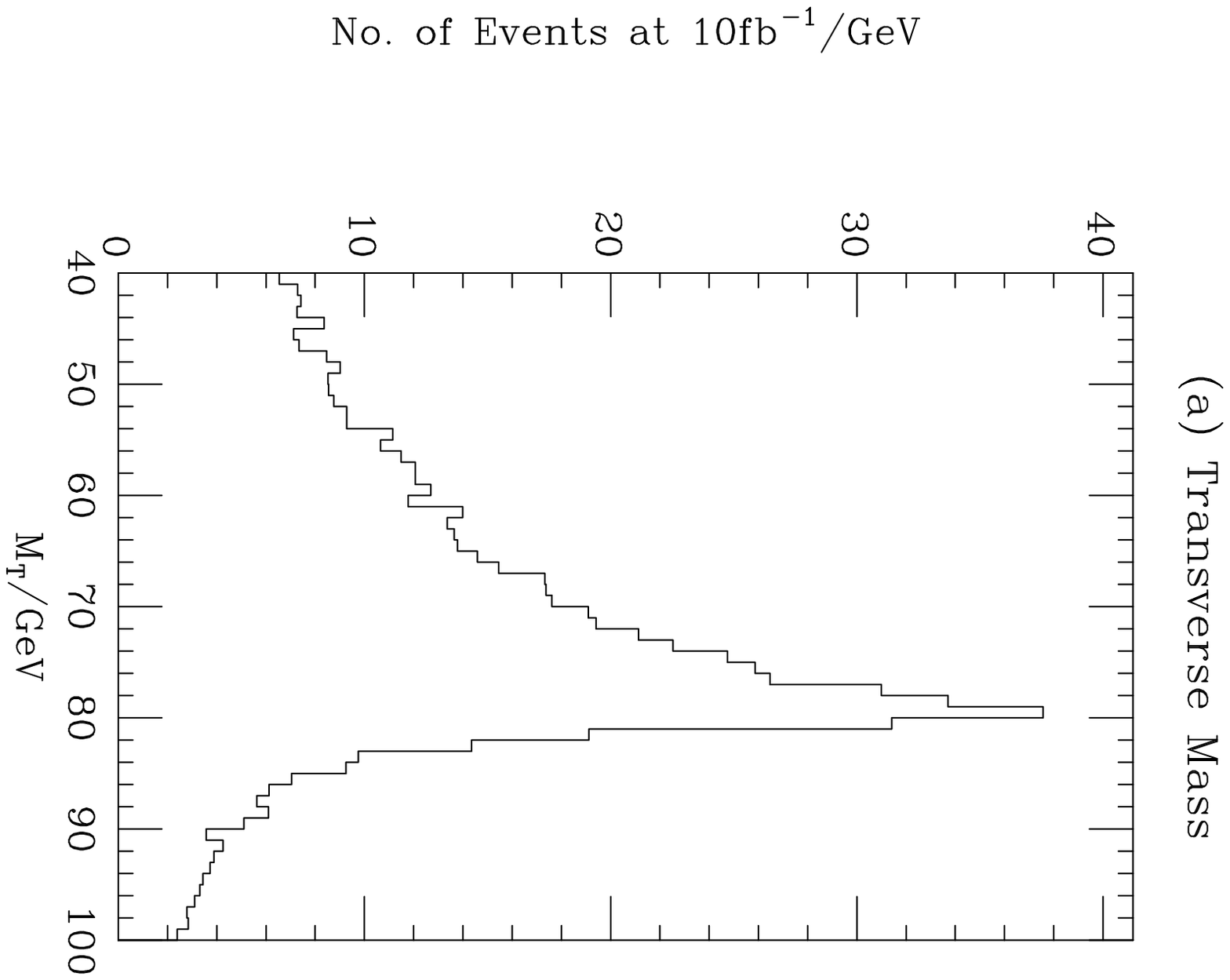}
\hfill
\includegraphics[angle=90,width=0.48\textwidth]{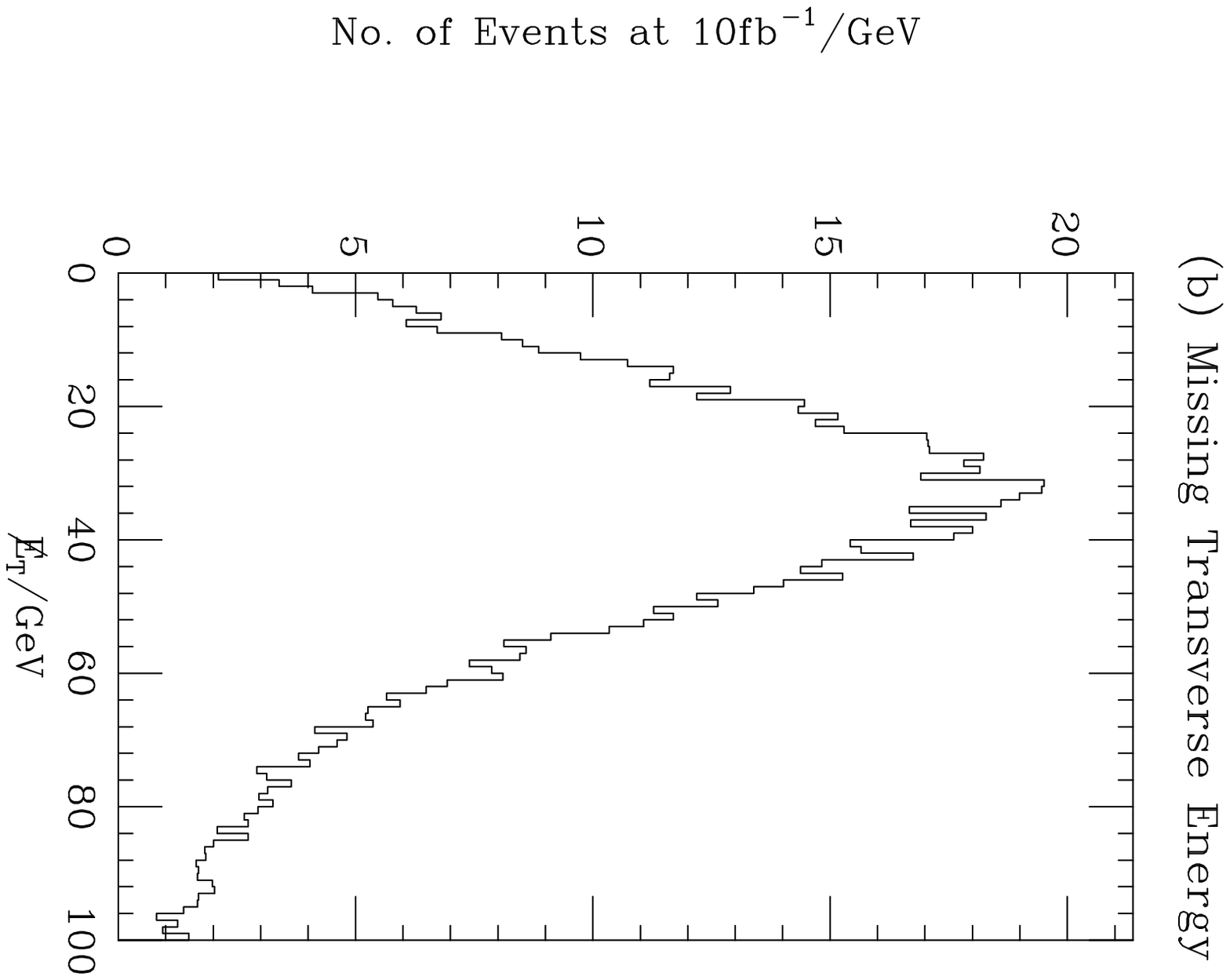}\\
\caption{The transverse mass and missing transverse energy in WZ events
	 at the LHC. The distributions are normalized to
	$10\  \mr{fb}^{-1}$ luminosity.} 
\label{fig:lhcemiss}
\end{center}
\begin{center}
\includegraphics[angle=90,width=0.48\textwidth]{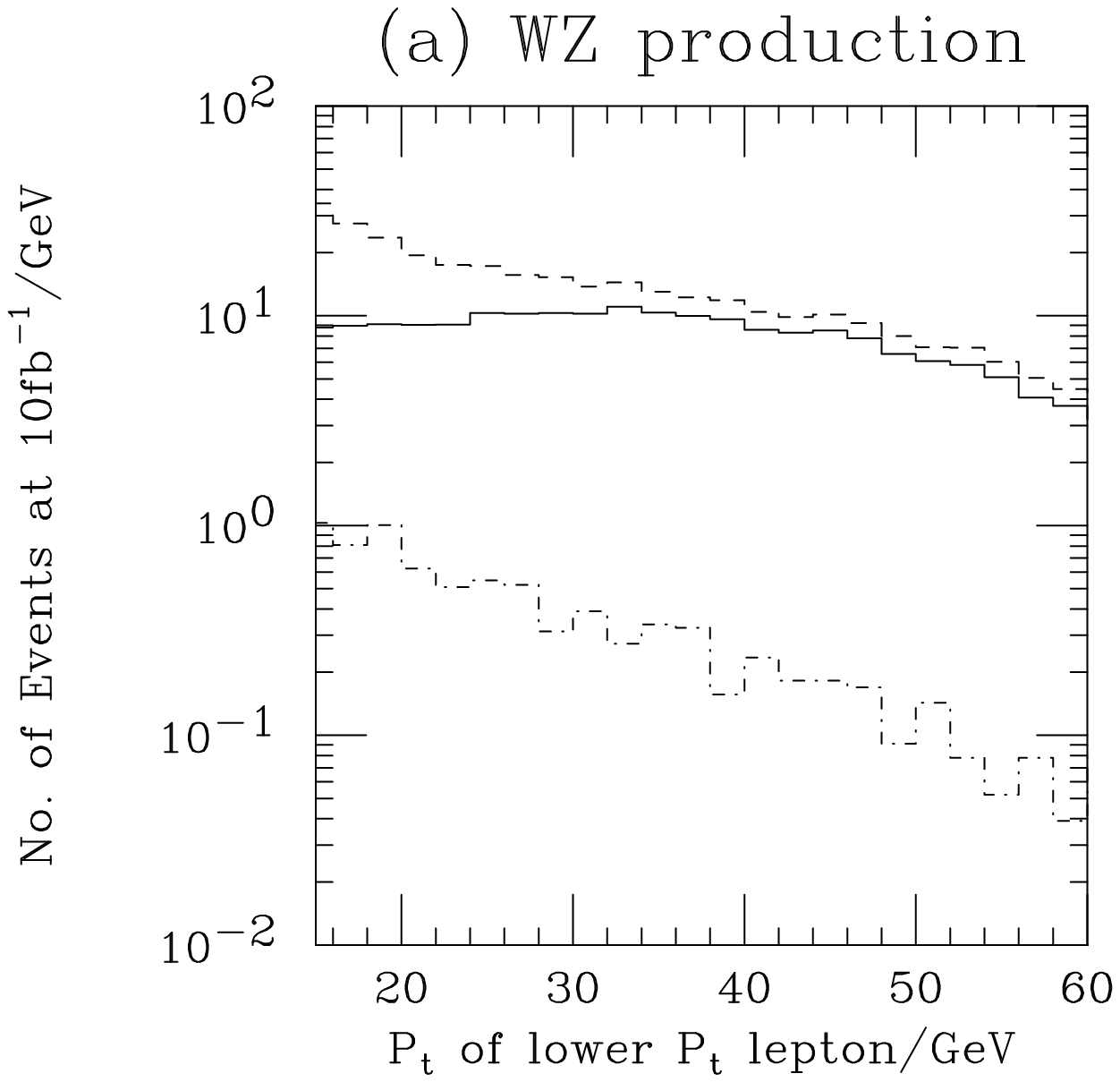}
\hfill
\includegraphics[angle=90,width=0.48\textwidth]{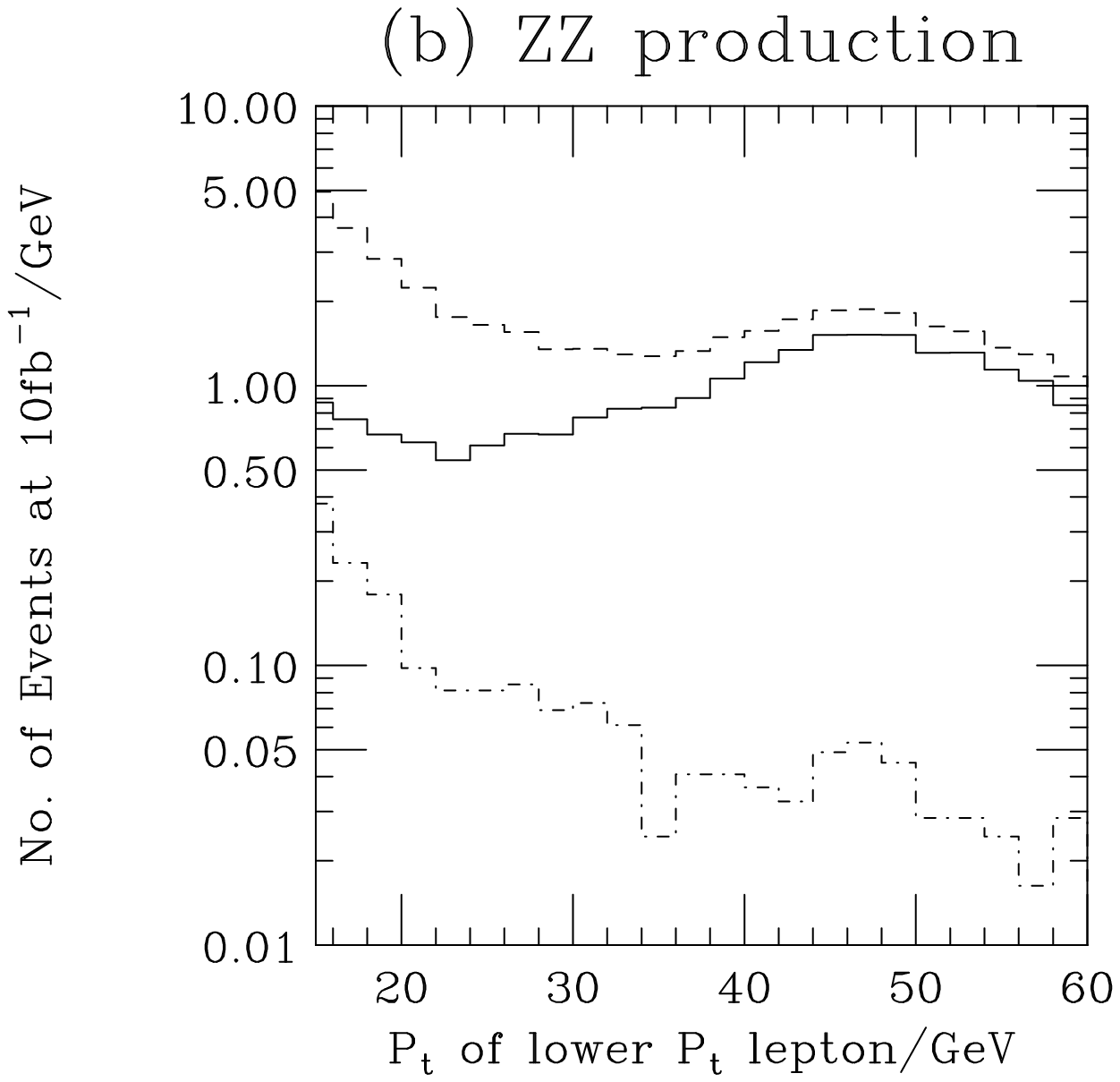}\\
\caption{Effect of the isolation cuts on the WZ and ZZ backgrounds at 
	the LHC. The dashed line gives the background before any cuts,
	the solid line shows the effect of the isolation cut described in the
	text. The dot-dash line gives the effect of all the cuts, including
	the cut on the number of jets. The distributions are normalized to
	$10\  \mr{fb}^{-1}$ luminosity.} 
\label{fig:lhcgaugeiso}
\end{center}
\end{figure}

  The remaining cuts reduce the background from gauge boson pair,
  particularly WZ, production which dominates the Standard Model 
  background after the imposition of the isolation and $p_T$ cuts.
  Fig.\,\ref{fig:lhcemiss}a
  shows that the cut on the transverse mass, \ie removing the region
  $60\, \mr{\gev} < M_T < 85\, \mr{\gev}$, for each of the 
  like sign leptons will reduce the background
  from WZ production, which is the largest of the gauge boson pair
  production backgrounds. Similarly the cut on the missing transverse
  energy, $\not\!\!E_T<20\, \mr{\gev}$, 
  will significantly reduce the background from WZ production as
  can be seen in Fig.\,\ref{fig:lhcemiss}b. The effect of these cuts is
  shown in Fig.\,\ref{fig:lhcgaugeiso}. Again the simulation of the gauge
  boson pair production backgrounds does not include $\mr{W\gamma}$ 
  production which
  may be an important source of background, but should be significantly
  reduced by the cuts.

%
%

%
%

  The effect of all these cuts on the background is shown in 
  Table~\ref{tab:lhcback}.
  While the dominant background is from WZ
  production,
  the dominant contribution to the error comes
  from $\mr{b\bar{b}}$ production. 
  This can only be reduced with a significantly more elaborate simulation.

  This gives a total background after all the cuts of
  $4.9\pm1.6$ events, for 10~$\mr{fb}^{-1}$ integrated luminosity.
  If we take a conservative approach and take a background of
  6.5 events, \ie a $1\sigma$ fluctuation above the central value of our 
  calculation a $5\sigma$ statistical 
  fluctuation would correspond to 16 events, for
  an integrated luminosity of 10~$\mr{fb}^{-1}$. As can be seen
  from Fig.\,\ref{fig:lhcheavyiso} this is a conservative upper bound.

\begin{figure}
\begin{center}
\includegraphics[angle=90,width=0.48\textwidth]{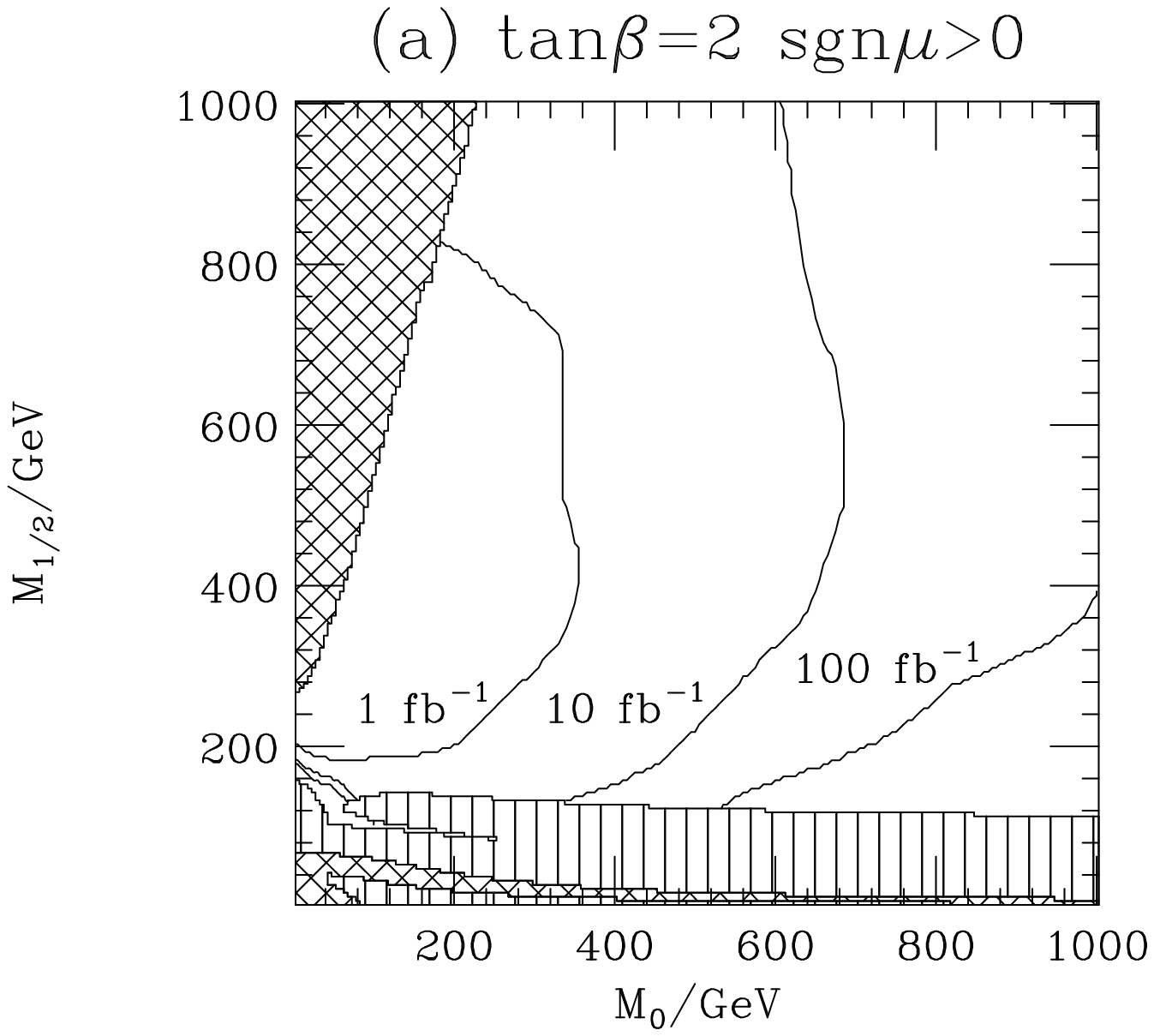}
\hfill
\includegraphics[angle=90,width=0.48\textwidth]{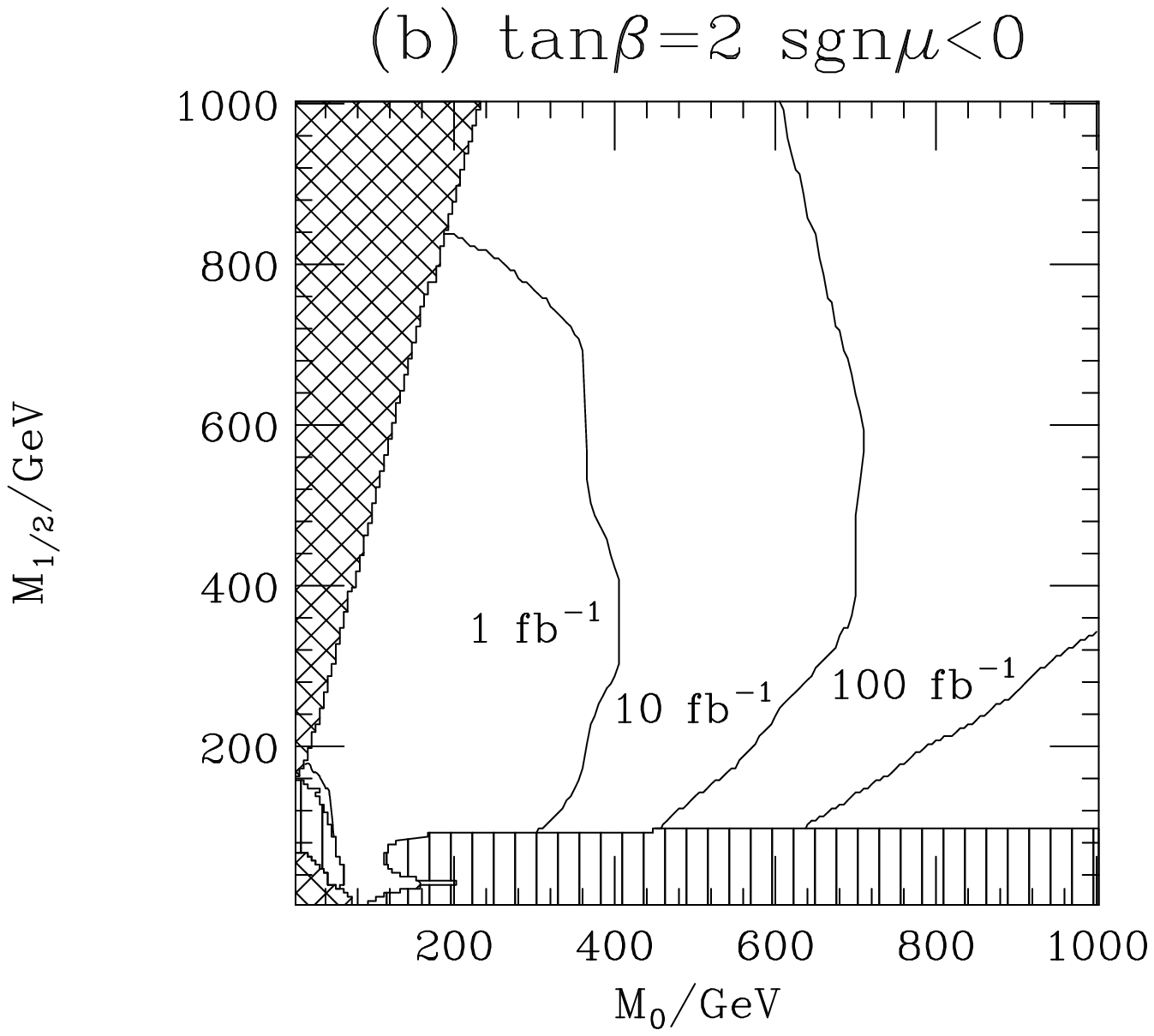}\\
\vskip 15mm
\includegraphics[angle=90,width=0.48\textwidth]{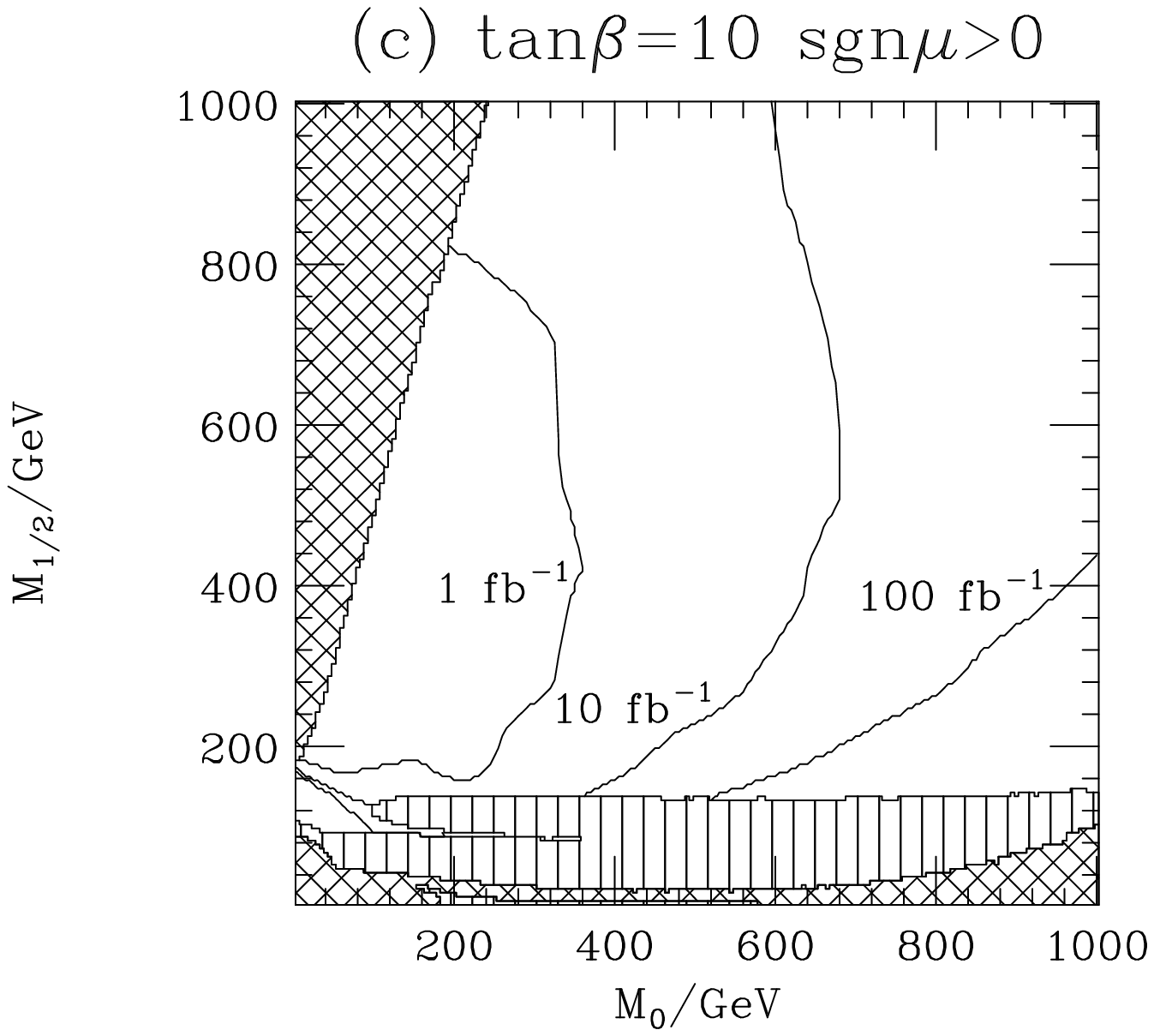}
\hfill
\includegraphics[angle=90,width=0.48\textwidth]{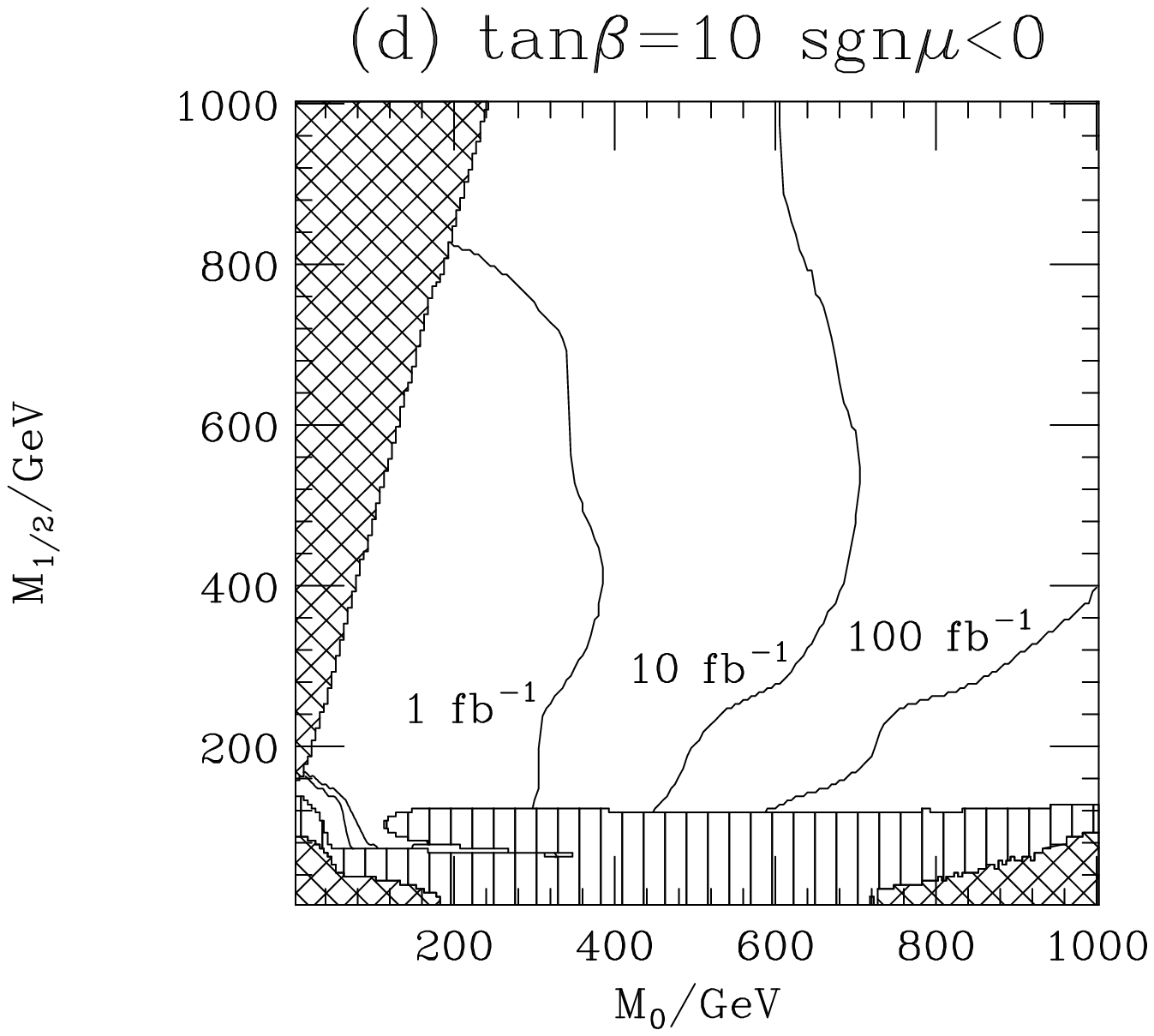}\\
\caption{Contours showing the discovery potential of the LHC in the $M_0$,
	 $M_{1/2}$ plane for ${\lam'}_{211}=10^{-2}$ and $A_0=0\, \mr{\gev}$.
	 These are a $5\sigma$ excess of the signal above the
	 background. Here we have imposed cuts on the isolation and $p_T$
  	 of the leptons, the transverse mass
	 and the missing transverse energy
	 described in the text, and a veto on the presence of OSSF leptons.
	 We have only considered the Standard Model background.
	 The striped and hatched regions are
	 described in the caption of Fig.\,\ref{fig:SUSYmass}.} 
\label{fig:lhcSMnojet}
\end{center}
\end{figure}

\begin{figure}
\begin{center}
\includegraphics[angle=90,width=0.48\textwidth]{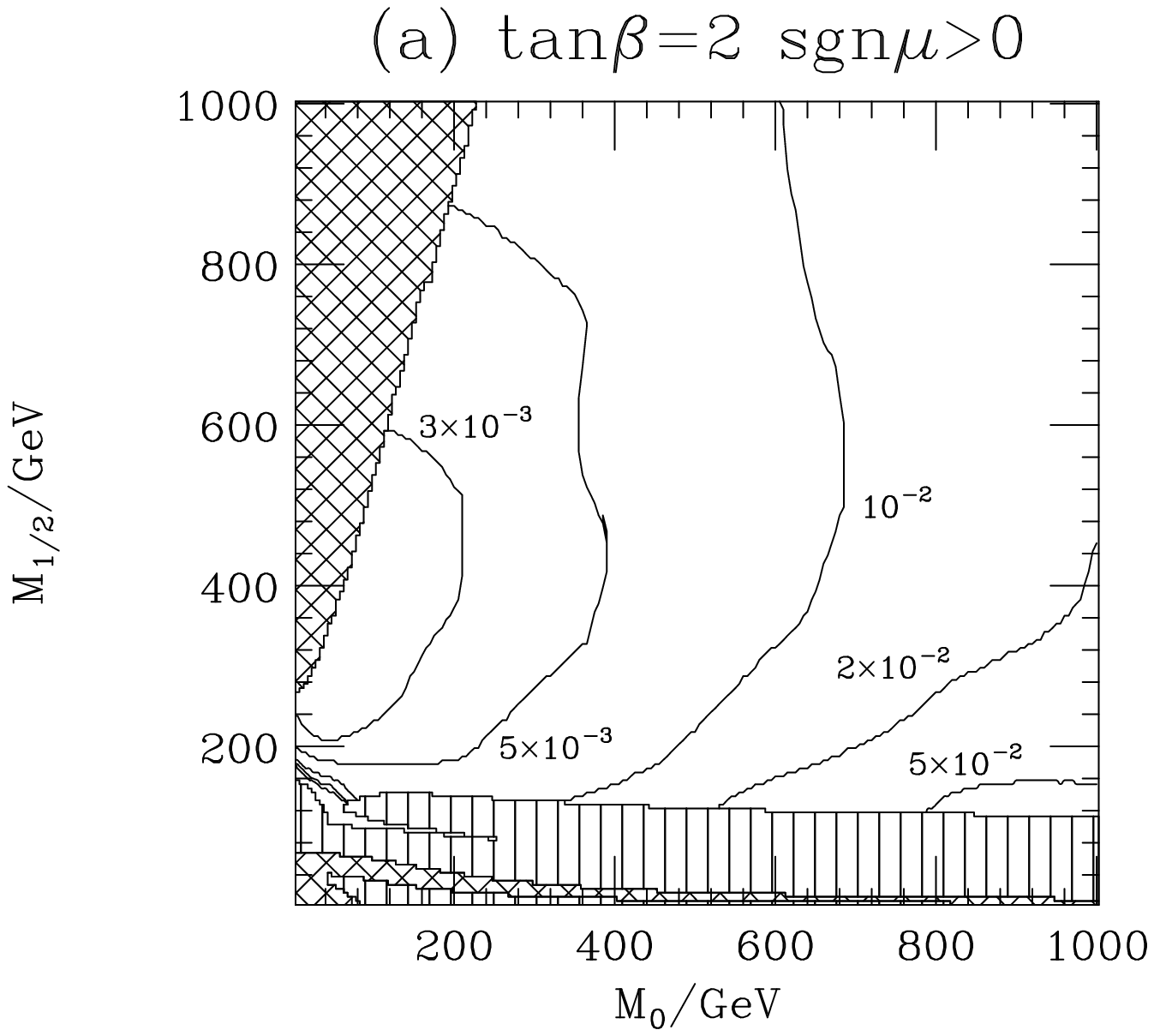}
\hfill						
\includegraphics[angle=90,width=0.48\textwidth]{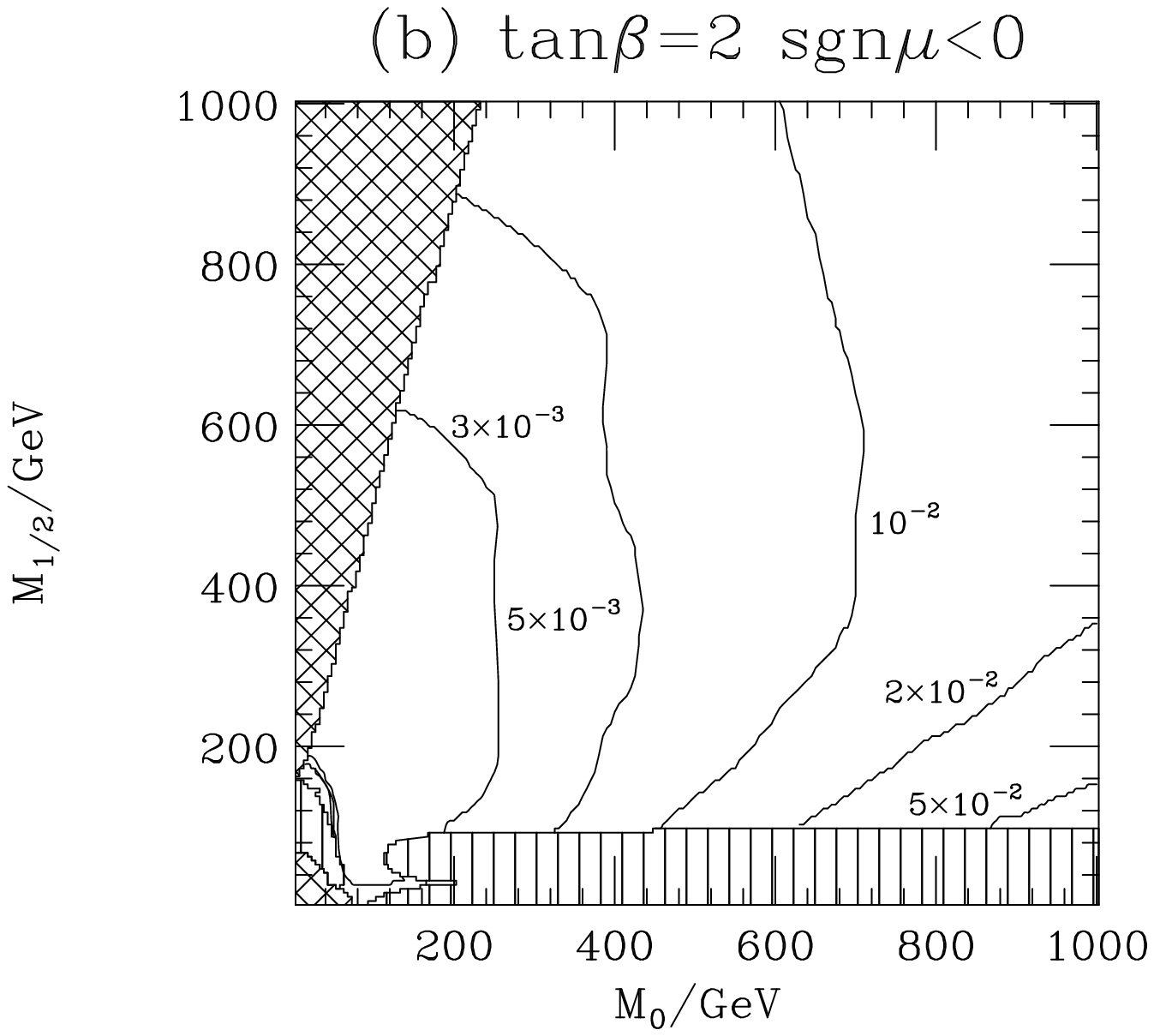}\\
\vskip 15mm					
\includegraphics[angle=90,width=0.48\textwidth]{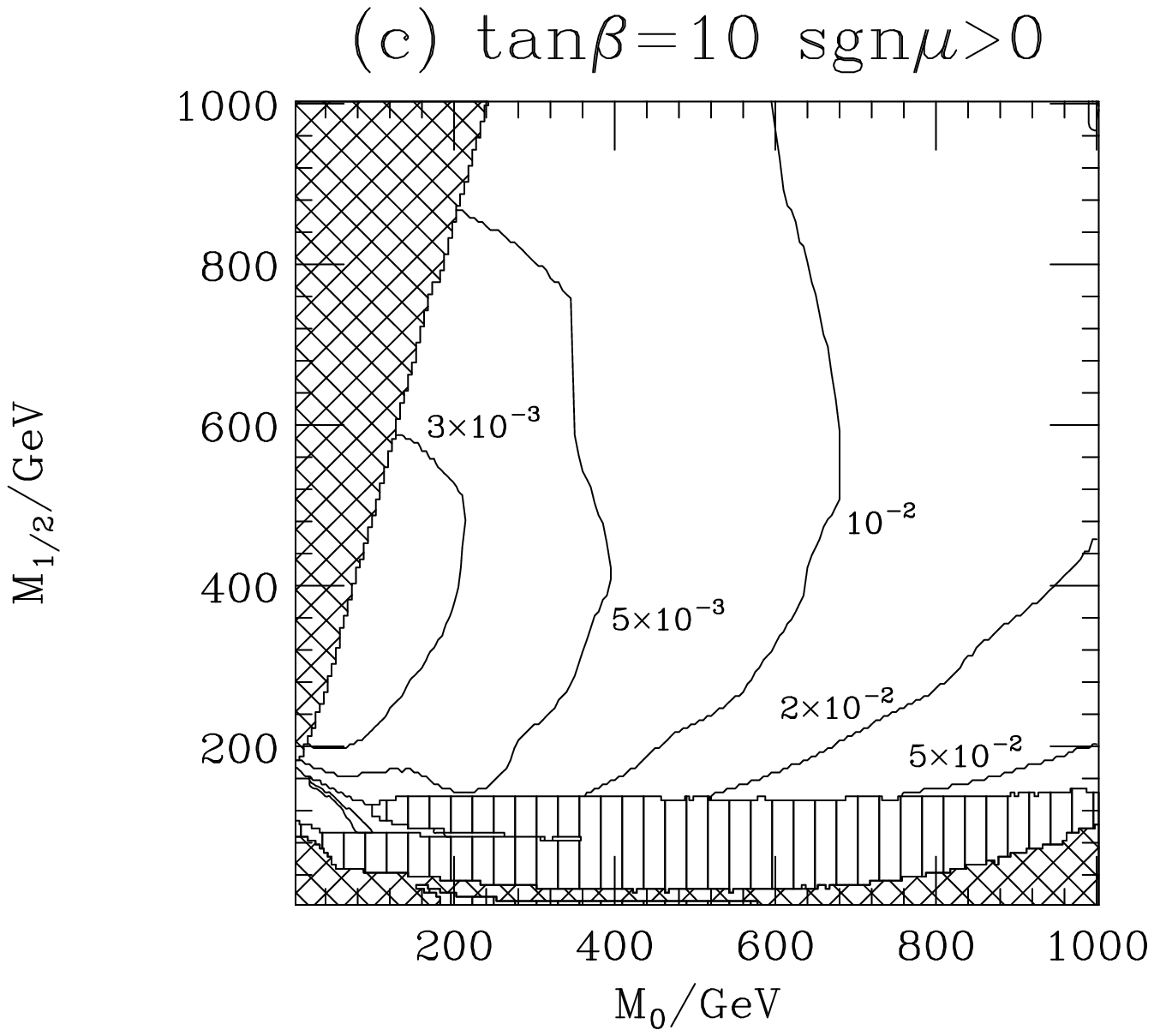}
\hfill						
\includegraphics[angle=90,width=0.48\textwidth]{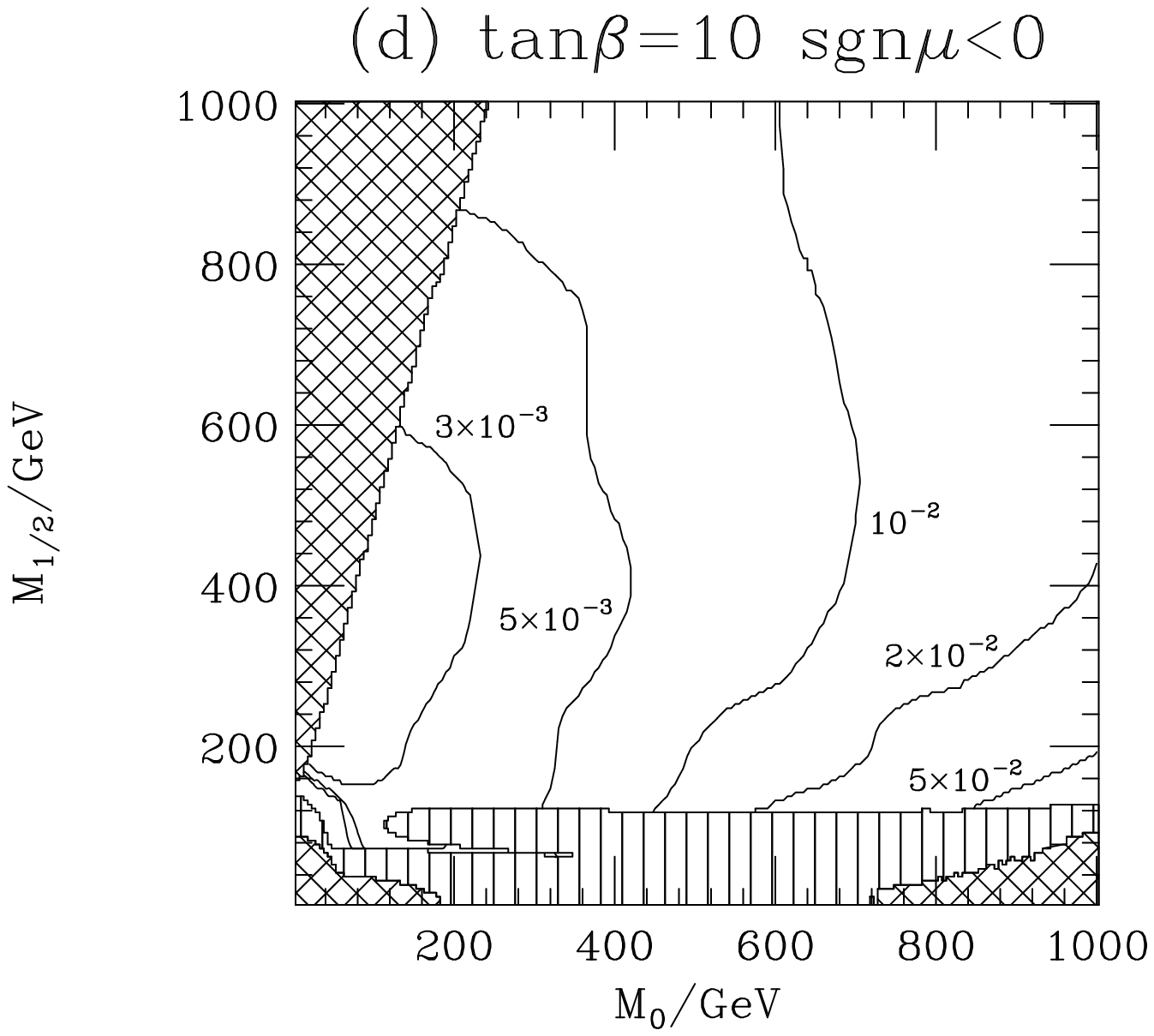}\\
\caption{Contours showing the discovery potential of the LHC in the $M_0$,
	 $M_{1/2}$ plane for $A_0=0\, \mr{\gev}$  and an integrated
	 luminosity of
	 $10\  \mr{fb}^{-1}$ with different values of ${\lam'}_{211}$.
	 These are a $5\sigma$ excess of the signal above the
	 background. Here we have imposed cuts on the isolation and $p_T$
  	 of the leptons, the transverse mass and the missing
	 transverse energy
	 described in the text, and a veto on the presence of OSSF leptons.
	 We have only considered the Standard Model background.
	 The striped and hatched regions are
	 described in the caption of Fig.\,\ref{fig:SUSYmass}.} 
\label{fig:lhcSMnojetb}
\end{center}
\end{figure}

  We adopted the same procedure described in the previous section to 
  obtain the acceptance for the \rpv\  signal given the cuts we have imposed.
  The discovery potential of the
  LHC is shown in Fig.\,\ref{fig:lhcSMnojet}, for ${\lam'}_{211}=10^{-2}$
  with different integrated luminosities, and in Fig.\,\ref{fig:lhcSMnojetb},
  for an integrated luminosity of $10\  \mr{fb}^{-1}$ with different
  values of ${\lam'}_{211}$.
  This is considerably greater than the discovery potential
  of the Tevatron at high $M_0$ and $M_{1/2}$ due
  to the larger centre-of-mass energy of the LHC and hence the larger
  cross sections. In particular the search potential
  with one years running at
  high luminosity, \ie $100\ \mr{fb^{-1}}$, covers large regions of the 
  $M_0$, $M_{1/2}$ plane. At very large values of $M_{1/2}$ this extends
  to regions where the sparticle pair production cross section is
  small due to the high masses of the SUSY particles.

%
%
\begin{table}
\begin{center}
\begin{tabular}{|c|c|c|c|c|}
\hline
	& \multicolumn{4}{c|}{Number of Events} \\
\cline{2-5}
 		& 	     	  &  		    & After isolation,&\\
Background      & After $p_T$ cut & After isolation & $p_T$, $M_T$, \met\
						   cuts & 
		 After all cuts \\
 Process        &                 & and $p_T$ cuts  & and OSSF lepton & \\
	        & 		  &		    & veto.  & \\
\hline
WW 		& $3.6\pm0.5$			& $0.0\pm0.06$ 
	        & $0.0\pm0.06$	 		& $0.0\pm0.06$ \\
\hline		                
WZ 		& $239\pm2.5$			& $198.6\pm2.3$
		& $3.8\pm0.3$ 			& $3.8\pm0.3$ \\
\hline				                
ZZ 		& $55.4\pm0.7$			& $45.2\pm0.6$	
		& $1.04\pm0.09$ 		& $1.04\pm0.09$ \\
\hline		                
$\mr{t\bar{t}}$ & $(4.4\pm0.2)\times10^3$ 	& $0.28\pm0.13$
		& $0.06\pm0.06$ 		& $0.06\pm0.06$ \\
\hline		                
$\mr{b\bar{b}}$ & $(4.4\pm0.9)\times10^4$	& $0.0\pm1.6$
		& $0.0\pm1.6$ 			& $0.0\pm1.6$ \\
\hline	                
Single Top 	& $36.6\pm1.5$ 		& $0.0\pm0.004$	
		& $0.0\pm0.004$ & $0.0\pm0.004$ \\
\hline		                
Total 	 	& $(4.9\pm0.9)\times10^4$	& $244.1\pm2.9$ 
		& $4.9\pm1.6$ 		& $4.9\pm1.6$ \\
\hline
\end{tabular}
\end{center}
\caption{Backgrounds to like-sign dilepton production at the LHC.
	 The numbers of events
	are based on an integrated luminosity of $10\  \mr{fb}^{-1}$.
	Again we have calculated an error on the cross section by varying the
	scale between half and twice the hard scale, apart from the
	gauge boson pair cross section where we do not have this
	information and the effect of varying the scale is expected
	to be small
	anyway. The error on the number of events is then the error on
	the cross section and the statistical error from the Monte
	Carlo simulation added in quadrature. If no events passed
	the cut the statistical error was taken to be the same
	as if one event had passed the cuts. }
\label{tab:lhcback}
\end{table}

%
%
%
%

  At small values of $M_0$ and $M_{1/2}$ there are regions of SUGRA parameter
  space which cannot be probed for any couplings due to the cuts we have
  applied.
  However these regions can be excluded by either LEP or the Tevatron and
  we will therefore ignore them in the rest of this analysis. If we
  neglect these regions the LHC can observe a resonant slepton with
  a mass of up to $510\,(710)\,\mr{\gev}$ 
  for a coupling of ${\lam'}_{211}=0.02$ with
  10\,(100)~$\mr{fb}^{-1}$ integrated luminosity and for a coupling
  ${\lam'}_{211}=0.05$ a resonant slepton can be observed with a mass
  of  $750\,(950)\,\mr{\gev}$  with
  10\,(100)~$\mr{fb}^{-1}$ integrated luminosity. 

  As with the Tevatron analysis we have neglected the background from
  sparticle pair production which is reasonable in an initial search for an
  excess of like-sign dilepton pairs over the Standard Model expectation.
  If such an excess were observed it would then be necessary to establish
  which process was producing the effect. In the next section, we will
  present the cuts necessary to reduce the background from sparticle pair 
  production and enable a resonant slepton signature to be established
  over all the backgrounds.

\subsubsection{SUSY backgrounds}

  The background from sparticle pair production is much more important at the
  LHC than the Tevatron given the much higher cross sections for
  sparticle pair production. The nature of the sparticles produced is also
  different due to the higher energies.
  In the regions of SUGRA parameter space
  where the pair production cross section at the Tevatron is large the
  lightest SUSY particles, \ie the electroweak
  gauginos, are predominately produced. This
  is because the production of the heavier squarks and gluinos
  is suppressed by
  the higher parton-parton centre-of-mass energies required. However given
  the higher centre-of-mass energy of the LHC the production of the
  coloured sparticles which occurs via the strong interaction dominates the
  cross section. This means that a cut on the number of jets in an event
  will be more effective in reducing the background from sparticle pair
  production. The following cut was applied:
\begin{itemize}
\item 	Vetoing all events when there are more than two jets each with
	$p_T>50\, \mr{\gev}$. 
\end{itemize}

%
%
\begin{figure}[htp]
\begin{center}
\includegraphics[angle=90,width=0.48\textwidth]{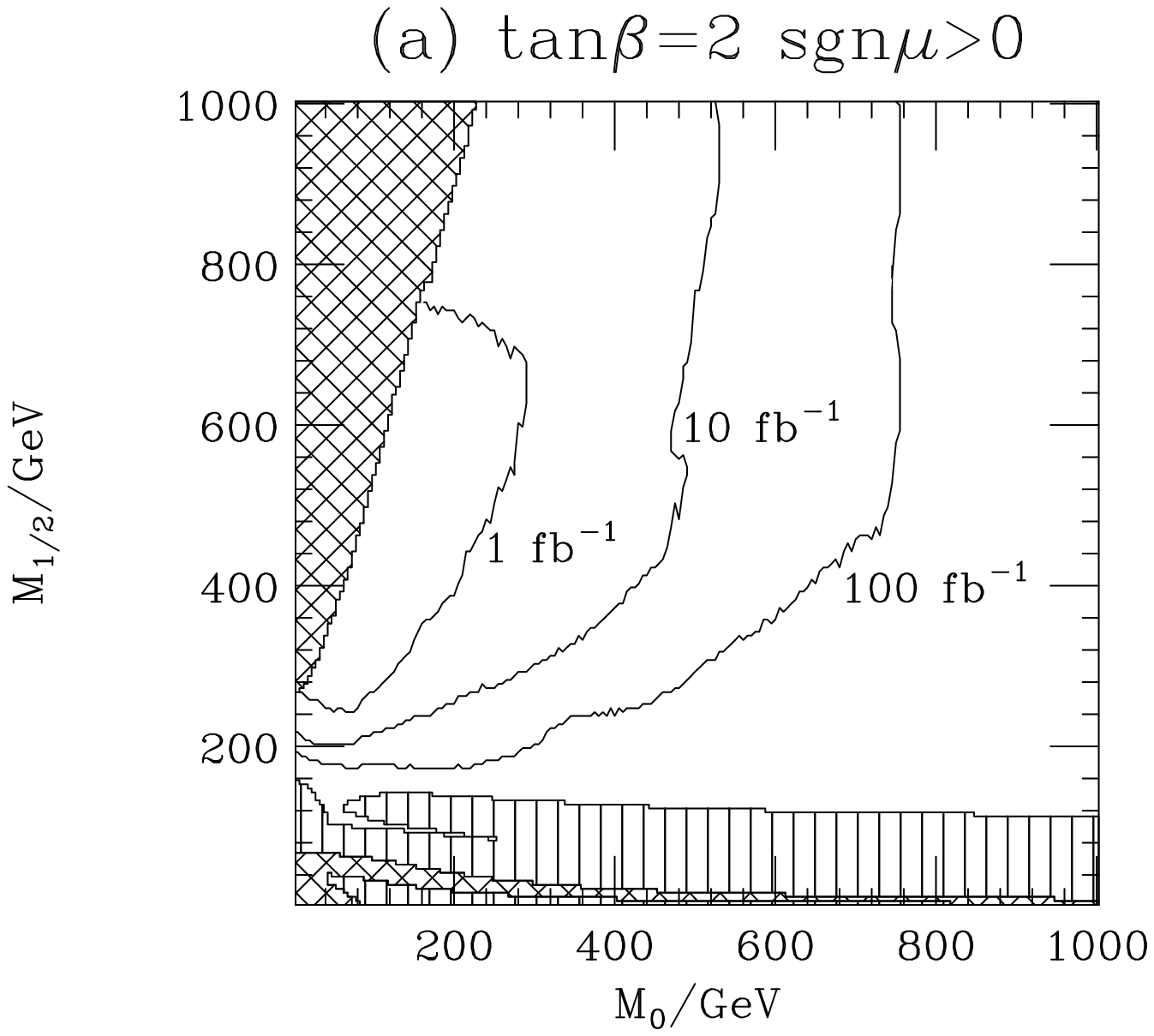}
\hfill
\includegraphics[angle=90,width=0.48\textwidth]{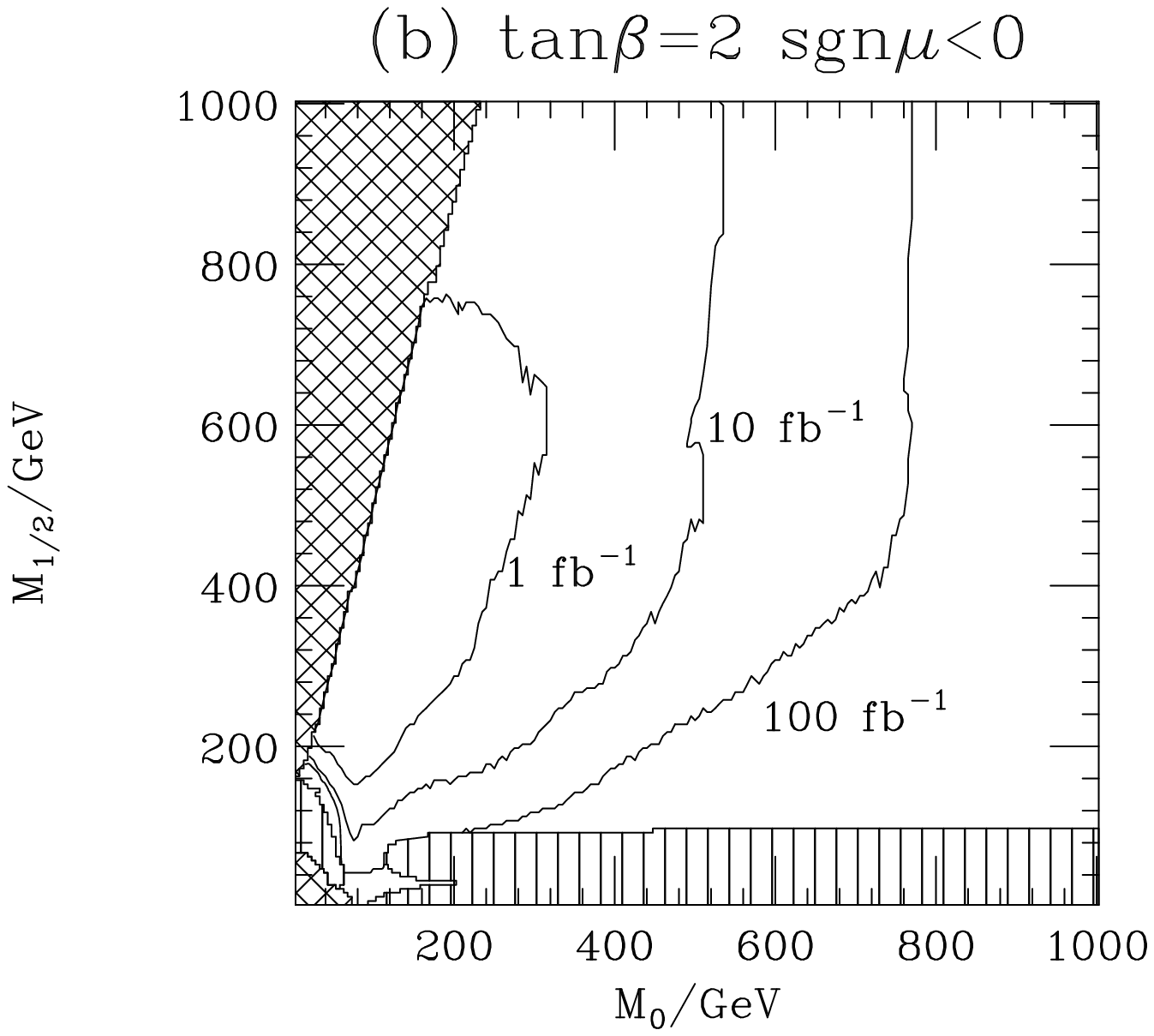}\\
\vskip 7mm
\includegraphics[angle=90,width=0.48\textwidth]{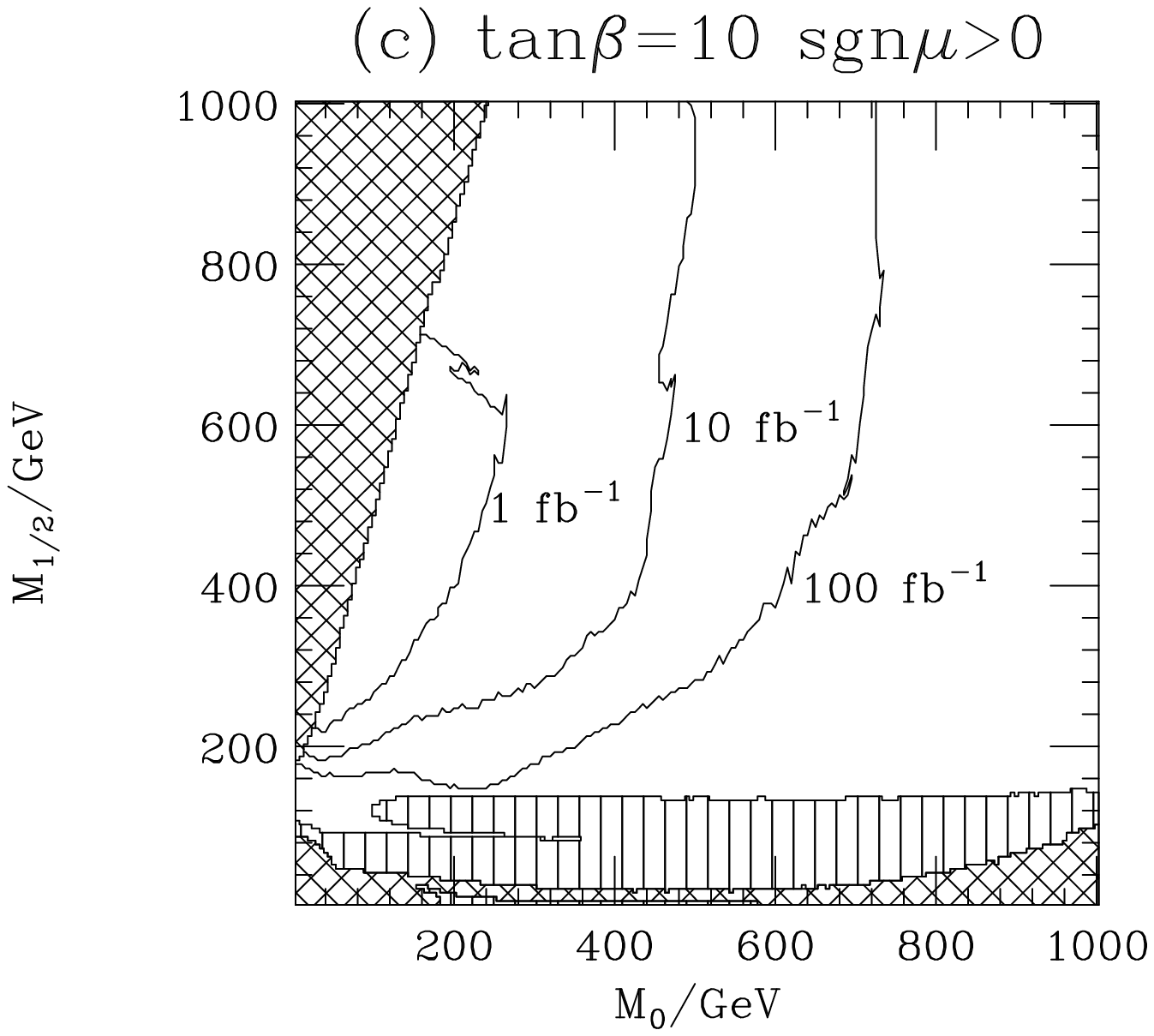}
\hfill
\includegraphics[angle=90,width=0.48\textwidth]{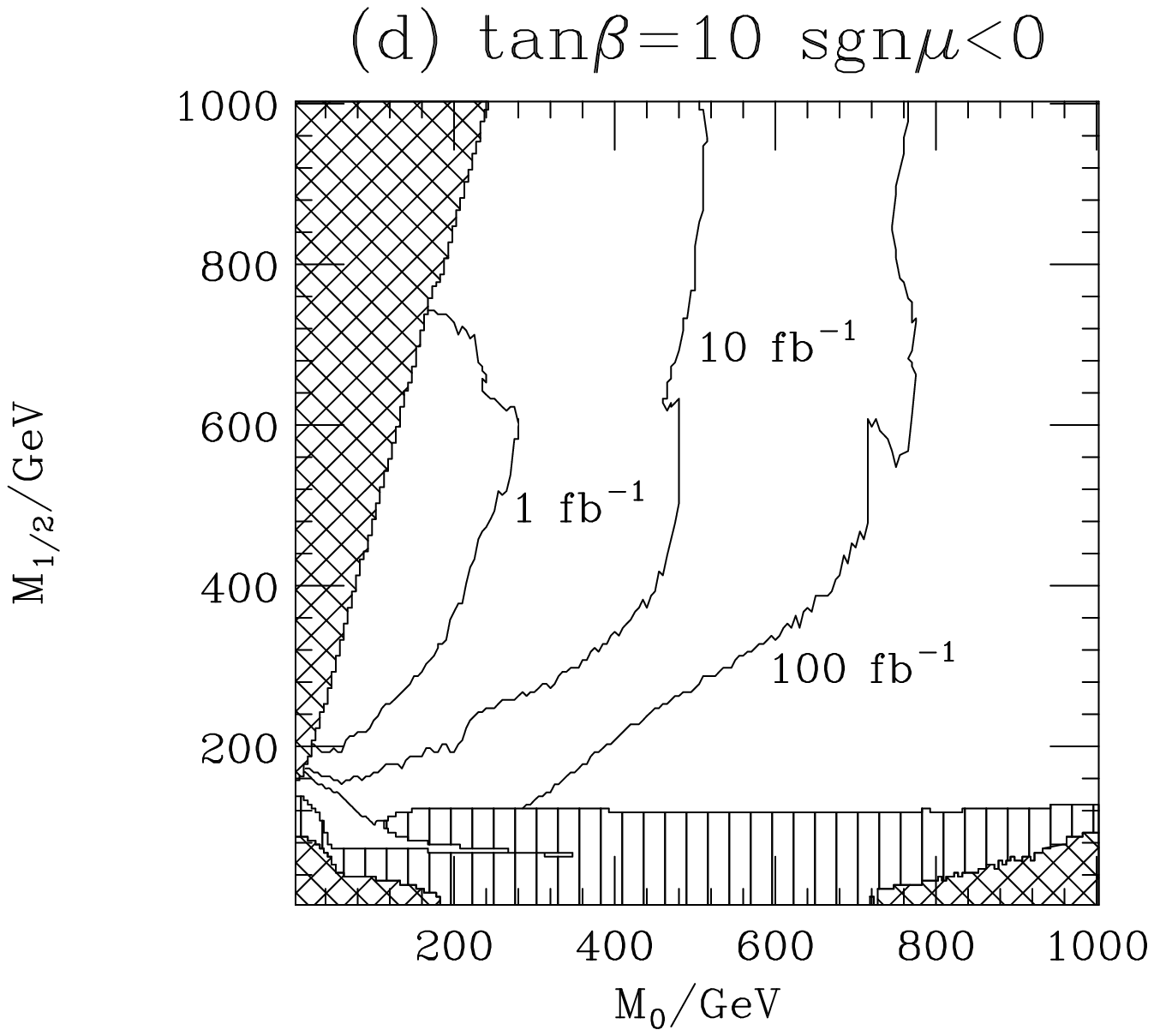}\\
\caption{Contours showing the discovery potential of the LHC in the $M_0$,
	 $M_{1/2}$ plane for ${\lam'}_{211}=10^{-2}$ and $A_0=0\, \mr{\gev}$.
	 These are a $5\sigma$ excess of the signal above the
	 background. Here in addition to the cuts on the isolation and $p_T$
  	 of the leptons, the transverse mass and
	 the missing transverse energy
	 described in the text, and a veto on the presence of OSSF leptons
	 we have imposed a cut on the presence of more than two jets. Here we
	 have included the sparticle pair production background as
	 well as the Standard Model backgrounds.
	 The striped and hatched regions are
	 described in the caption of Fig.\,\ref{fig:SUSYmass}.} 
\label{fig:lhcSUSYjet}
\end{center}
\end{figure}
%
%
\begin{figure}[htp]
\begin{center}
\includegraphics[angle=90,width=0.48\textwidth]{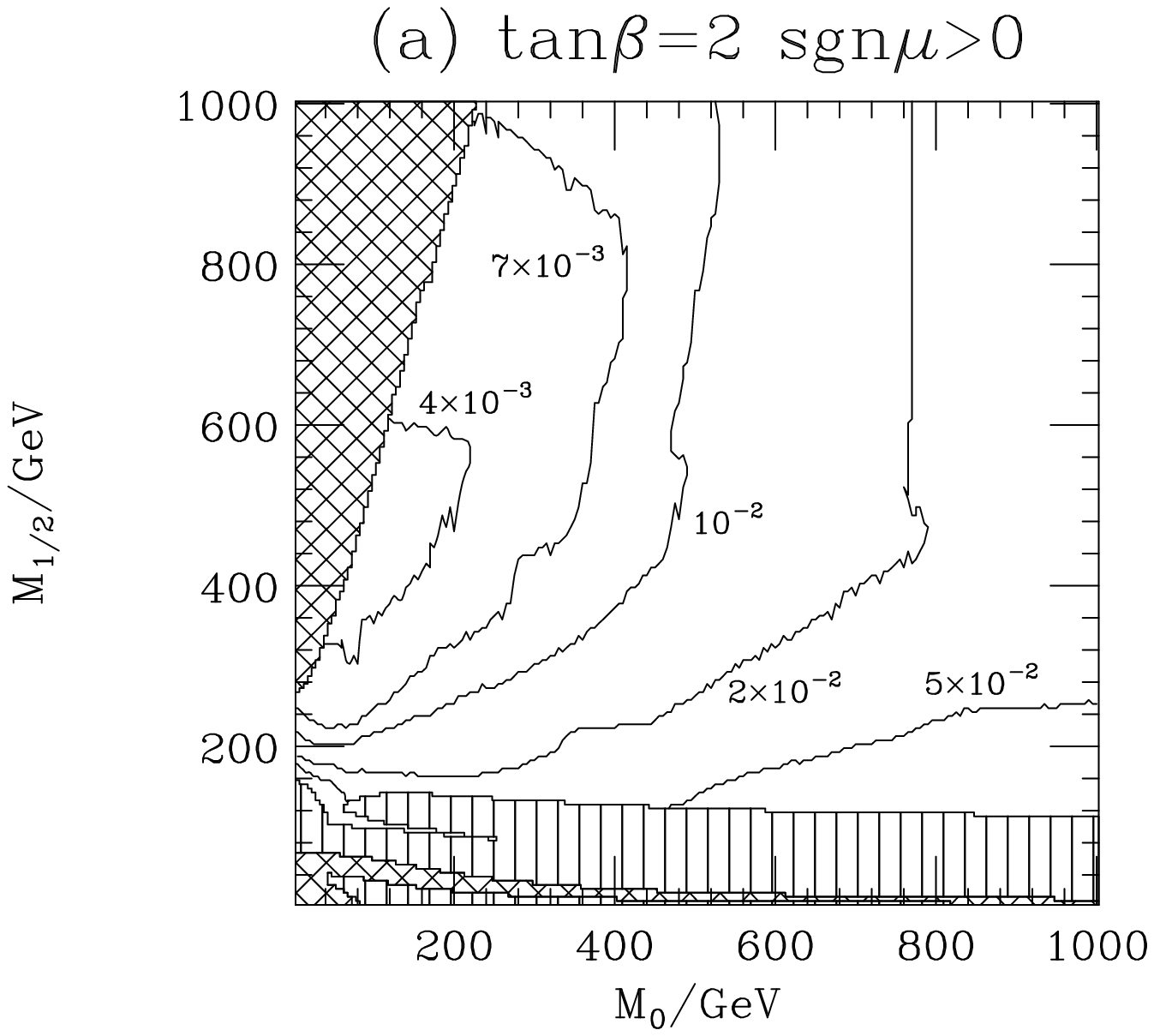}
\hfill						
\includegraphics[angle=90,width=0.48\textwidth]{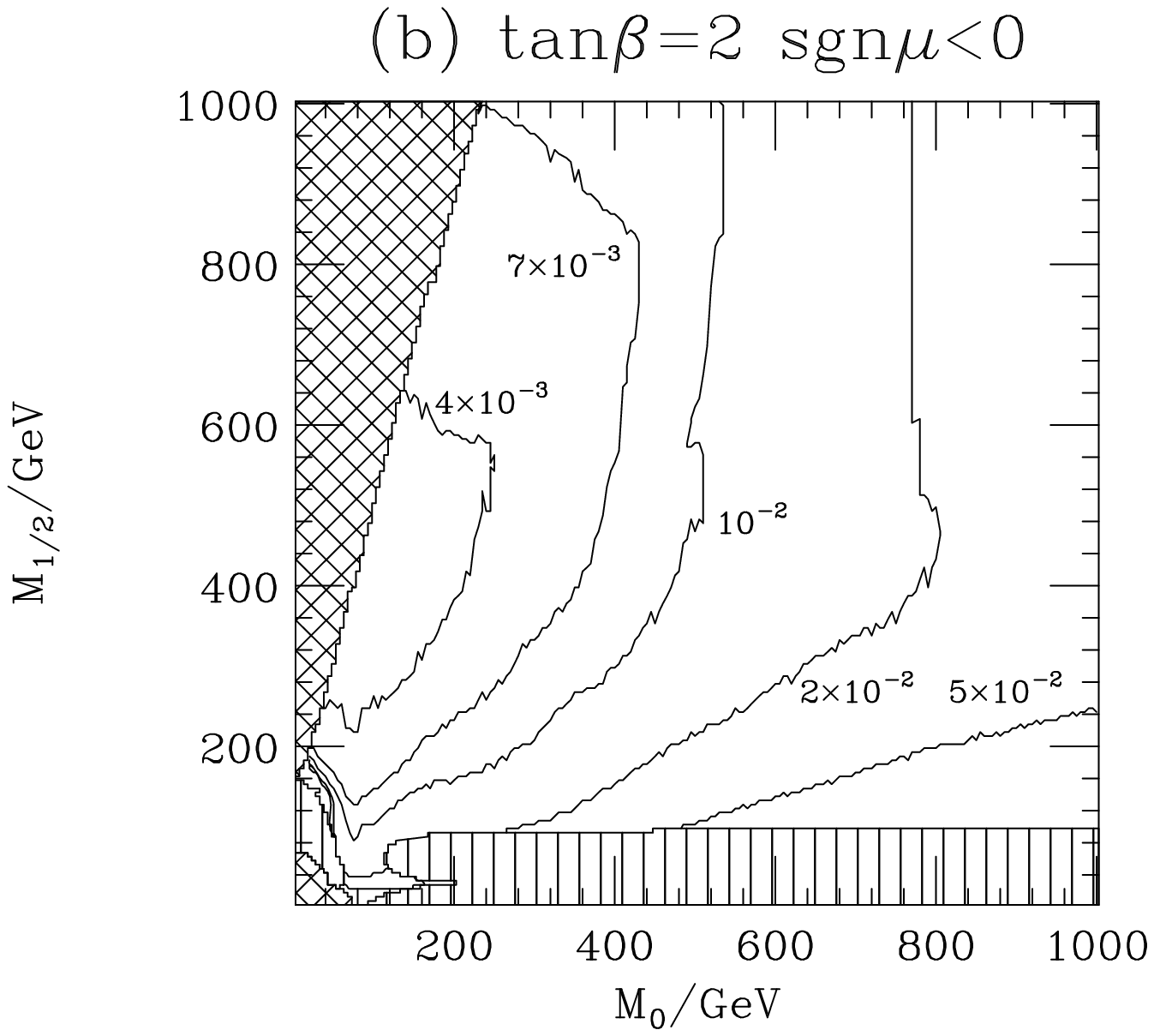}\\
\vskip 7mm					
\includegraphics[angle=90,width=0.48\textwidth]{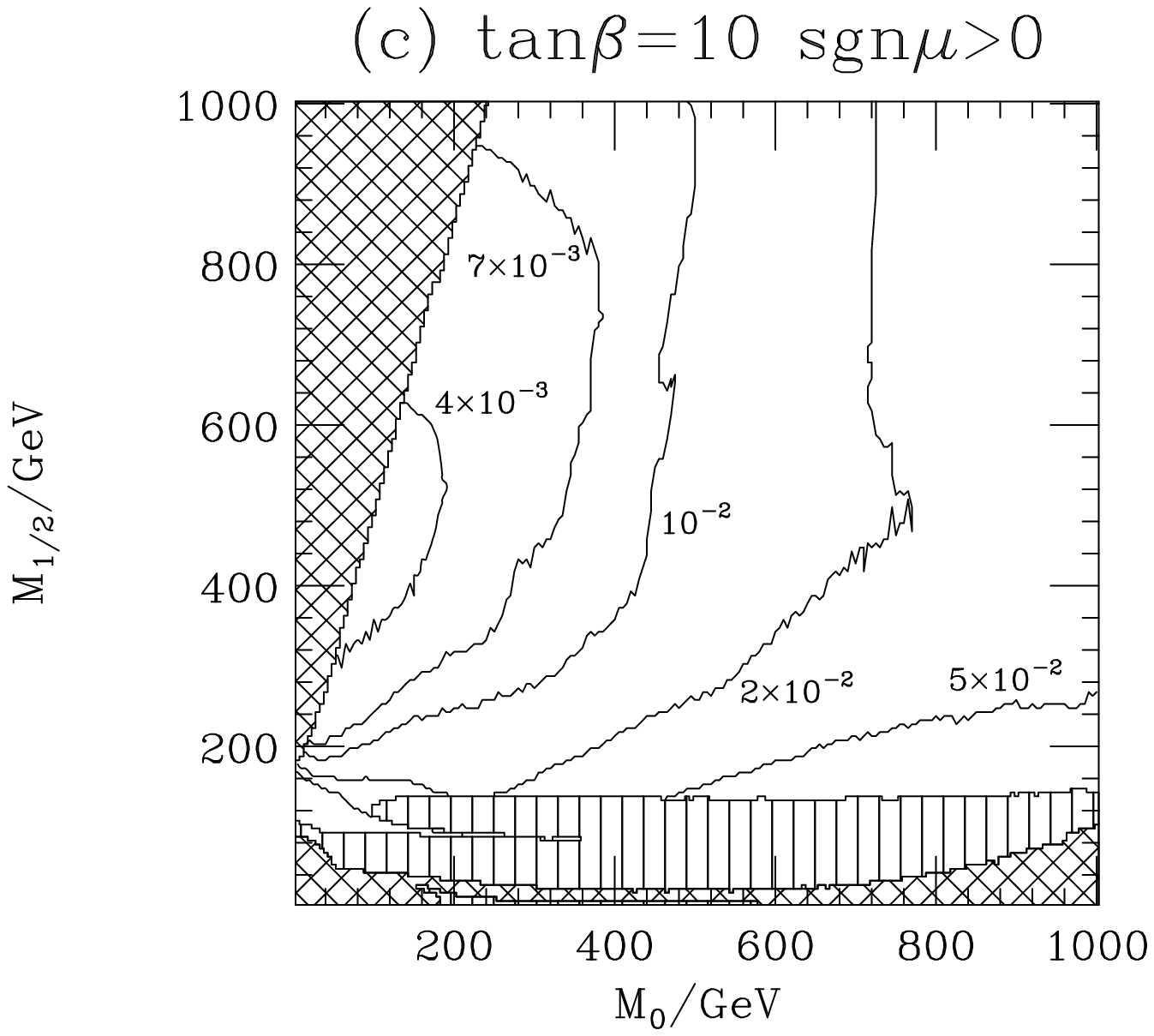}
\hfill						
\includegraphics[angle=90,width=0.48\textwidth]{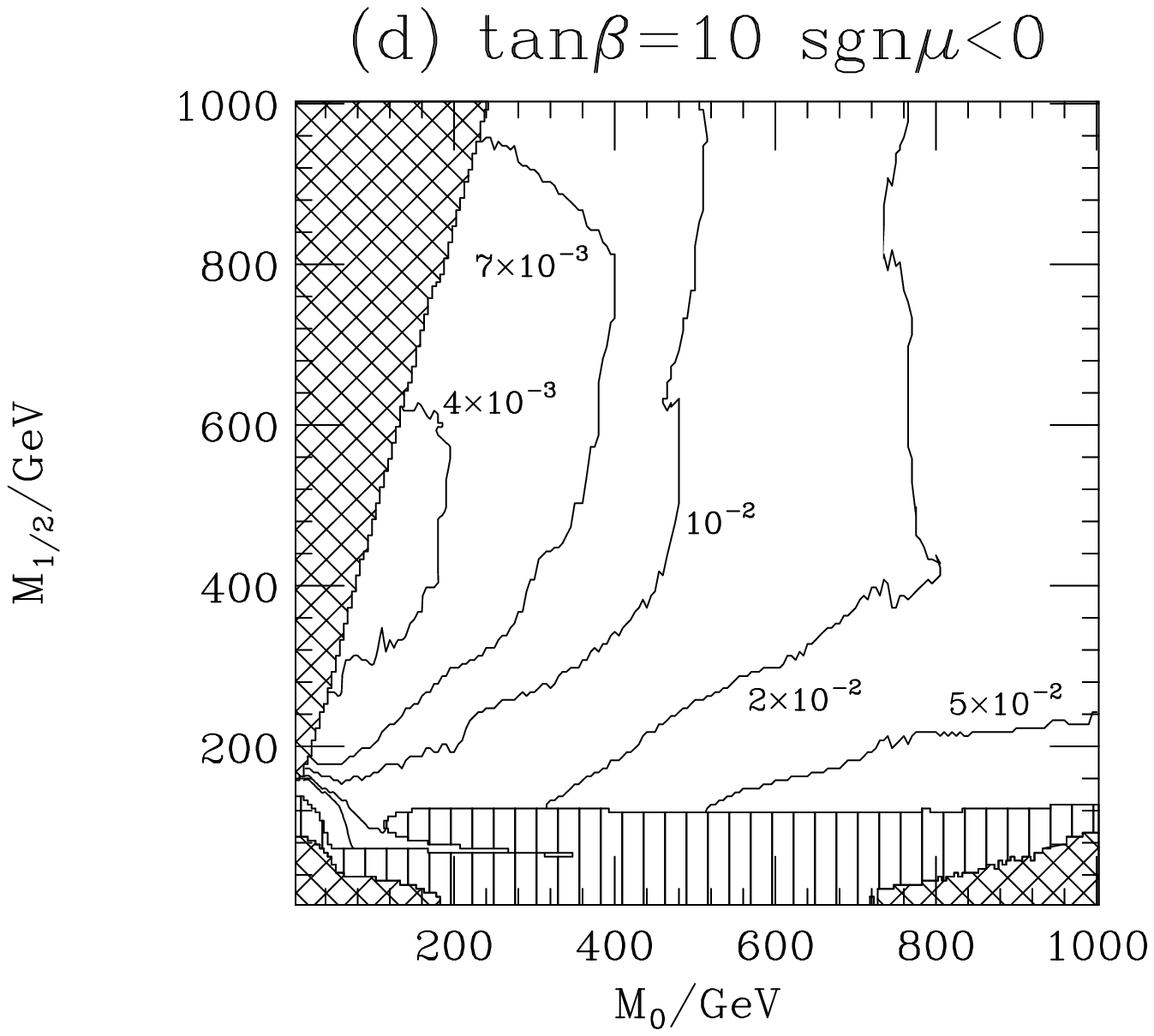}\\
\caption{Contours showing the discovery potential of the LHC in the $M_0$,
	 $M_{1/2}$ plane for $A_0=0\, \mr{\gev}$ 
	 and an integrated luminosity of
  	 $10\  \mr{fb}^{-1}$ with different values of ${\lam'}_{211}$.
	 These are a $5\sigma$ excess of the signal above the
	 background. Here in addition to the cuts on the isolation and $p_T$
  	 of the leptons, the transverse mass and the
	 missing transverse energy
	 described in the text, and a veto on the presence of OSSF leptons
	 we have imposed a cut on the presence of more than two jets. Here we
	 have included the sparticle pair production background
	 as well as the Standard Model backgrounds.
	 The striped and hatched regions are
	 described in the caption of Fig.\,\ref{fig:SUSYmass}.} 
\label{fig:lhcSUSYjetb}
\end{center}
\end{figure}

  As the sparticle pair production background at the LHC is larger than at
  the Tevatron we needed to simulate more events in order to 
  obtain a reliable
  estimate of the acceptance for this background. This meant that
  with the available resources we were forced to use a coarser scan of
  the $M_0$, $M_{1/2}$ plane. We used a 16 point grid and simulated a 
  different number of events at each point 
  depending on the value of $M_{1/2}$ as
  the sparticle pair production cross section is decreases as $M_{1/2}$
  increases. We simulated $10^5$, $10^5$, $10^6$,  and $10^7$ events
  at each of four points for $M_{1/2}=875\, \mr{\gev}$,
  $M_{1/2}=625\,\mr{\gev}$,
  $M_{1/2}=375\,\mr{\gev}$ and $M_{1/2}=125\,\mr{\gev}$, respectively.

  Our estimate of the discovery potential of the LHC after this cut,
  including all the
  backgrounds is given in Fig.\,\ref{fig:lhcSUSYjet},
  for ${\lam}_{211}=10^{-2}$
  with different integrated luminosities,
  and in Fig.\,\ref{fig:lhcSUSYjetb}, for
  an integrated luminosity of $10\  \mr{fb}^{-1}$
  with different values of the \rpv\  Yukawa couplings. As with the
  Tevatron the discovery potential 
  is reduced in two regions relative to that shown in 
  Figs.\,\ref{fig:lhcSMnojet} and \ref{fig:lhcSMnojetb}.
  The reduction at high $M_{1/2}$ is due to the smaller signal after the
  imposition of the jet cut, whereas the reduction at small $M_{1/2}$ is
  due to the larger background.
  However there are still large regions of SUGRA
  parameter space in which this process is visible above the background,
  particularly at large $M_{1/2}$ where
  there is less sensitivity to sparticle
  pair production. Due to the larger backgrounds from sparticle pair
  production there are large regions where the signal is detectable above
  the background although the $S/B$ is small. In general there is a region
  extending around $200\,\mr{\gev}$  in $M_{1/2}$ above the bottom of the
  $5\sigma$ discovery contour for 100~$\mr{fb}^{-1}$ where $S/B<1$.

  If we again neglect the region at small $M_0$ and $M_{1/2}$, which
  cannot be probed for any \rpv\  Yukawa couplings given our cuts, we 
  can obtain a mass reach for the LHC, with a given \rpv\  Yukawa coupling.
  Slepton masses of $460\,(600)\,\mr{\gev}$  can be
  discovered with 10\,(100)~$\mr{fb}^{-1}$
  integrated luminosity for a coupling ${\lam'}_{211}=0.05$ and
  slepton masses of $610\,(820)\,\mr{\gev}$  can be observed with
  10\,(100)~$\mr{fb}^{-1}$
  integrated luminosity for a coupling ${\lam'}_{211}=0.1$.

%
%
\subsection{Mass Reconstruction}

   There are many possible models which lead to an excess of like-sign
   dilepton pairs, over the prediction of the Standard Model. Indeed, we
   have seen that within the \rpv\  extension of the MSSM such an excess
   could be due to either sparticle pair production followed by \rpv\  
   decays of the LSPs, or resonant charged slepton production followed
   by a supersymmetric gauge decay of the slepton. The cut
   on the number of jets described above gives one way of discriminating
   between these two scenarios. 

\begin{figure}[htp]
\vspace{-15mm}
\begin{center}
\includegraphics[angle=90,width=0.48\textwidth]{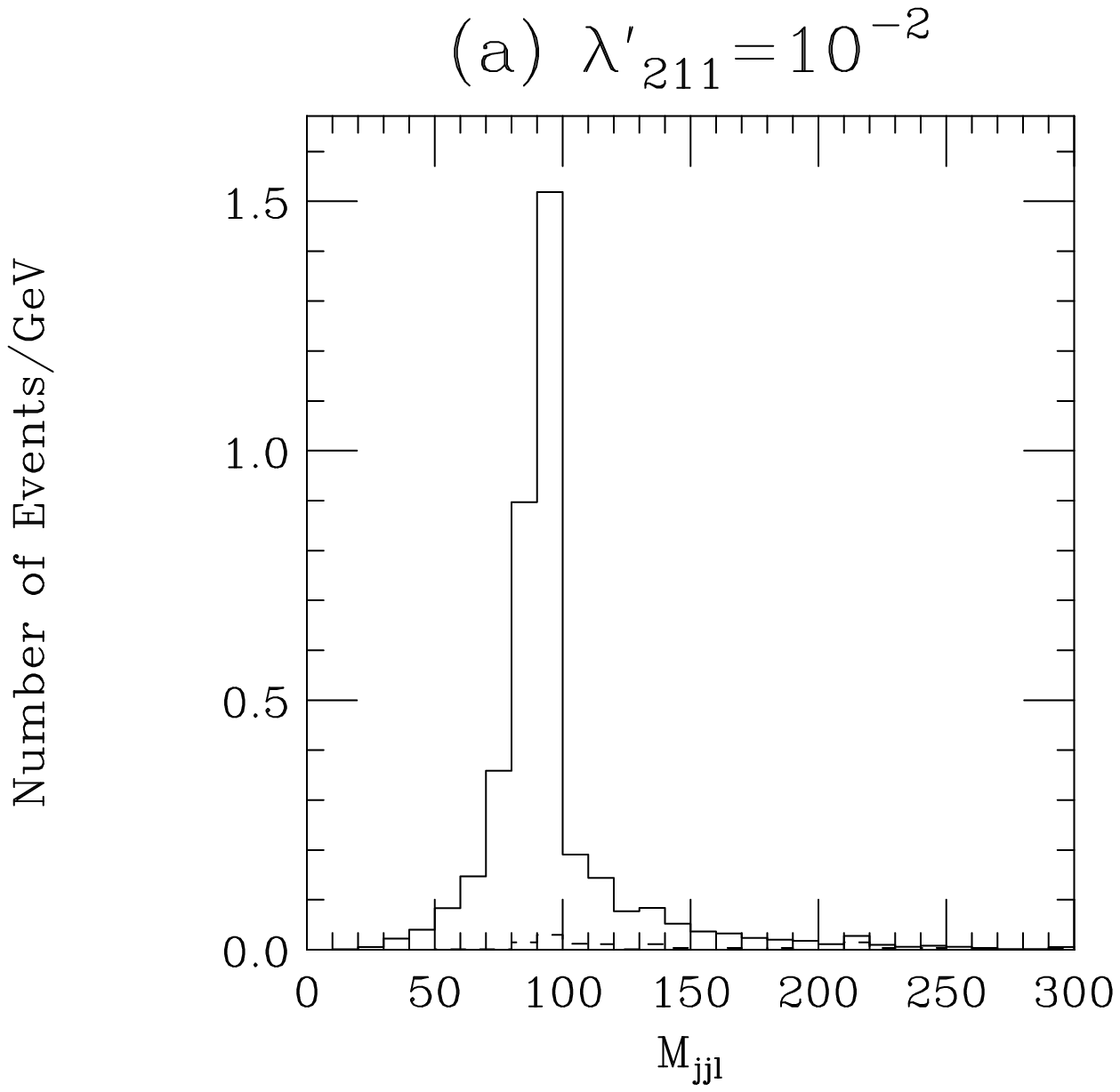}
\hfill
\includegraphics[angle=90,width=0.48\textwidth]{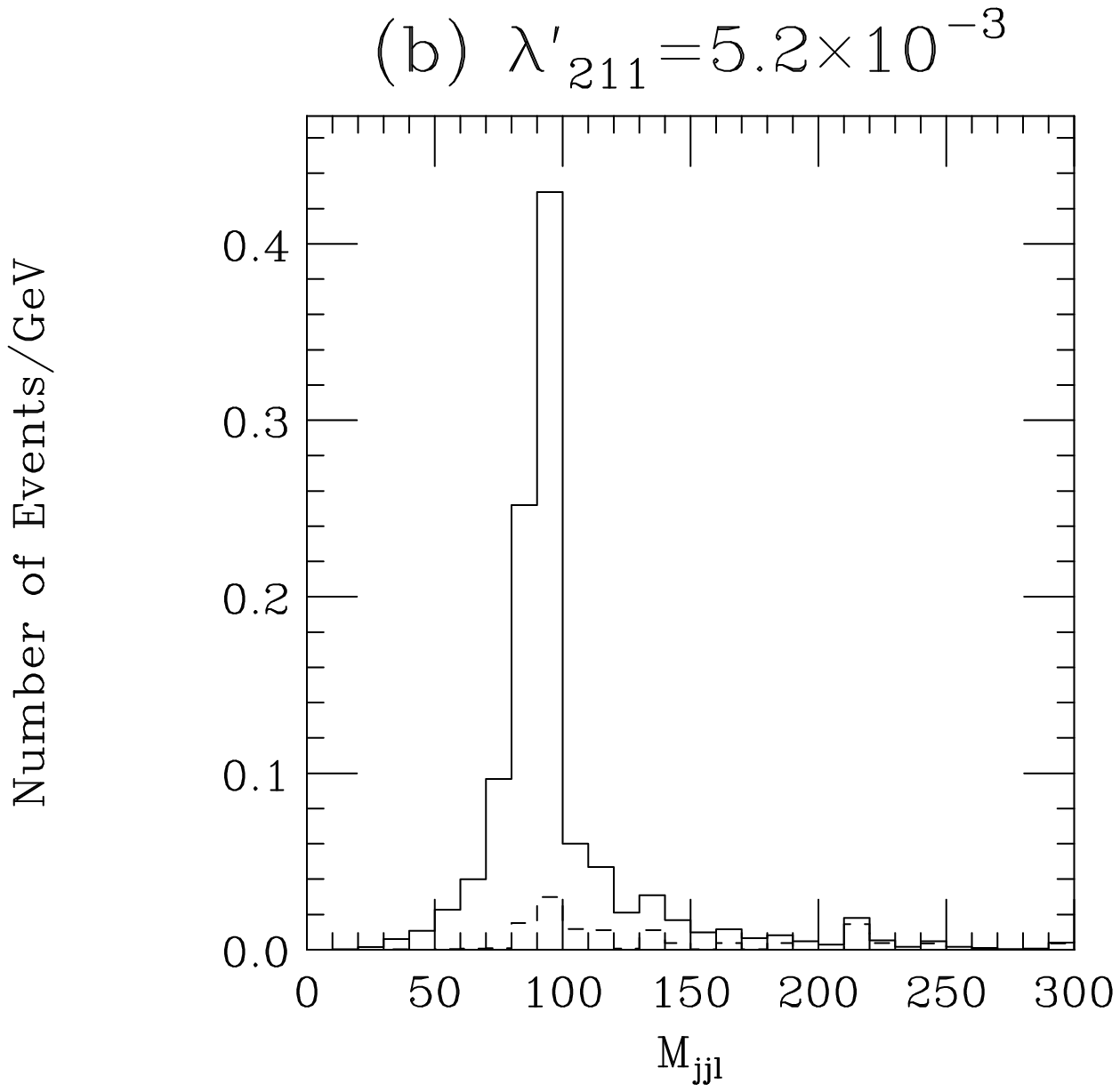}\\
\caption{The reconstructed neutralino mass at the Tevatron 
	for $M_0=50\,\mr{\gev}$, \mbox{$M_{1/2}=250\,\mr{\gev}$},
	$\tan\beta=2$, $\sgn\mu>0$ and $A_0=0\,\mr{\gev}$. The value of the
	coupling in
     	(b) is chosen such that after the cuts applied in 
	Section~\ref{sub:tevatron}
	the signal is $5\sigma$ above the background. At this point the 
	lightest neutralino mass is $M_{\cht^0_1}=98.9\,\mr{\gev}$.
	We have again normalized the
	distributions to an integrated luminosity of $2\ \mr{fb}^{-1}$.
	The dashed line shows the background and the solid line
	the sum of the signal and the background.} 
\label{fig:tevsusyneutmass}
\end{center}
%
%
\begin{center}
\includegraphics[angle=90,width=0.48\textwidth]{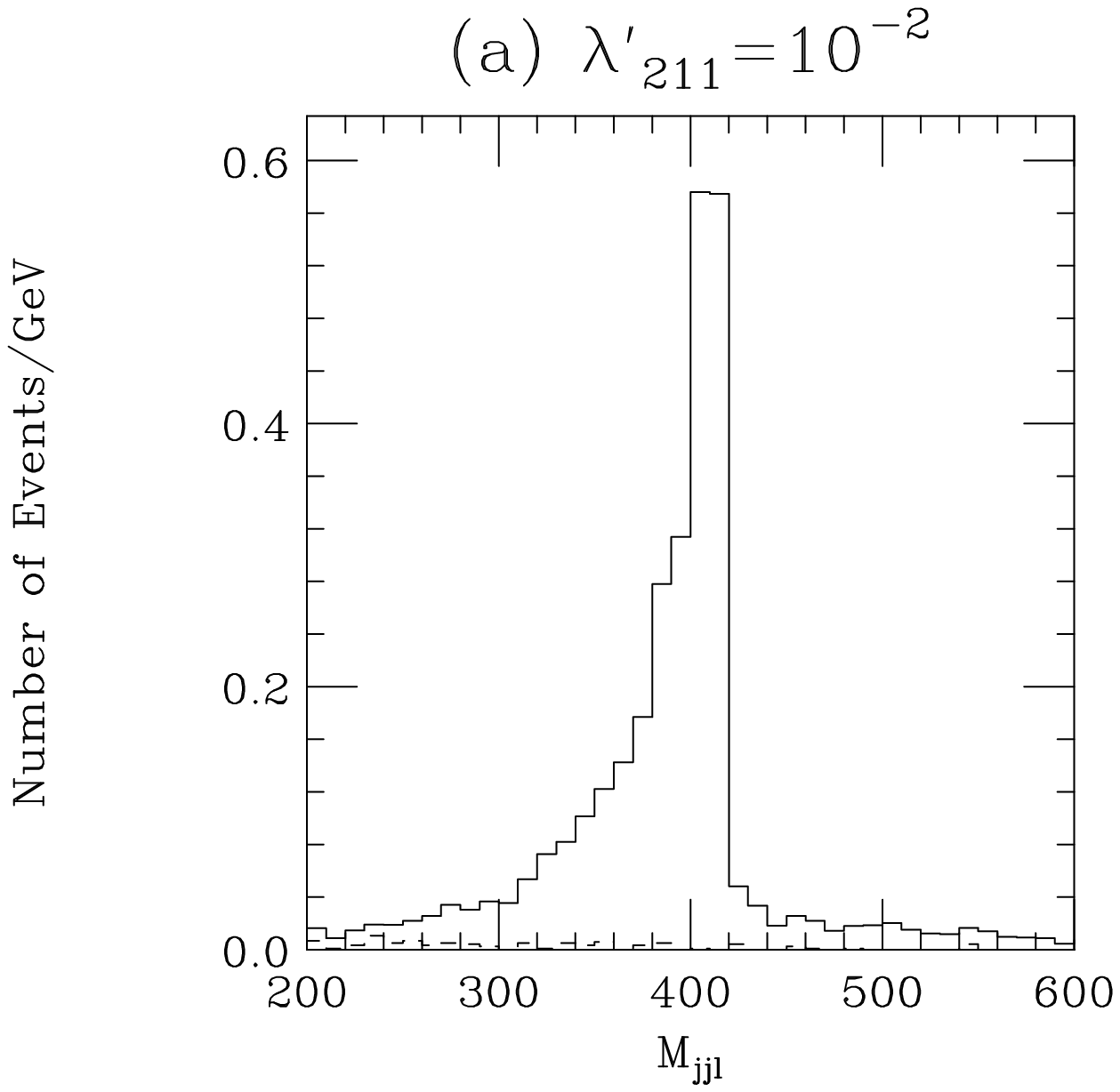}
\hfill
\includegraphics[angle=90,width=0.48\textwidth]{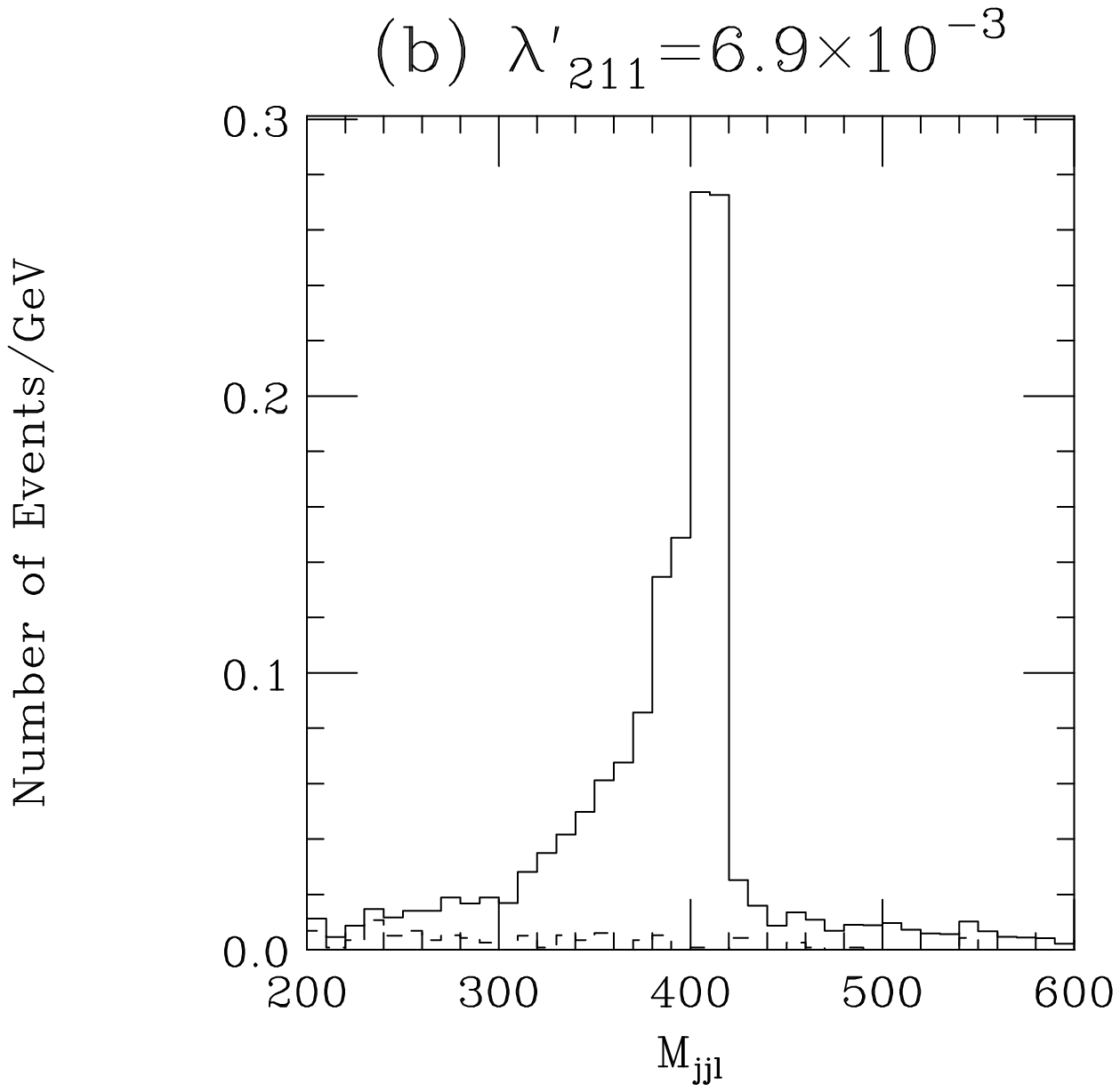}\\
\caption{The reconstructed neutralino mass at the LHC 
	for  $M_0=350\,\mr{\gev}$, \mbox{$M_{1/2}=950\,\mr{\gev}$},
	$\tan\beta=10$, $\sgn\mu<0$ and $A_0=0\,\mr{\gev}$. The value of the
	coupling in
     	(b) is chosen such that after
	the cuts applied in Section~\ref{sub:LHC}
	the signal is $5\sigma$ above the background. At this point the
	lightest
	neutralino mass is $M_{\cht^0_1}=418.0\,\mr{\gev}$.
	 We have again normalized the
	distributions to an integrated luminosity of $10\ \mr{fb}^{-1}$.
	The dashed line shows the background and the solid line
	the sum of the signal and the background.} 
\label{fig:lhcsusyneutmass}
\end{center}
\end{figure}

   An additional method of distinguishing between these two scenarios 
   is to try and reconstruct the masses of the
   decaying sparticles for the resonant slepton production. In principle this
   is
   straightforward. The neutralino decay to a quark-antiquark pair
   and a charged lepton, will give two jets 
   (or more after the emission of QCD
   radiation) and a charged lepton. These decay products should be relatively
   close
   together. Therefore to reconstruct the neutralino we took the highest two 
   $p_T$ jets in the event and combined them with the charged lepton which
   was closest
   in $(\eta,\phi)$ space. We only used events in which both jets had 
   $p_T>10\,\mr{\gev}$
   in addition to passing all the cuts described, 
   \linebreak \ie both the cuts
   required to suppress the Standard Model and SUSY backgrounds,
   in the previous sections.
   This then gives a neutralino candidate, the masses of these candidates are
   shown for
   a sample point in SUSY parameter space for both the Tevatron, 
   Fig.\,\ref{fig:tevsusyneutmass}, and the LHC, 
   Fig.\,\ref{fig:lhcsusyneutmass}.
   In both cases in addition to showing the result for the
   coupling ${\lam'}_{211}=10^{-2}$, we have shown a coupling such that the
   signal is 
   exactly $5\sigma$ above the background at this point to show what can be
   seen if the
   signal is only just detectable. As can be seen in both figures the
   reconstructed
   neutralino mass is in good agreement with the simulated value, although
   the situation
   may be worse once detector effects have been included.

%
%
%
%
\begin{figure}
\vspace{-15mm}
\begin{center}
\includegraphics[angle=90,width=0.48\textwidth]{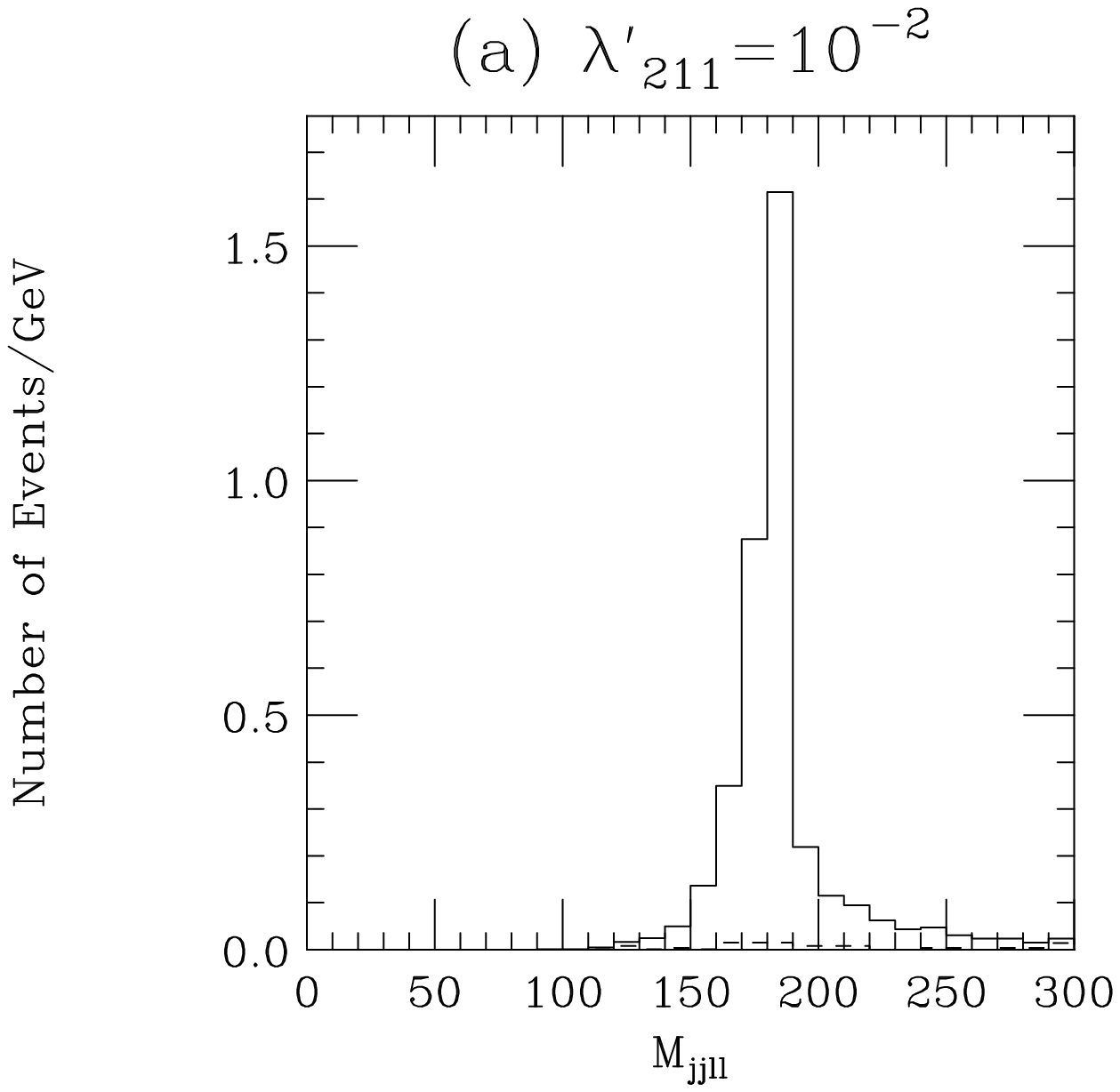}
\hfill
\includegraphics[angle=90,width=0.48\textwidth]{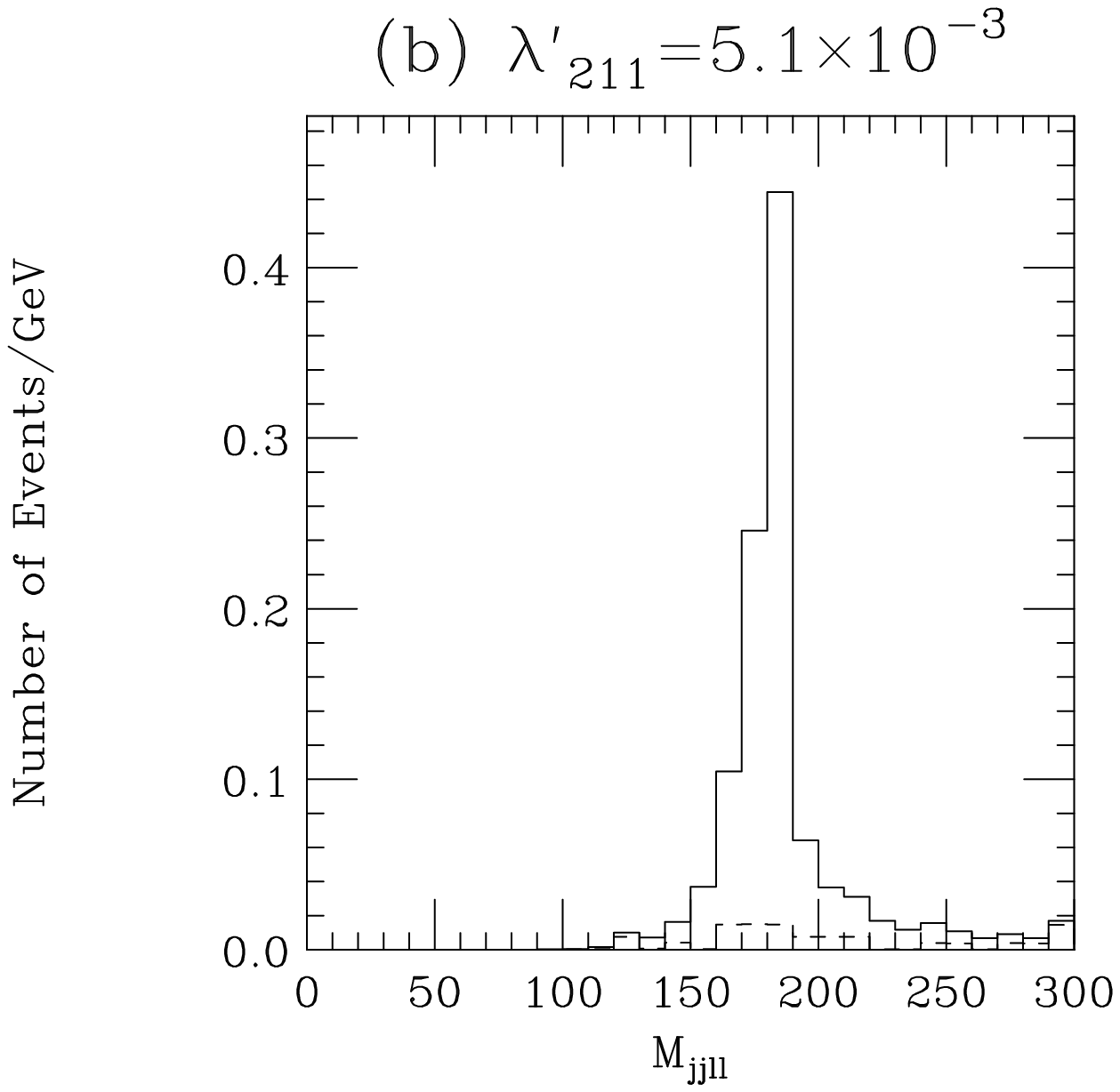}\\
\caption{The reconstructed slepton mass at the Tevatron 
	for  $M_0=50\,\mr{\gev}$, \mbox{$M_{1/2}=250\,\mr{\gev}$},
	$\tan\beta=2$, $\sgn\mu>0$ and $A_0=0\,\mr{\gev}$. The value of the
	coupling in
     	(b) is chosen such that after the cuts applied in
	Section~\ref{sub:tevatron}
	the signal is $5\sigma$ above the background. At this point the
	smuon mass
  	is $M_{\mr{\mut_L}}=189.1\,\mr{\gev}$. We have again normalized the
	distributions to an integrated luminosity of $2\ \mr{fb}^{-1}$.
	The dashed line shows the background and the solid line
	the sum of the signal and the background.} 
\label{fig:tevsusyslepmass}
\end{center}
%
%
\begin{center}
\includegraphics[angle=90,width=0.48\textwidth]{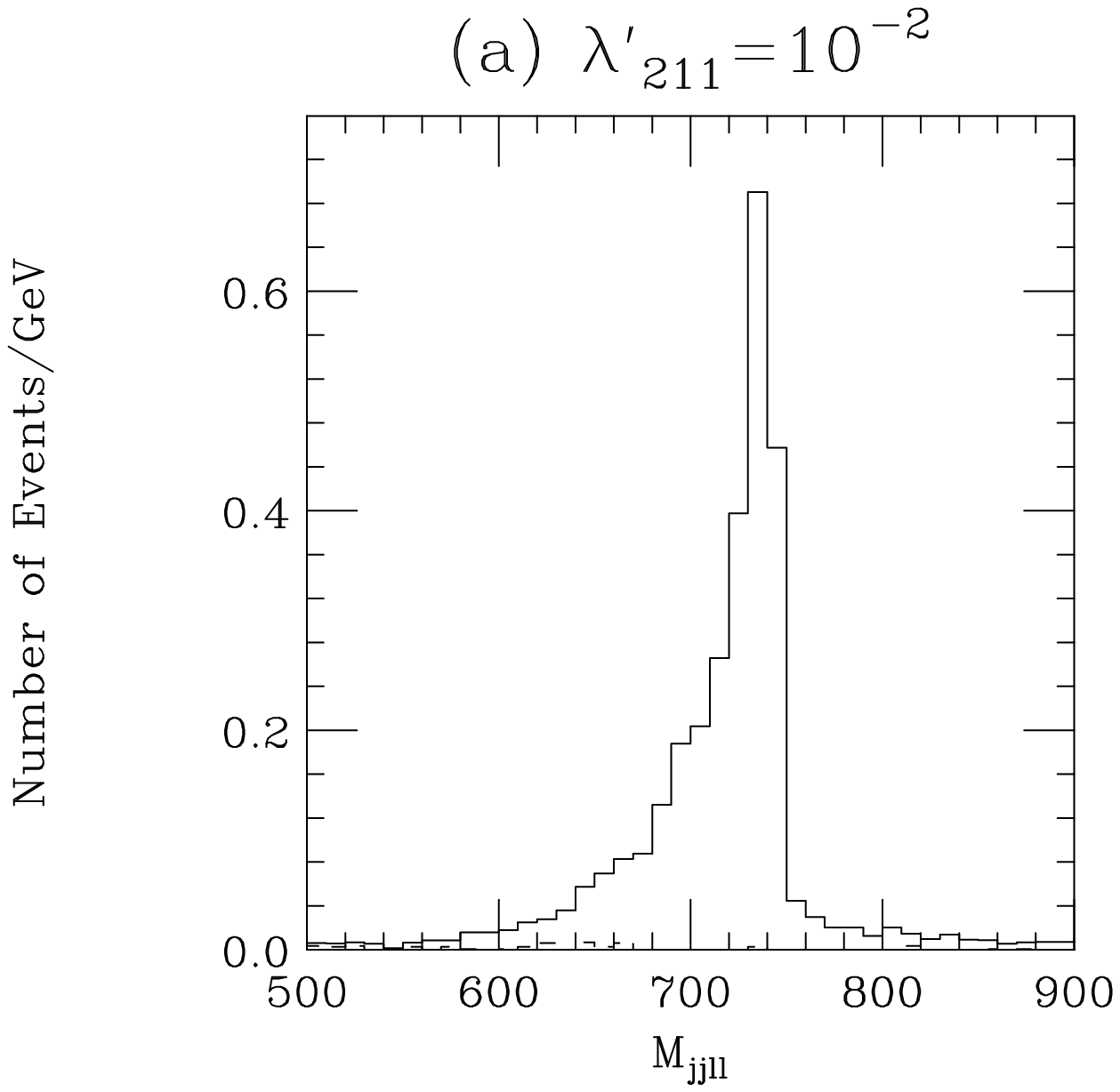}
\hfill
\includegraphics[angle=90,width=0.48\textwidth]{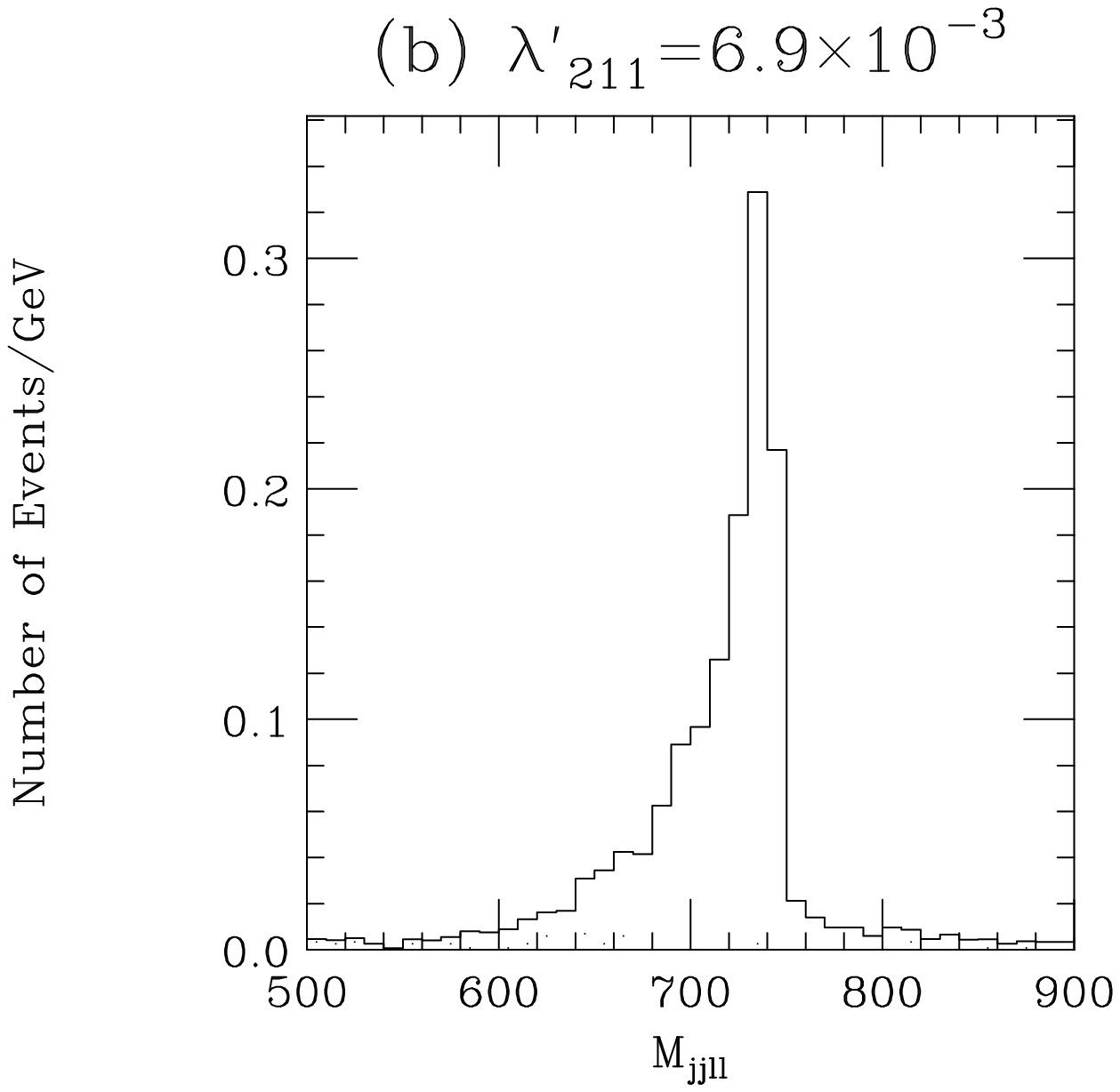}\\
\caption{The reconstructed slepton mass at the LHC 
	for  $M_0=350\,\mr{\gev}$, \mbox{$M_{1/2}=950\,\mr{\gev}$},
	$\tan\beta=10$, $\sgn\mu<0$ and $A_0=0\,\mr{\gev}$. The value of the
	coupling in
     	(b) is chosen such that after the cuts
	applied in Section~\ref{sub:LHC}
	the signal is $5\sigma$ above the background. At this point the smuon
	mass is $M_{\mr{\mut_L}}=745.9\,\mr{\gev}$.
	We have again normalized the
	distributions to an integrated luminosity of $10\ \mr{fb}^{-1}$.
	The dashed line shows the background and the solid line
	the sum of the signal and the background.} 
\label{fig:lhcsusyslepmass}
\end{center}
\end{figure}
    
   We can then combine this neutralino candidate with the remaining lepton in
   the event
   to give a slepton candidate, under the assumption that the like-sign
   leptons were produced in the process $\mr{\elt^+\ra\ell^+\cht^0_1}$.
   The mass distribution of these slepton
   candidates is shown
   in Fig.\,\ref{fig:tevsusyslepmass} for the Tevatron and 
   Fig.\,\ref{fig:lhcsusyslepmass} for the LHC. Again there is good agreement
   between
   the position of the peak in the distribution and the value of the smuon
   mass used in the simulation.

   The data for both the neutralino and smuon mass reconstructions is
   binned in $10\,\mr{\gev}$  bins.
   We have used the events in the central bin and
   the two bins on either side to  reconstruct the neutralino and smuon
   masses. These reconstructed masses are given in 
   Table~\ref{tab:reconstruct}. As can be seen for both the points
   we have shown the reconstructed mass lies between $5\,\mr{\gev}$
   and $15\,\mr{\gev}$  
   below the simulated sparticle mass. This is due to the loss of
   some of the energy of the jets produced
   in the neutralino decay from the cones used to define the jets.
   It is common
   to include this effect in the jet energy correction, so this shift would
   probably not be observed in a full experimental simulation.

   The agreement between the results of the simulation and the input values 
   is good provided that the Standard
   Model background is dominant over the background from sparticle pair 
   production and the lightest neutralino is predominantly produced in the
   smuon decay. This is the case at the points used in 
   Figs.\,\ref{fig:tevsusyneutmass}-\ref{fig:lhcsusyslepmass}. At the
   point $M_0=50\,\mr{\gev}$, $M_{1/2}=250\,\mr{\gev}$,
   $\tan\beta=2$, $\sgn\mu>0$ and
   $A_0=0\,\mr{\gev}$  the branching ratio for the decay of the smuon to the
   lightest neutralino is $\mr{BR}(\mr{\mut_L}\ra\cht^0_1\mu^+)=98\%$. 
   Similarly at the point  $M_0=350\,\mr{\gev}$, $M_{1/2}=950\,\mr{\gev}$, 
   $\tan\beta=10$, $\sgn\mu<0$ and $A_0=0\, \mr{\gev}$  the dominant 
   decay mode of the smuon is to the lightest neutralino, 
   $\mr{BR}(\mr{\mut_L}\ra\cht^0_1\mu^+)=99\%$.

%
%
%
\begin{table}
\begin{center}
\begin{tabular}{|c|c|c|c|c|c|c|c|}
\hline
  Experiment & ${\lam'}_{211}$ & Cuts & Point & 
  \multicolumn{2}{c|}{Neutralino mass/\gev} & 
  \multicolumn{2}{c|}{Slepton mass/\gev}\\
\cline{5-8}
  & & &  & Actual & Recon. & Actual & Recon. \\
\hline
 Tevatron & $10^{-2}$          & no & A & 98.9 & 90.3  & 189.1 & 181.6 \\
\hline					                
 Tevatron & $5.2\times10^{-3}$ & no & A & 98.9 & 91.8  & 189.1 & 182.3 \\
\hline					                
 LHC      & $10^{-2}$          & no & B & 418.0 & 404.1 & 745.0 & 734.1 \\
\hline					                
 LHC      & $6.9\times10^{-2}$ & no & B & 418.0 & 405.1 & 745.0 & 732.6 \\
\hline					                
 LHC      & $10^{-2}$          & no & C & 147.6 & 142.7 & 432.0 & 421.3 \\
\hline					                
 LHC      & $10^{-2}$ 	       &yes & C & 147.6 & 143.4 & 432.0 & 423.3 \\
\hline
\end{tabular}
\caption[Reconstructed neutralino and slepton masses.]
	{Reconstructed neutralino and slepton masses.
	 The following SUGRA points were used in these simulations: point A
	 has 
         $M_0=50\, \mr{\gev}$, $M_{1/2}=250\, \mr{\gev}$, $\tan\beta=2$, 
	 $\sgn\mu>0$ and $A_0=0\, \mr{\gev}$;
	 point B has
    	 $M_0=350\, \mr{\gev}$, $M_{1/2}=950\, \mr{\gev}$,     
   	 $\tan\beta=10$, $\sgn\mu<0$ and $A_0=0\, \mr{\gev}$;
	 point C has 
	 $M_0=350\, \mr{\gev}$, $M_{1/2}=350\, \mr{\gev}$,
  	 $\tan\beta=10$, $\sgn\mu<0$ and $A_0=0\, \mr{\gev}$.
         The Tevatron  and LHC results are based on $2\  \mr{fb}^{-1}$
	 and $10\  \mr{fb}^{-1}$ integrated luminosity, respectively.}
\label{tab:reconstruct}
\end{center}
\end{table}

\begin{figure}
\begin{center}
\includegraphics[angle=90,width=0.48\textwidth]{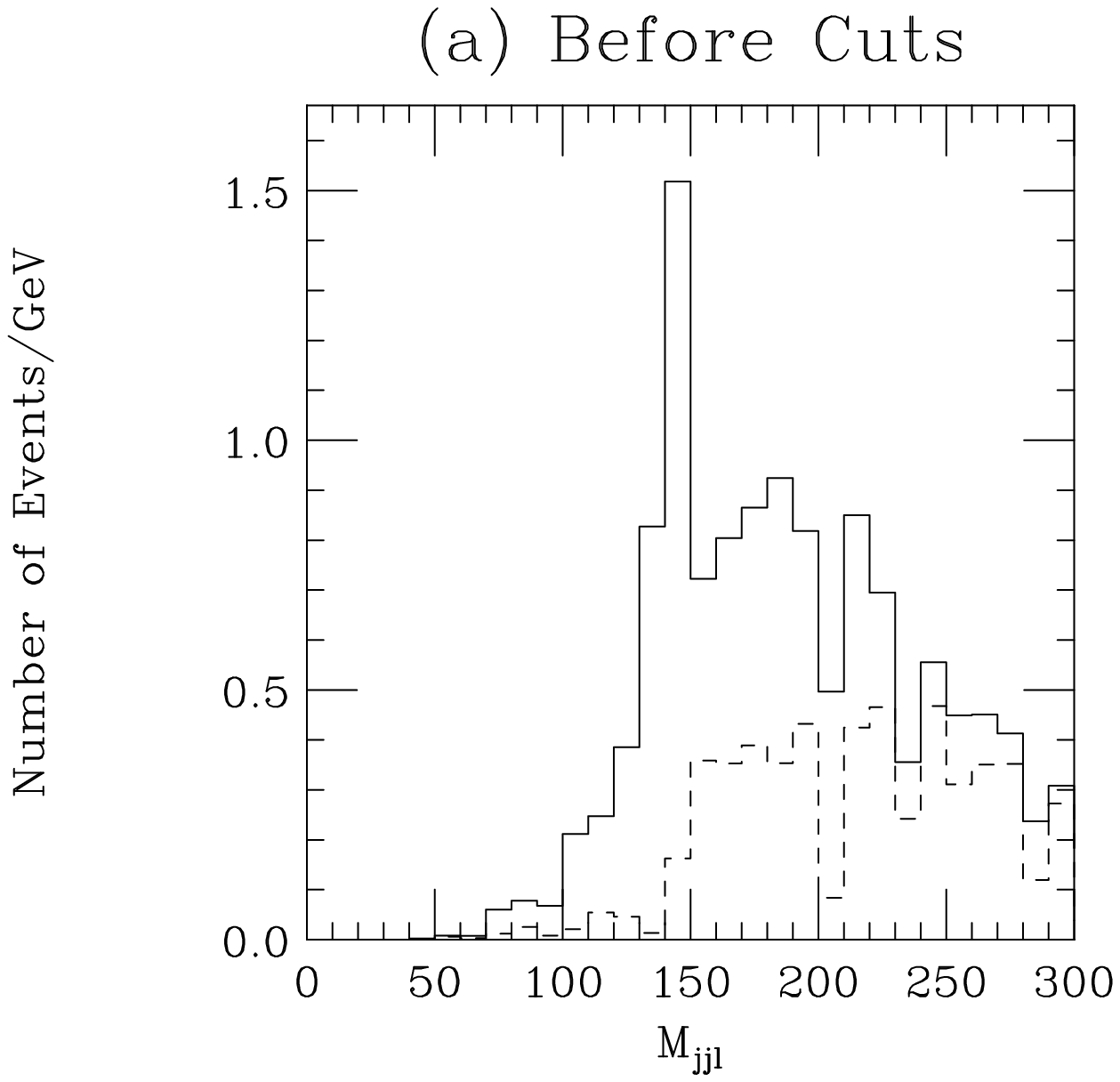}
\hfill
\includegraphics[angle=90,width=0.48\textwidth]{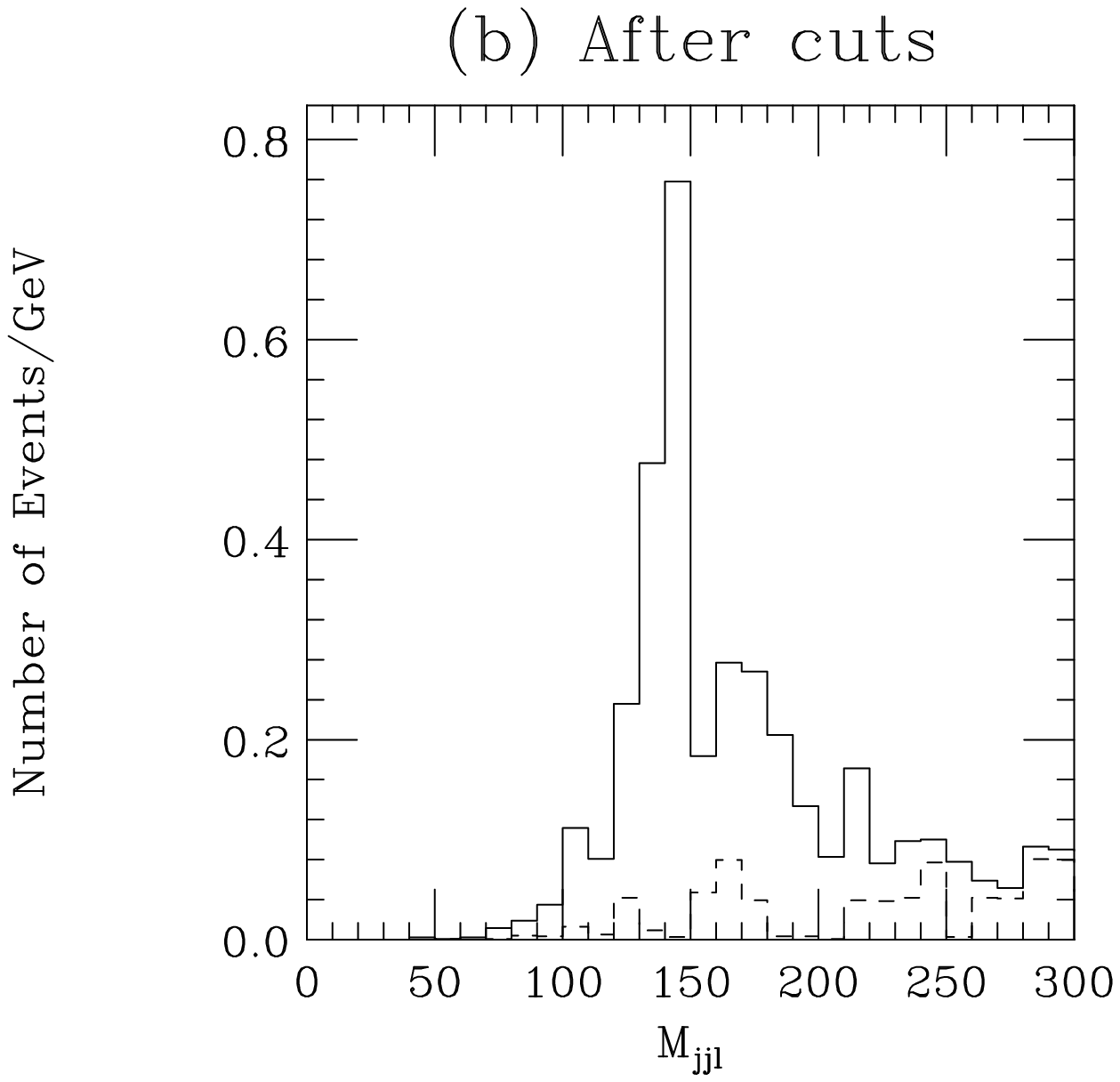}\\
\caption{The reconstructed neutralino mass at the LHC 
	for  ${\lam'}_{211}=10^{-2}$, $M_0=350\, \mr{\gev}$,
        $M_{1/2}=350\, \mr{\gev}$,
	$\tan\beta=10$, $\sgn\mu<0$ and $A_0=0\, \mr{\gev}$.
	At this point the
	lightest neutralino mass is $M_{\cht^0_1}=147.6\, \mr{\gev}$.
	We have again normalized the
	distributions to an integrated luminosity of $10\ \mr{fb}^{-1}$. The
	cuts used are described in the text.
	The dashed line shows the background and the solid line
	the sum of the signal and the background.} 
\label{fig:lhcsusyneutmasscut}
\end{center}
%
%
\begin{center}
\includegraphics[angle=90,width=0.48\textwidth]{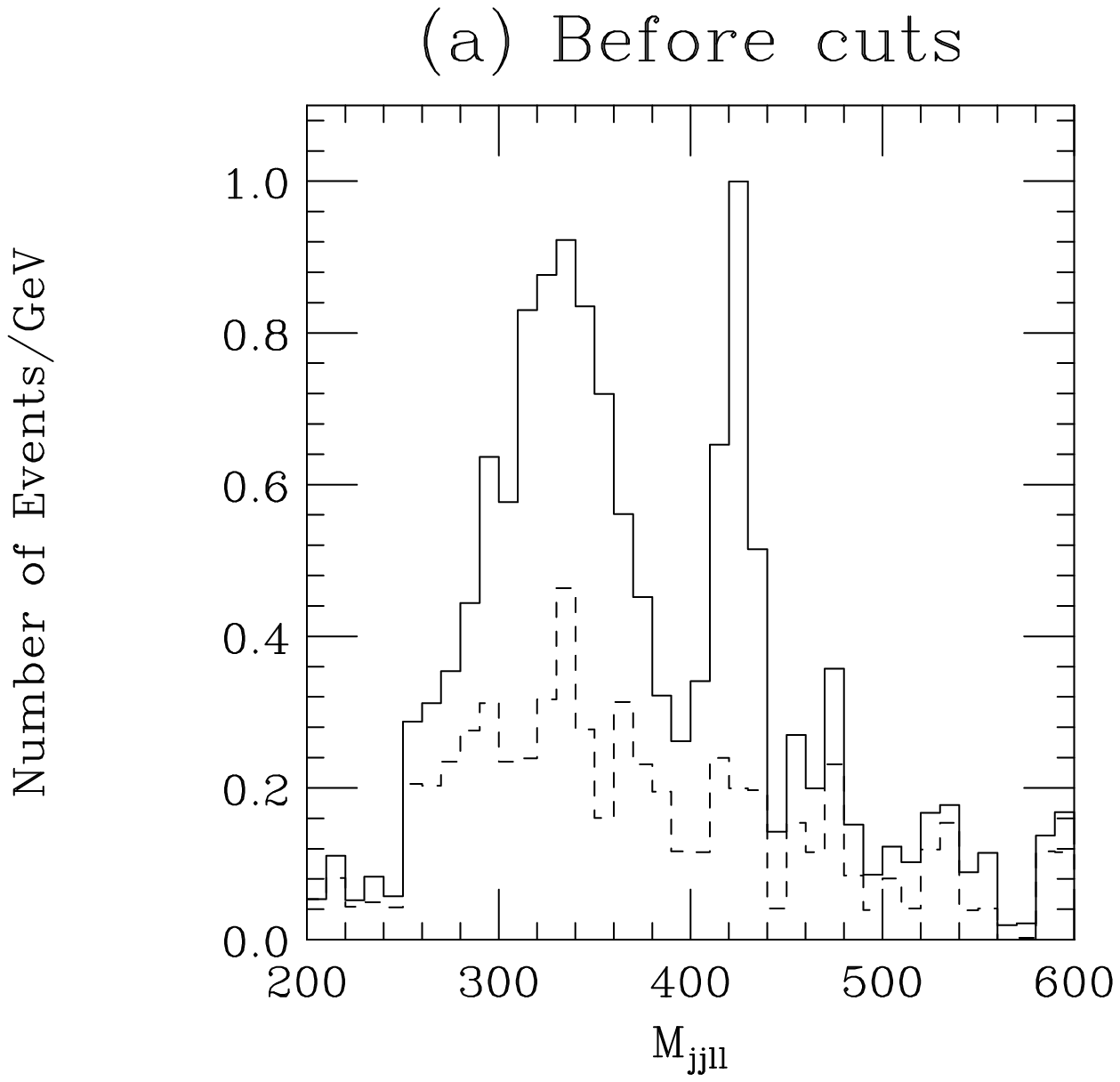}
\hfill
\includegraphics[angle=90,width=0.48\textwidth]{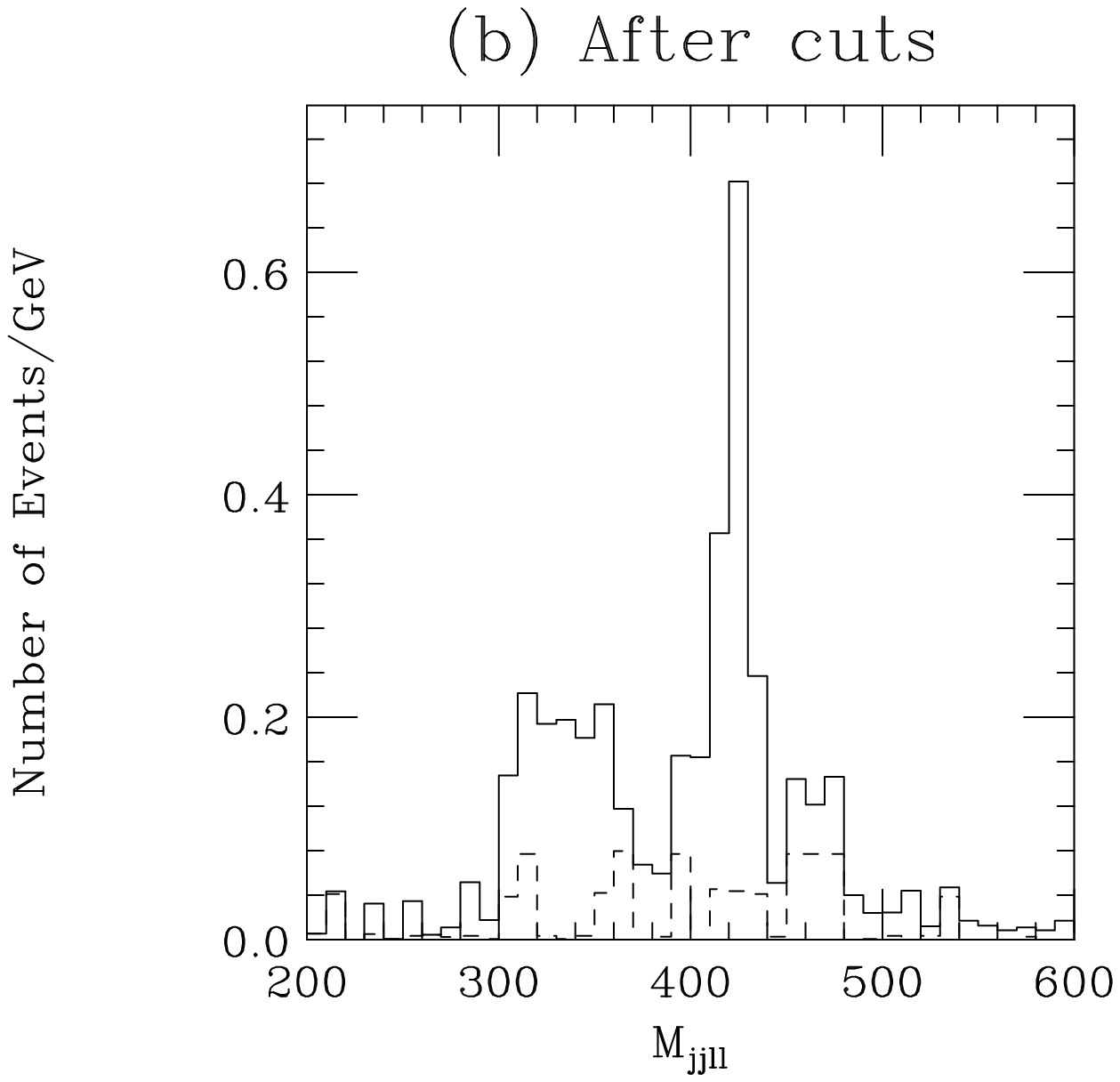}\\
\caption{The reconstructed slepton mass at the LHC 
	for  ${\lam'}_{211}=10^{-2}$, $M_0=350\, \mr{\gev}$, $M_{1/2}=350\,
	\mr{\gev}$,
	$\tan\beta=10$, $\sgn\mu<0$ and $A_0=0\, \mr{\gev}$.
	At this point the smuon
	mass is $M_{\mr{\mut_L}}=432.0\, \mr{\gev}$.
	We have again normalized the
	distributions to an integrated luminosity of $10\ \mr{fb}^{-1}$. The
	cuts used are described in the text.
	The dashed line shows the background and the solid line
	the sum of the signal and the background.} 
\label{fig:lhcsusyslepmasscut}
\end{center}
\end{figure}

  It can however be the case that there is a significant background from
  sparticle pair production and a substantial contribution from the
  production 
  of charginos and heavier neutralinos. This is shown in
  Fig.\,\ref{fig:lhcsusyneutmasscut}a for
  the neutralino mass reconstruction and Fig.\,\ref{fig:lhcsusyslepmasscut}a
  for
  the smuon mass reconstruction. As can be seen in the 
  Figs.\,\ref{fig:lhcsusyneutmasscut}a and \ref{fig:lhcsusyslepmasscut}a
  there is
  a significant background in both distributions. At this point, \ie
  ${\lam'}_{211}=10^{-2}$, \mbox{$M_0=350\, \mr{\gev}$},
  $M_{1/2}=350\, \mr{\gev}$,
  $\tan\beta=10$, $\sgn\mu<0$ and $A_0=0\, \mr{\gev}$, the smuon dominantly 
  decays to the lightest chargino,
  $\mr{BR}(\mr{\mut_L\ra\cht^-_1\nu_\mu})=50.9\%$,
  with the other important decay modes being to the next-to-lightest
  neutralino,
  $\mr{BR}(\mr{\mut_L}\ra\cht^0_2\mu^+)=28.0\%$, and the lightest neutralino
  $\mr{BR}(\mr{\mut_L}\ra\cht^0_1\mu^+)=20.9\%$.
  In the neutralino distribution there is still a peak at the simulated
  neutralino
  mass, although there is a large tail at higher masses. This tail is mainly
  due to
  the larger sparticle pair production background. The production 
  of charginos and the
  heavier
  neutralinos does not significantly effect this distribution as the heavier
  gauginos will cascade decay to the LSP due to the small \rpv\  coupling.

  In the slepton distribution in addition to the larger background there is
  also a spurious peak in the mass distribution due to the production of the
  the lightest chargino and the $\cht^0_2$.
  As we are not including all the decay products of the chargino or 
  $\cht^0_2$ in the mass
  reconstruction the reconstructed slepton mass in signal events where a 
  chargino or 
  heavier neutralino is produced is below the true value.

  We can improve the extraction of both the neutralino and slepton masses by
  imposing some additional cuts.
  The aim of these cuts is to require that the 
  neutralino candidate and the second lepton are produced back-to-back,
  because in most of the signal events the resonant smuon will only
  have a small transverse momentum due to the initial-state parton shower.
  We therefore 
  require the transverse momenta of the neutralino candidate and the second
  lepton
  to satisfy $|p_T^{\mr{jj\ell_1}}-p_T^{\ell_2}|<20\, \mr{\gev}$, and the
  azimuthal angles to satisfy 
  $||\phi_{\mr{jj\ell_1}}-\phi_{\ell_2}|-180^0|<15^0$.
  $p_T^{\mr{jj\ell_1}}$ is the transverse momentum of the combination of
  the hardest
  two jets in the event and the lepton closest to the jets in $(\eta,\phi)$
  space, \ie the transverse momentum of the neutralino candidate.
  Similarly $\phi_{\mr{jj\ell_1}}$ is the azimuthal
  angle of the combination of the hardest
  two jets in the event and the lepton closest to the jets in $(\eta,\phi)$
  space, \ie the azimuthal angle of the neutralino candidate.

%
%

  As can be seen in both Figs.\,\ref{fig:lhcsusyneutmasscut}b and 
  \ref{fig:lhcsusyslepmasscut}b this significantly reduces the background and
  the spurious peak in the slepton mass distribution. At these points it is
  also possible to reconstruct the lightest 
  neutralino, chargino and sneutrino masses using the
  the decay chain \mbox{$\mr{\nut\ra\cht^+_1\ell^+}$} followed
  by the decay of the chargino 
  \mbox{$\mr{\cht^+_1\ra\ell^+\nu_\ell\cht^0_1}$}
  and the \rpv\  decay of the lightest neutralino to a lepton and two jets 
  \cite{Moreau:1999bt:Moreau:2000ps:Moreau:2000bs,Abdullin:1999zp}.
  The reconstructed
  neutralino and slepton masses, before and after the imposition of the
  new cuts, are given in Table~\ref{tab:reconstruct}. The same procedure
  as before was used to extract the sparticle masses. There is
  reasonable agreement been the simulated and reconstructed sparticle
  masses  although again the reconstructed values lie between 5
  and $15\, \mr{\gev}$  
  below the values used in the simulations, due to the loss of energy
  from the cones used to define the jets in the neutralino decay.

%
%
\section{Conclusions}
\label{sec:conclusions}

  We have performed a detailed analysis of the background
  to like-sign dilepton
  production at both Run II of the Tevatron and the LHC. We find a background
  from Standard Model processes of $0.43\pm0.16$ events for $2\ \mr{fb}^{-1}$
  integrated luminosity at the Tevatron and $4.9\pm1.6$ events for 
  $10\  \mr{fb}^{-1}$ integrated luminosity at the LHC after a set of
   cuts. If we only consider
  this background there are large regions of SUGRA parameter space where
  resonant slepton production followed by a supersymmetric
  gauge decay of the slepton is
  visible above the SM background even for the small values of the \rpv\  
  couplings we considered.

  This is presumably the strategy which would be adopted in any initial
  experimental search, \ie looking for an excess of a given type of event
  over the Standard Model prediction. If such an excess where
  observed it would
  then be necessary to identify which of the many possible models of beyond
  the Standard Model physics was correct.

  In the \rpv\  MSSM 
  such an excess of like-sign dileptons can come from two
  possible sources, from sparticle pair production followed by the decay
  of the LSP, and from resonant sparticle production. We have considered the
  background to resonant slepton production from sparticle pair production
  and found that after an additional cut on the number of jets the signal
  from resonant slepton production is visible
  above the combined Standard Model
  and SUSY pair production background for large ranges of SUGRA parameter
  space.  Finally we have studied
  the possibility of measuring the mass of the
  resonant slepton and the neutralino into which it decays. Our results
  suggest that this should be possible even if the signal is only just 
  detectable above the background.

%
%
\section*{Acknowledgments}

  We would like to thank the organizers of the Tevatron Run II workshop
  on supersymmetry and Higgs physics for motivating us to start this
  work. We thank R. Thorne for helpful discussions.
  P. Richardson would like to thank PPARC for a research studentship,
  number PPA/S/S/1997/02517.

\end{document}